\documentclass[11pt,preprintnumbers,nofootinbib,amsmath,amssymb]{revtex4}
\usepackage{float}
\usepackage{indentfirst}
\usepackage{amsmath,amssymb,epsfig,subfigure,bm}
\usepackage{graphicx}
\usepackage[export]{adjustbox}
\usepackage{color}
\usepackage{multirow}
\usepackage{booktabs}
\usepackage{changes}
\usepackage{footnote}
\usepackage{mathrsfs} 
\usepackage{subfloat} 
\usepackage{hhline}
\usepackage{mathtools}
\usepackage[shortcuts]{extdash}

\usepackage{hyperref}
\hypersetup{colorlinks=true,linkcolor=blue,citecolor=magenta}

\begin{document}
\renewcommand{\baselinestretch}{1.15}
\makeatletter
\renewcommand{\thesubfigure}{\alph{subfigure}}
\renewcommand{\@thesubfigure}{(\thesubfigure)\hskip\subfiglabelskip}
\makeatother

\title{Boson Stars in Teleparallel Gravity with a Nonminimally Coupled Field: The Violation of Energy Conditions and Gravitational Waveforms from EMRIs}


\preprint{}

\author{Long-Xing Huang$^{a,b}$, Ke Yang$^{c}$, and Yong-Qiang Wang$^{a,b}$\footnote{E-mail: yqwang@lzu.edu.cn, corresponding author}}

\affiliation{$^{a}$ Lanzhou Center for Theoretical Physics, Key Laboratory of Theoretical Physics of Gansu Province, School of Physical Science and Technology, Lanzhou University, Lanzhou 730000, China\\
 $^{b}$ Institute of Theoretical Physics $\&$ Research Center of Gravitation, Lanzhou University, Lanzhou 730000, China\\
$^{c}$ School of Physical Science and Technology, Southwest University, Chongqing 400715, China}


\begin{abstract}

 In this work, we investigate boson star models within the framework of teleparallel gravity with non-minimal coupling, and obtain static, spherically symmetric solutions for both the ground state and excited states. The results indicate that the energy density of the excited-state solutions can become negative. For these solutions, the four commonly used energy conditions are no longer satisfied. In contrast, for all the ground-state solutions we have studied, the energy density remains positive and all four energy conditions are consistently satisfied. Moreover, considering the importance of astrophysical observations, the gravitational-wave signals from Extreme-Mass-Ratio Inspirals (EMRIs) composed of these boson stars are   investigated. Our results reveal that the frequency-domain characteristic strain of these waveforms falls within the detectability range of LISA, which can provide potential evidence for distinguishing compact astrophysical objects.
 
\end{abstract}


\maketitle
\newpage

\section{INTRODUCTION}\label{sec: introduction}

    In the last ten years, with the rapid advancement of technology, two key predictions in general relativity (GR) have been observationally verified. One is the detection of gravitational waves (GWs) from the binary black hole merger~\cite{LIGOScientific:2016aoc}, and the other is the image of black hole shadows~\cite{EventHorizonTelescope:2019dse, EventHorizonTelescope:2019uob,EventHorizonTelescope:2019jan,EventHorizonTelescope:2019ths,EventHorizonTelescope:2019pgp,EventHorizonTelescope:2019ggy}. These landmark observations not only continue the success of GR but also open a new era in the study of strong gravitational fields. In GR, energy conditions  often serve as necessary requirements for the proofs of various important theorems, such as the famous Penrose’s singularity theorem~\cite{Penrose:1964wq,Hawking:1966sx}, black hole area theorem~\cite{Hawking:1971vc, Bardeen:1973gs, wald1994quantum} and positive mass theorem~\cite{Schoen:1979zz,Schon:1979rg,Schon:1982re,Penrose:1993ud}. In fact, from a historical perspective, the original energy conditions~\cite{Penrose:1964wq,Hawking:1966sx} were formulated by physicists driven by the pursuit of proving powerful results~\cite{Kontou:2020bta}.
    

    It should be noted that the Einstein field equations themselves do not impose any constraints on the energy-momentum tensor $T_{\mu\nu}$. The energy conditions impose certain restrictions on the energy-momentum tensor (i.e., on matter), which exclude unphysical solutions with exotic phenomena including closed timelike curves~\cite{Lobo2003}, faster-than-light travel~\cite{Morris:1988tu}, etc.
    
    Currently, there exist several different energy conditions~\cite{Kontou:2020bta}, each presenting distinct advantages and disadvantages regarding its scope of applicability, physical implications, and interpretation. However, precisely for this reason, there is a lack of a single preferred energy condition, which has drawn certain criticisms~\cite{Barcelo:2002bv,Curiel:2014zba}, particularly as physicists often invoke conditions that appear arbitrary and ad hoc, tailored specifically for certain proofs. Despite these shortcomings, energy conditions are still considered indispensable and continue to receive extensive attention in GR. Among these, the four most widely used energy conditions are the null (NEC), weak (WEC), strong (SEC), and dominant (DEC) energy conditions~\cite{Hawking1973,wald1984}.

    Going beyond standard GR, numerous modified gravity theories have been proposed. These developments are motivated both by the need to explain cosmological observations~\cite{SupernovaSearchTeam:1998fmf,SupernovaCosmologyProject:1998vns,Bull:2015stt,Koyama:2015vza,Nojiri:2017ncd,CANTATA:2021asi}, and by the persistent ambition to refine our fundamental physical theories~\cite{Petrov:2020wgy,Odintsov:2022cbm,Shankaranarayanan:2022wbx,Yunes:2024lzm}.
    One common way to extend GR is to introduce a nonlinear function of the scalar curvature $R$ into the Einstein–Hilbert action, and the resulting modified theories of gravity are known as $f(R)$ gravity theories~\cite{Sotiriou:2008rp}. In addition, it is also well known that one can construct a theory of gravity fully equivalent to GR by using torsion $T$. In this formulation, the tetrad field serves as the fundamental dynamical degree of freedom, and the resulting theory is known as teleparallel gravity~\cite{Maluf:2013gaa}. Similar to $f(R)$ gravity theories, a nonlinear function of $T$ can also be used to modify the teleparallel gravity theory. The resulting $f(T)$ gravity theory~\cite{Cai:2015emx,Bahamonde:2021gfp} can be used to explain the current problem of cosmic expansion~\cite{Ferraro:2006jd}.
    
    In 2014, D.~Horvat et al.~\cite{Horvat:2014xwa} extended this framework by investigating a massive complex scalar field non-minimally coupled to the torsion scalar $T$. A distinctive feature of this model is that the non-minimal coupling breaks local Lorentz invariance, making the equations of motion tetrad-dependent. By identifying a specific tetrad for a static spherically symmetric spacetime that leads to self-consistent equations, the authors constructed boson star configurations that satisfy the DEC. Notably, they observed that at sufficiently large coupling, the energy density increases radially before a sharp drop toward its asymptotic tail. Such a feature is absent in models where the scalar field is coupled to the curvature scalar. Generally, boson stars~\cite{Kaup:1968zz,Ruffini:1969qy} are exotic compact objects that serve as black hole mimickers~\cite{Guzman:2009zz,Bambi:2025wjx} (see~\cite{Schunck:2003kk,Liebling:2012fv} for reviews), exhibiting various configurations characterized by the node number $n$. The study in Ref.~\cite{Horvat:2014xwa}, however, was restricted to the case of nodeless ground-state configurations.
    
    In this work,  we extend the analysis to encompass both ground-state and excited-state boson stars ($n \geq 0$) within the framework of teleparallel gravity, exploring a broader parameter space that includes both positive and negative values of the field-to-torsion coupling parameter $\xi$. In contrast to the ground state results reported in Ref.~\cite{Horvat:2014xwa}, we find that for excited states at certain scalar field frequencies, a large non-minimal coupling can lead to negative energy densities, resulting in the violation of all four standard energy conditions. Furthermore, to investigate the observational signatures that distinguish these teleparallel boson stars from their GR counterparts, we examine the gravitational-wave signals emitted from extreme-mass-ratio inspirals (EMRIs) where these boson stars serve as the central compact objects.

    This paper is organized as follows. In Section~\ref{sec: model}, we introduce the general framework of the nonminimally coupled scalar field in teleparallel gravity. We then present numerical solutions for the field equations in Section~\ref{sec: results}. Subsequently, Section~\ref{sec:emri} explores the gravitational-wave signatures of EMRIs consisting of a central boson star. Finally, Section~\ref{sec:conclusion} provides a brief summary of our main findings. 
    
\section{THE GENERAL FRAMEWORK}\label{sec: model}
\subsection{The action and field equations}
This section provides a brief setup of a complex scalar field $\Phi$ nonminimally coupled to the torsion scalar within the framework of teleparallel gravity. Following Ref.~\cite{Horvat:2014xwa}, we adopt the simplest form of non-minimal coupling, which closely resembles the scalar-curvature coupling in GR~\cite{vanderBij:1987gi}. This form has been widely employed in recent literature, e.g., Ref.~\cite{Geng:2011aj,Wei:2011yr,Otalora:2013tba,Kehal:2023rhc}. The action of the model, in units of $\hbar = c = G = 1$, is expressed as:
    \begin{equation}
        S=\int \mathrm{d}^4 x h\left[-\frac{T}{2\kappa}- \xi \Phi^* \Phi T+\mathcal{L}_\text{M}\right],
  \label{eq:action}
    \end{equation}	
where $\kappa=8\pi$, $\mathcal{L}_\text{M}=-\frac{1}{2} g^{\alpha \beta}\left(\Phi_{, \alpha}^* \Phi_{, \beta}+\Phi_{, \beta}^* \Phi_{, \alpha}\right)-\mu^2 \Phi^* \Phi$, $\xi$ is the field-to-torsion coupling parameter, and $h$ is the determinant of the tetrad $h_{\mu}^{a}$, in which the Latin and Greek indices correspond to the Lorentz frame and spacetime coordinates, respectively. The metric and the tetrad fields satisfy the following relations:
\begin{equation}
    g_{\mu\nu}=h^{a}{}_{\mu}h^{b}{}_{\nu}\eta_{ab},\quad \eta_{ab}=h_{a}{}^{\mu}h_{b}{}^{\nu}g_{\mu\nu},\quad h^{a}{}_{\mu}h_{b}{}^{\mu}=\delta^a_b,\quad h^{a}{}_{\mu}h_{a}{}^{\nu}=\delta_\mu^\nu,
\end{equation}
in which $\eta_{ab}=\rm{diag}(-1,1,1,1)$ is the Minkowski metric and $h_a{}^{\mu}$ is the inverse of the tetrads. $T$ is the torsion scalar, and it can be written as
\begin{equation}
    T =\frac{1}{2} S_{\alpha}{}^{\beta\gamma} T^{\alpha}{}_{\beta\gamma},
\end{equation}
where $T^{\alpha}{}_{\beta\gamma}$ is the torsion tensor defined by the tetrad fields and the inertial spin connection $\omega^a{}_{b\alpha}$~\cite{Krssak:2015oua}:
\begin{equation}
    T^{\alpha}{}_{\beta\gamma}=h_a{}^{\alpha}\left(\partial_{\beta} h^a{}_{\gamma}-\partial_{\gamma} h^a{}_{\beta}+\omega^a{}_{b\beta}h^b{}_{\gamma}-\omega^a{}_{b\gamma}h^b{}_{\beta}\right).
\end{equation}
The spin connection can be obtained by requiring each component of the torsion tensor tends to zero in the flat space limit (see Ref.~\cite{Krssak:2015rqa} for more technical details). It plays a crucial role in compensating for inertial effects~\cite{Krssak:2018ywd} and is essential for maintaining invariance under local Lorentz transformations and ensuring the frame-independence of the entire theory~\cite{Krssak:2015oua}. The tensor $S_{\alpha}{}^{\beta\gamma}$ is the so-called superpotential torsion tensor, which can be expressed in terms of the torsion tensor $T^{\alpha}{}_{\beta\gamma}$ and the contorsion tensor $K^{\alpha}{}_{\beta\gamma}=\frac{1}{2}(T_{\beta}{}^\alpha{}_\gamma+T_{\gamma}{}^\alpha{}_\beta-T^\alpha{}_{\beta\gamma})$:
\begin{equation}
    S_{\alpha}{}^{\beta\gamma}=K^{\beta\gamma}{}_\alpha+\delta_{\alpha}{}^{\beta}T^{\sigma\gamma}{}_{\sigma}-\delta_{\alpha}{}^{\gamma}T^{\sigma\beta}{}_{\sigma}.
\end{equation}

The Euler–Lagrange equation can be obtained by varying the action (\ref{eq:action}) with respect to the tetrad $h_{\mu}^{a}$ as~\cite{Horvat:2014xwa}:
\begin{align}
 &\partial_\mu\left[\frac{h}{2 \kappa}\left(1+2 \kappa \xi \Phi^* \Phi\right)\left(2 S_a{}^{\nu \mu}\right)\right]=h_a^\nu h\left[\mathcal{L}_\text{M}-\left(1+2\kappa \xi \Phi^* \Phi\right)\frac{T}{2\kappa}\right]\nonumber\\
 &+\frac{h}{2 \kappa}\left(1+2 \kappa \xi \Phi^* \Phi\right)\left(2 T_{\alpha \beta a} S^{\alpha \beta \nu}\right)+\frac{h}{2}\left(g^{\nu \beta} h_a^\alpha+g^{\nu \alpha} h_a^\beta\right)\left(\Phi_{, \alpha}^* \Phi_{, \beta}+\Phi_{, \beta}^* \Phi_{, \alpha}\right).
\end{align}
It is worth noting that by contracting with the tetrad, the above  equation can be rewritten in the form of the ``Einstein equation":
\begin{equation}
    G^\mu{}_\nu=\kappa \Theta^\mu{}_\nu,
    \label{eq:Einsteineq}
\end{equation}
where  
\begin{equation}
    G^\mu{}_\nu=\frac{1}{2} T \delta^\mu{}_\nu-S_{\alpha}{}^{\beta\mu}T^{\alpha}{}_{\beta\nu}+\frac{1}{h}h^a{}_\nu\partial_\rho(hS_a{}^{\mu\rho}),
\end{equation}
is exactly equal to Einstein tensor $R^\mu{}_\nu-\frac{1}{2} R \delta^{\mu}{}_{\nu}$ ($R$ denotes Ricci scalar and $R^\mu{}_\nu$ is Ricci tensor), and 
\begin{equation}
\Theta_{\mu \nu}=\frac{\Phi_{, \mu}^* \Phi_{, \nu}+\Phi_{, \nu}^* \Phi_{, \mu}+g_{\mu \nu}\mathcal{L}_\text{M}}{\left(1+2 \kappa \xi \Phi^* \Phi\right)},
\label{eq:emtensor}
\end{equation}
is the energy–momentum tensor.

Furthermore, the variation of the action with respect to the scalar field leads to the equation for the scalar field
\begin{equation}
    \nabla^2\Phi-(\xi T+\mu^2)\Phi=0.
    \label{eq:KGeq}
\end{equation}

From equations~(\ref{eq:Einsteineq}),~(\ref{eq:emtensor}) and~(\ref{eq:KGeq}), one can see that for the case of $\xi=0$, the obtained field equations reduce to the minimal coupling scenario and are equivalent to the field equations for mini boson stars in GR~\cite{Kaup:1968zz}. The word ``mini" is used because for typical masses of bosonic particle candidates, the mass of such boson stars without self-interaction is much smaller than that of fermionic stars like neutron stars and white dwarfs, which are supported by Fermi degeneracy pressure~\cite{Liebling:2012fv}. 

Given that the action is invariant under global $U(1)$ transformations of the scalar field, the system possesses a conserved Noether current, defined by $j^{\alpha}=ig^{\alpha\beta}\left[(\nabla_\beta\Phi^*)\Phi-(\nabla_\beta\Phi)\Phi^*\right]$. The corresponding charge $Q$ can be interpreted as the particle number and can be obtained by integrating the timelike component of $j^{\alpha}$ on a spacelike slice, i.e.
\begin{equation}
    Q=\int_\Sigma j^t.
    \label{eq:noetherchargedo}
\end{equation} 
\subsection{The ansatz}

In this paper, we consider a static, spherically symmetric configuration. The corresponding spacetime metric can be assumed in the following form
\begin{equation}
    \mathrm{d}s^2=-\mathrm{e}^{2F_0(r)}\mathrm{d}t^2+\mathrm{e}^{2F_1(r)}\mathrm{d}r^2+r^2\left(d \theta^2+\sin ^2 \theta \mathrm{d} \varphi^2\right),
        \label{eq:ansatzmetric}
\end{equation}
where the functions $F_0(r)$ and $F_1(r)$ are two metric profile functions dependent only on the radius $r$. 

For this line element~(\ref{eq:ansatzmetric}), we adopt the ``rotated tetrad"~\cite{Boehmer:2011gw}
\begin{equation}
h^a{}_\mu = 
\begin{pmatrix}
e^\Phi & 0 & 0 & 0 \\
0 & e^\Lambda \sin\vartheta \cos\varphi & r \cos\vartheta \cos\varphi & -r \sin\vartheta \sin\varphi \\
0 & e^\Lambda \sin\vartheta \sin\varphi & r \cos\vartheta \sin\varphi & r \sin\vartheta \cos\varphi \\
0 & e^\Lambda \cos\vartheta & -r \sin\vartheta & 0
\end{pmatrix}.
\label{eq:tetrad}
\end{equation}
It can be found that the inertial spin connection $\omega^a{}_{b\alpha}=0$ for this tetrad~\cite{Krssak:2015oua}. It is worth noting that in early studies, before the significance of the spin connection was fully recognized, the use of certain tetrads, such as the diagonal tetrad $h^a{}_\mu = \operatorname{diag}(\mathrm{e}^{F_0},\mathrm{e}^{F_1},r,r\sin\theta)$, would led to unwanted off-diagonal components in the field equations, thereby imposing redundant constraints. Consequently, these were frequently labeled ``bad tetrads'' in the literature~\cite{Ferraro:2011us,Meng:2011ne,Boehmer:2011gw,Ferraro:2011ks}. In contrast, the rotated tetrad avoids such issues and is often referred to as a ``good tetrad''~\cite{Tamanini:2012hg}.

In addition, for the scalar field $\Phi$, we employ the harmonic ansatz 
\begin{equation}
     \Phi=\frac{1}{\sqrt{\kappa}}\phi(r)\mathrm{e}^{-i\omega t},
    \label{eq:ansatzphi}
\end{equation}
here $\phi(r)$ is a real radial function, and $\omega$ is a real constant corresponding to the oscillation frequency of the scalar field.

Substituting the tetrad~(\ref{eq:tetrad}) and the ansatz~(\ref{eq:ansatzmetric}),~(\ref{eq:ansatzphi}) into motion equations~(\ref{eq:Einsteineq}) and~(\ref{eq:KGeq}), yields the following reduced system of coupled ordinary differential equations (ODEs):
\begin{gather}
    \frac{e^{-2F_1}(1-2rF_1^\prime)-1}{r^2}+\kappa \rho = 0, \label{eq:motioneq1} \\
    \frac{e^{-2F_1}(1+2rF_0^\prime)-1}{r^2}-\kappa p_r = 0, \label{eq:motioneq2} \\
    \frac{e^{-2F_1}\bigl((rF^\prime_{0}-rF^\prime_{1})(1+rF^\prime_{0})+r^2F^{\prime\prime}_{0}\bigr)}{r^2}-\kappa p_\bot = 0, \label{eq:motioneq3} \\
    \phi^{\prime \prime}+\frac{2+F^{\prime}_0-F^{\prime}_1}{r}\phi^\prime+\Bigl[\frac{4\xi(1+e^{F_1})F^{\prime}_0}{r}+\frac{2\xi(1-2e^{F_1}+e^{2F_1})}{r^2}+(\omega^2e^{-2F_0}-\mu^2)e^{2F_1}\Bigr]\phi = 0, \label{eq:motioneq4}
\end{gather}
with the energy density 
\begin{equation}
    \rho=-\Theta^t{}_t=\frac{\left(\mathrm{e}^{-2 {F_0}} \omega^2+\mu^2\right) \phi^2+\mathrm{e}^{-2 {F_1}} \phi^{\prime 2}-8 \xi r^{-1} \mathrm{e}^{-2F_1}\left(\mathrm{e}^{{F_1}}-1\right) \phi \phi^{\prime}}{\kappa\left(1+2 \xi \phi^2\right)},\label{eq:rho}
\end{equation}
the radial pressure
\begin{equation}
    p_r=\Theta^r{}_r=\frac{\left(\mathrm{e}^{-2 {F_0}} \omega^2-\mu^2\right) \phi^2+\mathrm{e}^{-2 {F_1}} \phi^{\prime 2}}{\kappa\left(1+2 \xi \phi^2\right)}
\end{equation}
and the transverse pressure
\begin{equation}
p_{\bot}=\Theta^{\vartheta}{}_{\vartheta}=\Theta^{\varphi}{}_{\varphi}=\frac{\left(\mathrm{e}^{-2{F_0}} \omega^2-\mu^2\right) \phi^2-\mathrm{e}^{-2 {F_1}} \phi^{\prime 2}+4 \xi r^{-1} \mathrm{e}^{-2 {F_1}}\left(\mathrm{e}^{F_1}-1-rF_0^{\prime}\right)\phi\phi^{\prime}}{\kappa\left(1+2 \xi \phi^2\right)}.
\end{equation}
As for the Noether charge~(\ref{eq:noetherchargedo}), it reads
\begin{equation}
    Q= \frac{8\pi}{\kappa}\int^{\infty}_0 \omega r^2 \phi(r)^2\mathrm{e}^{F_1-F_0}\mathrm{d}r.
    \label{eq:particlenumber2}
\end{equation}

\subsection{The boundary conditions and numerical method}
\label{sec: bound}
In order to solve the system of ODEs~(\ref{eq:motioneq1}) - (\ref{eq:motioneq4}), appropriate boundary conditions should be provided. They can be derived from the assumptions of regularity and asymptotic flatness of the solution. At spatial infinity ($r\rightarrow \infty$), the metric functions, $F_0(r)$, $F_1(r)$, and scalar function $\phi(r)$ satisfy: 
\begin{equation}
    F_0(\infty)=0,\quad F_1(\infty)=0,\quad \phi(\infty)=0,
\end{equation}
and at the origin ($r\rightarrow0$), the scalar function $\phi(r)$ satisfies: 
\begin{equation}
    \phi^{\prime}(0)=0.
\end{equation}

It is worth noting that the total mass $M$ can be derived from the asymptotic sub-leading behavior of the metric functions:
\begin{equation}
    1/g_{rr}(\infty)=e^{-2F_1(\infty)}=-1+\frac{2M}{r}.
\end{equation}

In order to facilitate numerical computations, we employ the following scaling transformations to render the variables dimensionless. 
\begin{equation}
    r\rightarrow \frac{r}{\mu},\quad \omega\rightarrow\mu\omega,\quad M\rightarrow\frac{M}{\mu},\quad Q\rightarrow \frac{Q}{\mu^2}.
\end{equation}
In the numerical implementation, this scaling is equivalent to taking $\mu=1$ in the motion equation. Then, the solution is controlled only by the number of the node $n$, frequency $\omega$ and the field-to-torsion coupling parameter $\xi$.

Moreover, to enable numerical treatment over the entire spatial domain, we introduce a new radial coordinate
\begin{equation}
    \bar{r}=\frac{r}{r+1},
\end{equation}
which maps the radial coordinate range from the semi-infinite region $\left[0,\infty\right)$ to the unit interval $\left[0, 1\right]$. After obtaining the numerical solution, the inverse transformation $r=\frac{\bar{r}}{1-\bar{r}}$ can be used to replace the $\bar{r}$ coordinates with the $r$ coordinates. 

We numerically solve the system of ODEs~(\ref{eq:motioneq1}) - (\ref{eq:motioneq4}) using the finite element method, with $1000$ grid points distributed over the integration interval $[0, 1]$. The Newton–Raphson method is used as the iterative scheme, and the relative error is required to be below $10^{-5}$ to ensure the accuracy of the computed results. The corresponding convergence test is presented in the appendix~\ref{apA}.

\section{NUMERICAL RESULTS}\label{sec: results}
\begin{figure}[!htbp]
		\begin{center}
		\subfigure[~$-g_{tt}$]{ 
			\includegraphics[width=6.5cm]{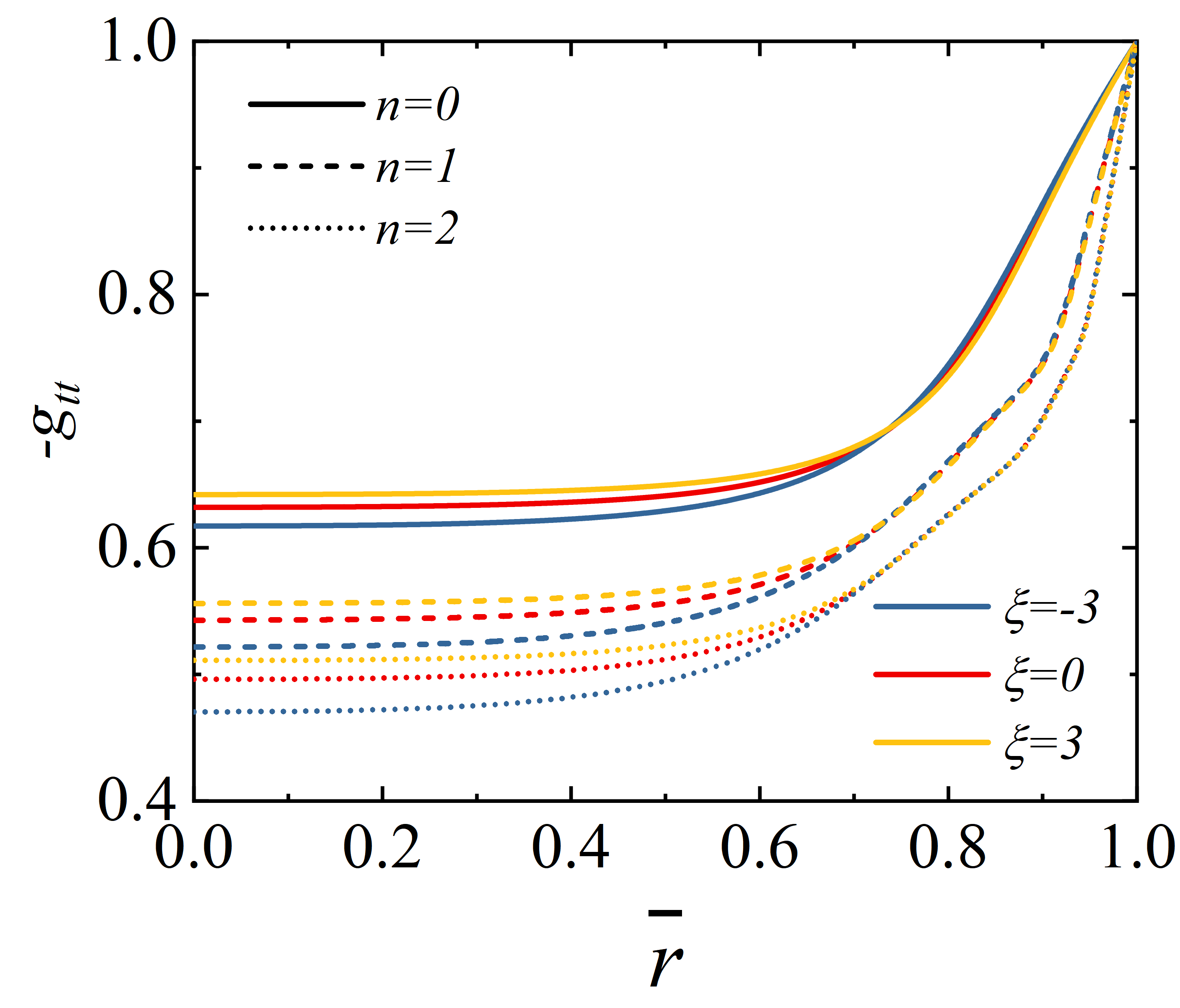}
			\label{fig:omeg09gtt}
		}	 
  		\subfigure[~$1/g_{rr}$]{  
			\includegraphics[width=6.5cm]{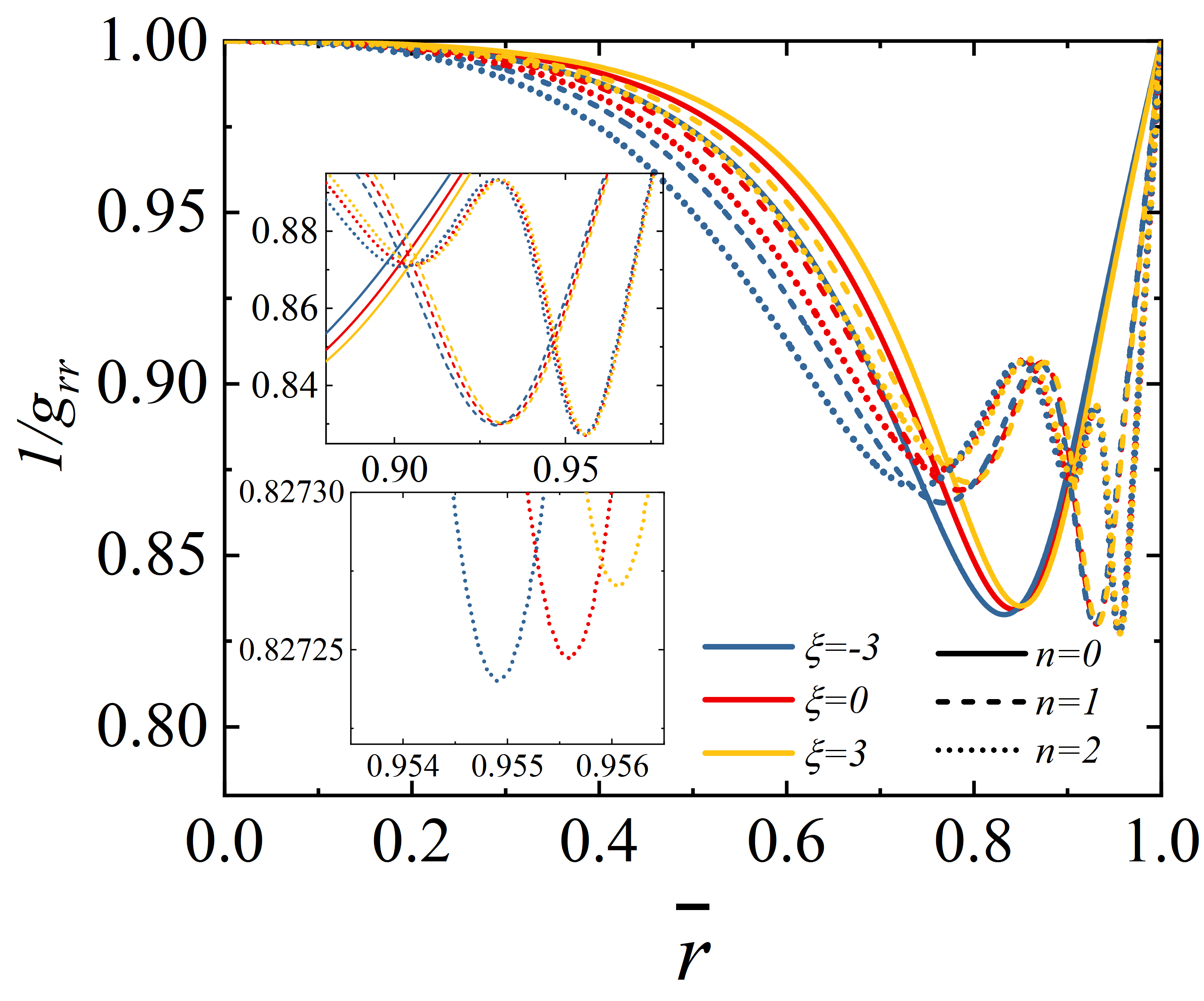}
			\label{fig:omeg09grr}
		}	
    \quad
        \subfigure[~$\phi$]{ 
			\includegraphics[width=6.5cm]{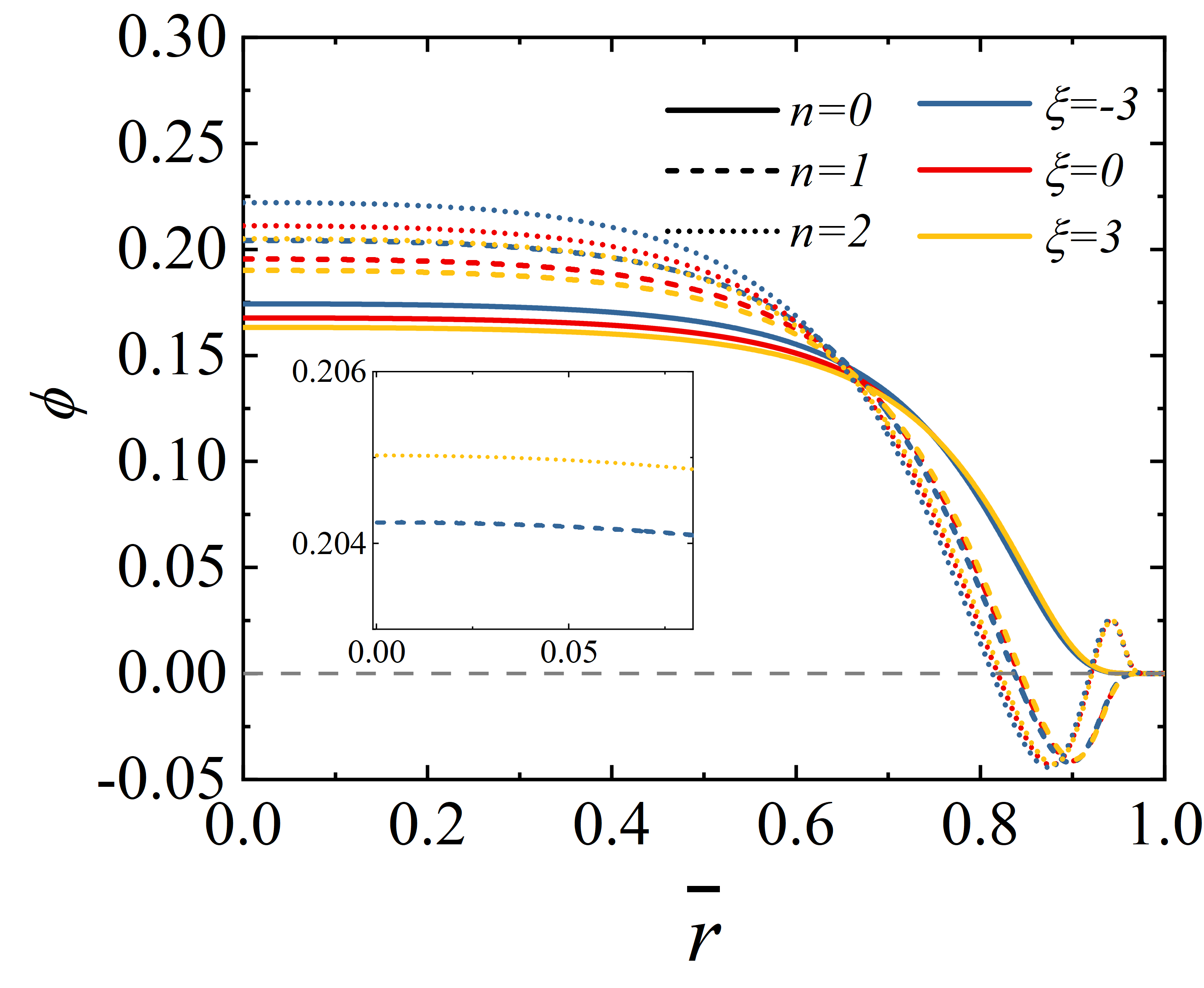}
			\label{fig:omega09phi}
		}	 	
  		\end{center}	
		\caption{The metric function $-g_{tt}=\mathrm{e}^{2F_0}$, $1/g_{rr}=\mathrm{e}^{-2F_1}$, and scalar field function $\phi$ as a function of $\bar{r}$ from the ground state (solid lines) to second excited state (dotted lines). All solutions have $\omega=0.9$. The horizontal axis $\bar{r}=r/(1+r)$ is the radial coordinate after the compactification transformation. }
	\label{fig:omega09function}	
		\end{figure}

In this section, we will present our numerical results. The numerical solutions of the metric function $-g_{tt}=\mathrm{e}^{2F_0}$ (top left panel), $1/g_{rr}=\mathrm{e}^{-2F_1}$ (top right panel) and the complex scalar field $\phi$ (bottom panel) for different values of the coupling parameter $\xi$ are shown in Fig.~\ref{fig:omega09function}. In each panel, the solid, dashed, and dotted lines represent the ground state, the first excited state, and the second excited state, respectively. For all solutions shown, the frequency of the scalar field is fixed at $\omega=0.9$.

As shown in Fig.~\ref{fig:omega09function}, as the node number $n$ increases or the coupling parameter $\xi$ decreases, the central value $\phi(0)$ increases and the field decays more rapidly, indicating enhanced radial confinement. For the ground state, the scalar field decreases monotonically toward the vacuum. In contrast, due to the presence of nodes, the excited states exhibit non-monotonic behavior with multiple local extrema. These structural differences in the scalar field directly affect the metric functions. Specifically, the metric component $g_{rr}$ exhibits a single minimum for the ground state, whereas it possesses multiple local minima for the excited states.

\begin{figure}[!htbp]
		\begin{center}
		\subfigure{ 
			\includegraphics[width=6.5cm, angle =0]{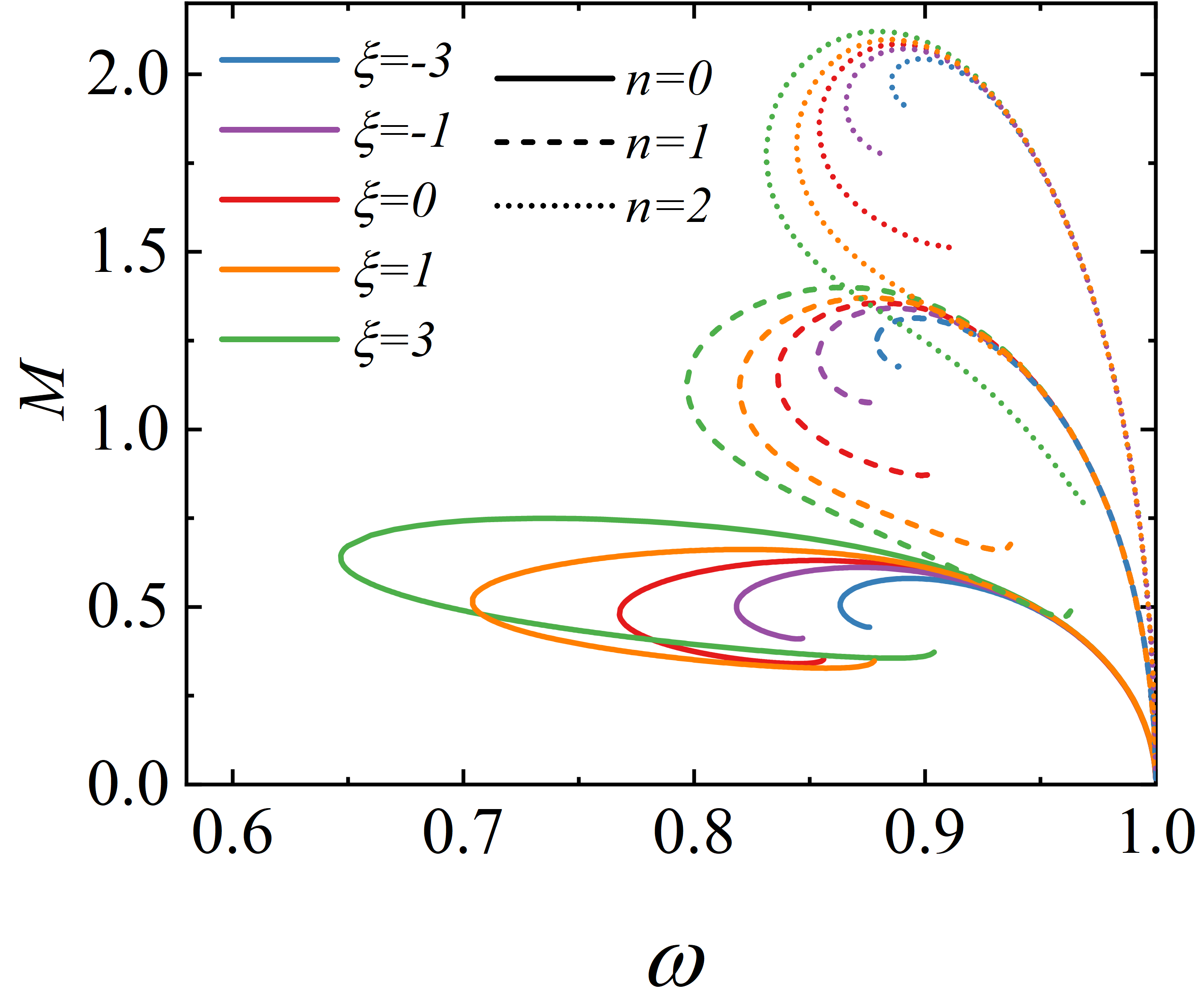}
			\label{fig:massm3top3}
		}	 
  		\subfigure{  
			\includegraphics[width=6.5cm, angle =0]{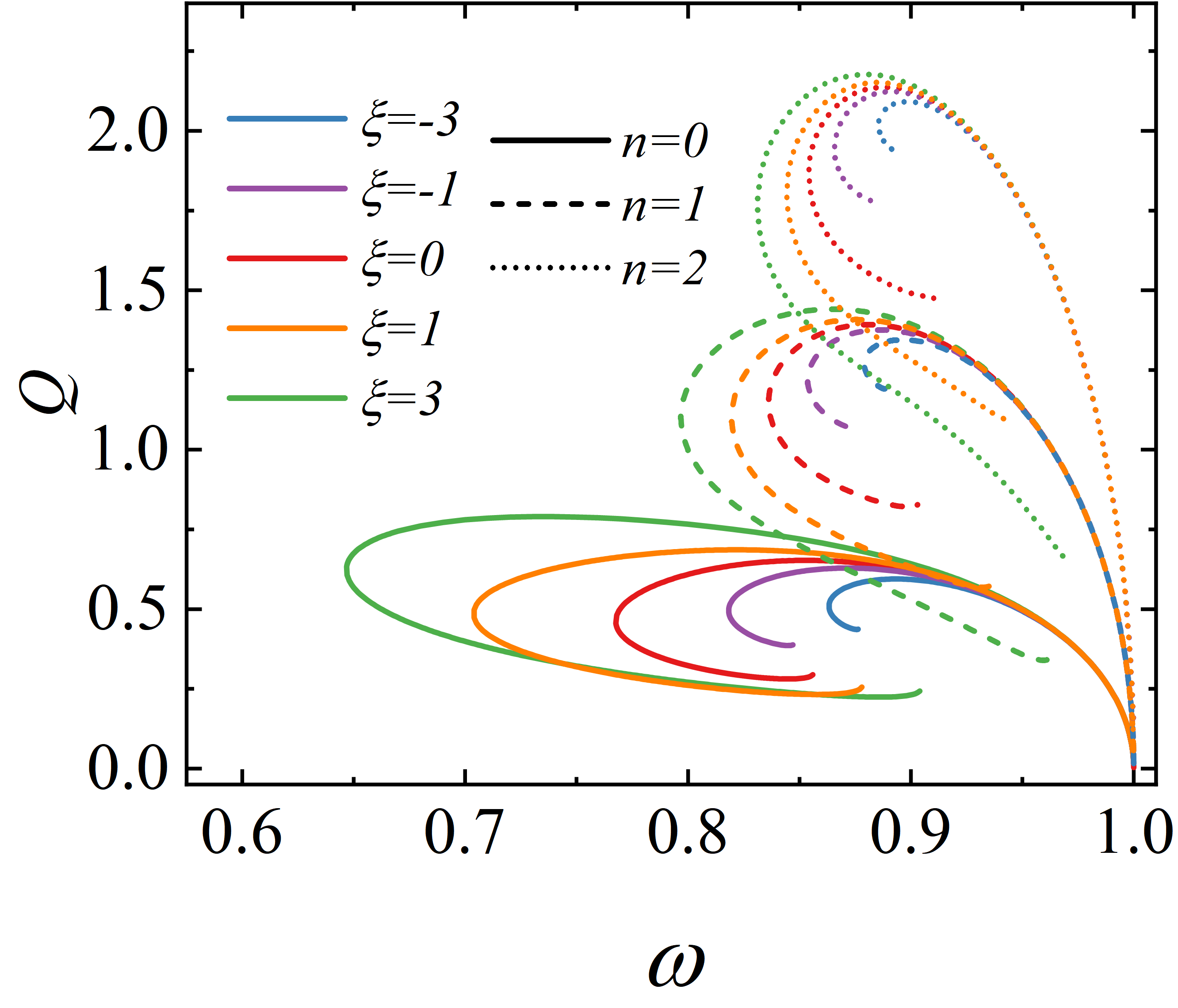}
			\label{fig:chargem3top3}
		}	 	
  		\end{center}	
		\caption{The mass $M$ and Noether charges $Q$ versus the frequency $\omega$ with the different values of the coupling parameter $\xi$ from the ground state (solid line) to second excited state (dotted line).}
	\label{fig:massandcharged}	
		\end{figure}

In Fig.~\ref{fig:massandcharged}, we show the mass $M$ and the Noether charge $Q$ of the boson stars versus the frequency $\omega$ for several values of $\xi$. For the solutions with $\xi = 0$ in the figure, as discussed in section~\ref{sec: model}, they are equivalent to the mini-boson stars in standard GR since their governing equations are identical. From these figures, many features can be observed. First, different coupling parameters $\xi$ correspond to a family of spiral curves. For any given one of these curves, the frequency $\omega$ of these solutions decreases from $\omega = 1$ to a minimum value, forming the first branch. Subsequently, the curve exhibits a backbending behavior, which gives rise to the second branch. Second, for each $n$, the larger $\xi$ broadens the solution space, and the lower $\xi$ narrows the solution space. Finally, for the same $\xi$ and $n$, both the mass $M$ and charge $Q$ possess a maximum value, and this maximum value increases as the coupling parameter $\xi$ or the number of nodes $n$ increases.

\begin{figure}[!htbp]
		\centering		
		\subfigure{  
			\includegraphics[height=.18\textheight,width=.22\textheight, angle =0]{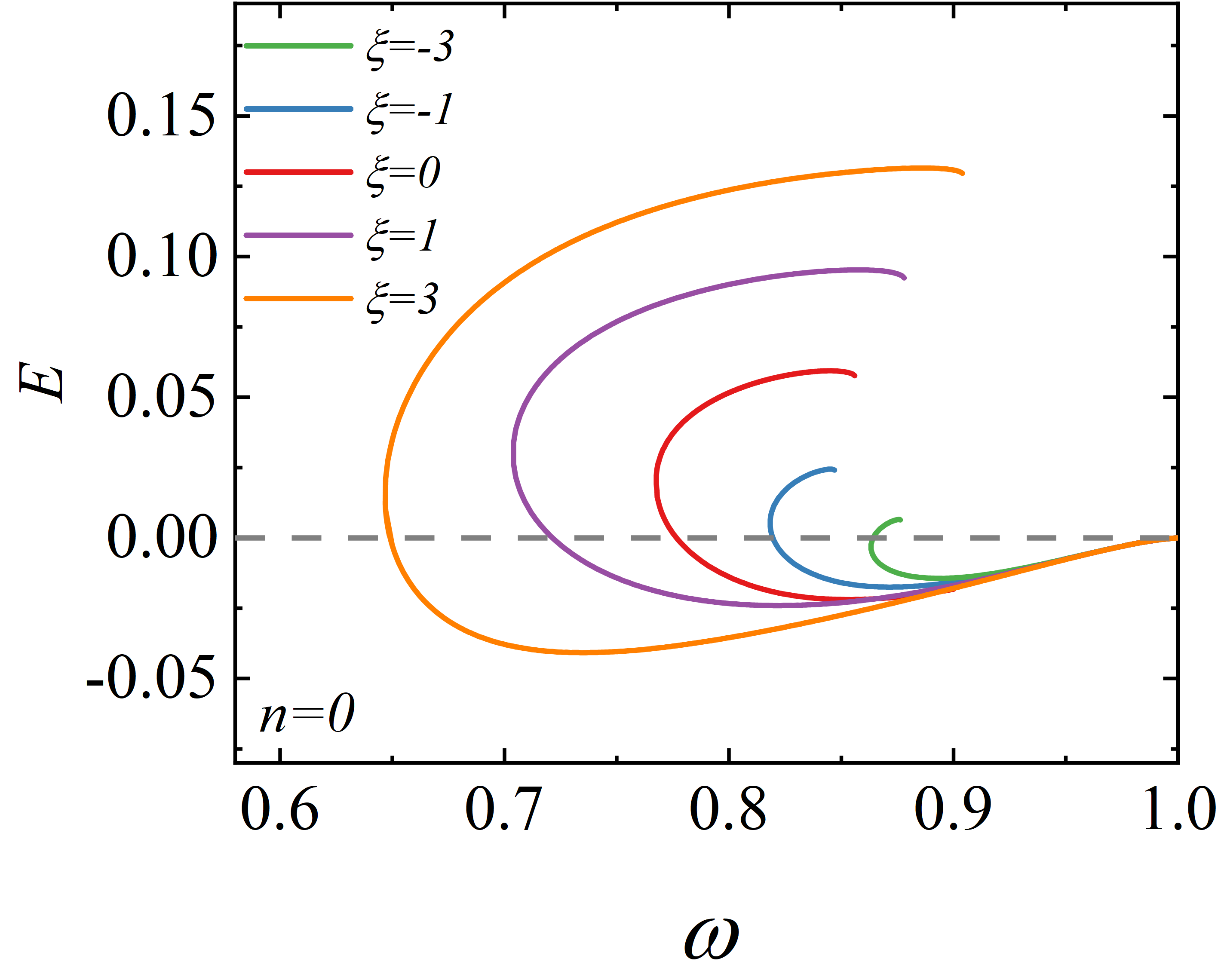}
		}
		\subfigure{
			\includegraphics[height=.18\textheight,width=.22\textheight, angle =0]{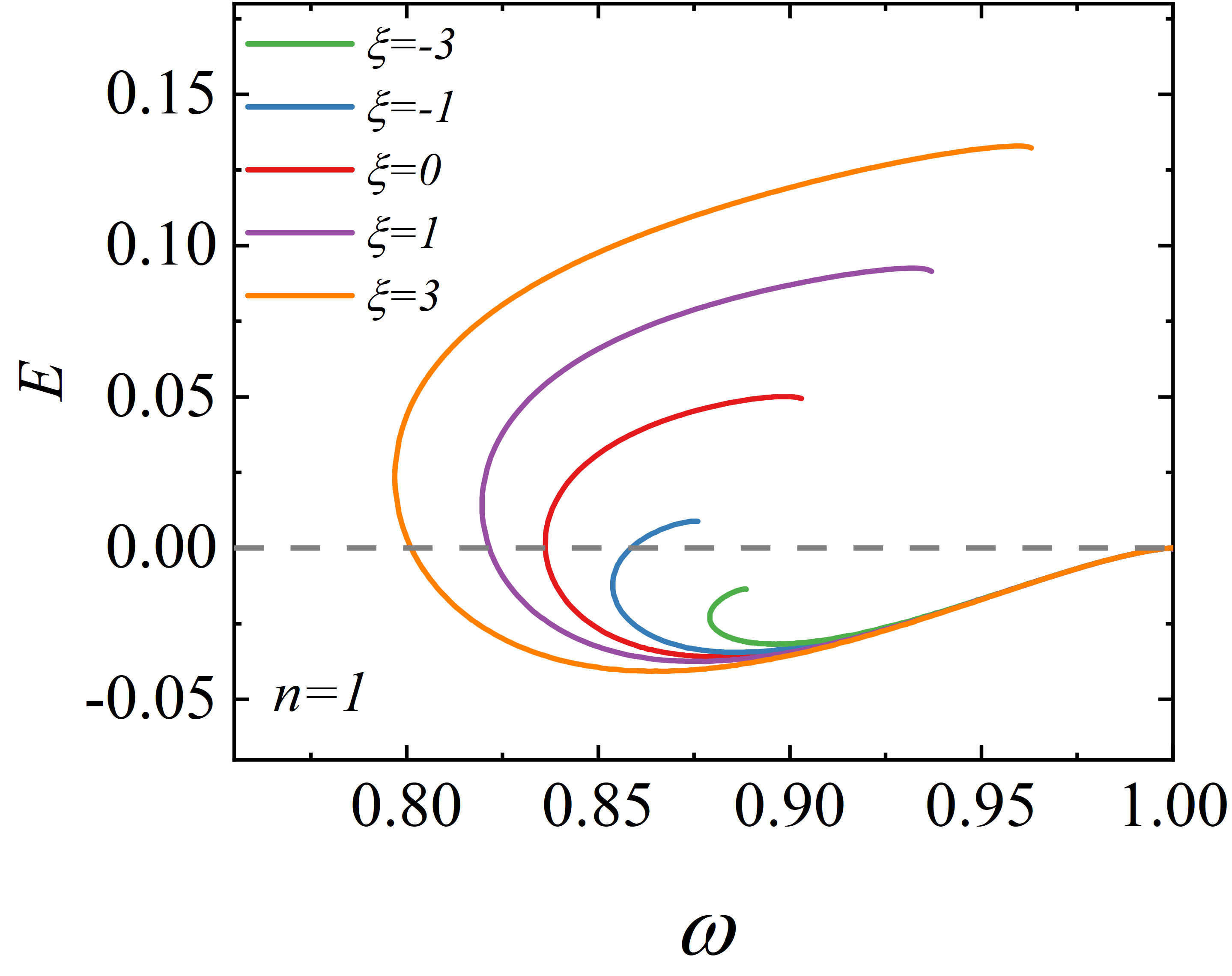}
   		} 	
     		\subfigure{
			\includegraphics[height=.18\textheight,width=.22\textheight, angle =0]{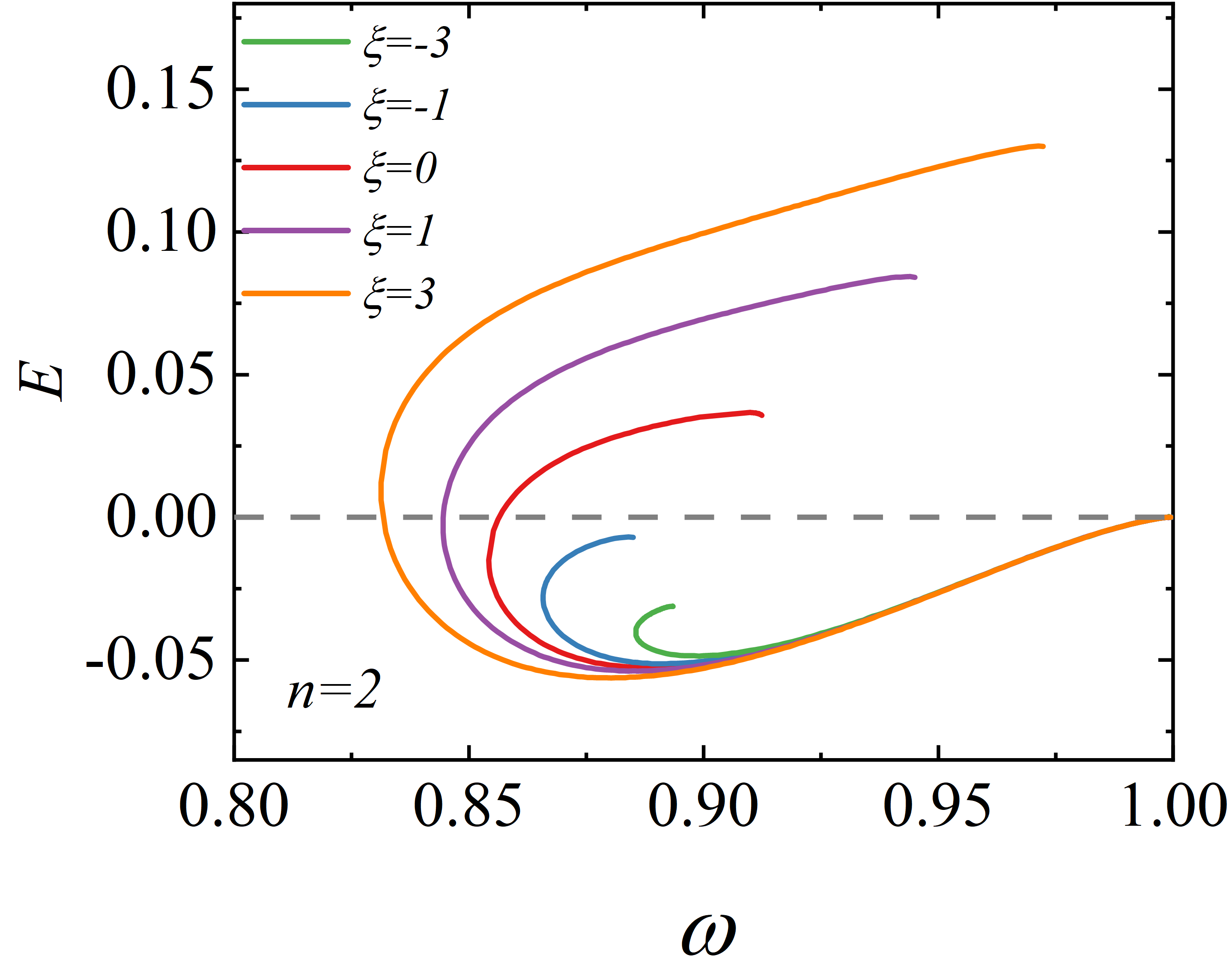}
   		} 	
		\caption{Binding energy $E$ curves as a function of scalar field frequency $\omega$ for different coupling parameter $\xi$ and number of node $n$.}
		\label{fig:BE}		
		\end{figure}

The binding energy $E = M - \mu Q$ is commonly employed in the analysis of the stability of boson stars in early studies~\cite{Schunck:2003kk}. The negative binding energy ($E < 0$) implies that the total mass of the star is less than the sum of the rest masses of its constituent particles, meaning the system is in a bound 
state and stable against fission. As seen in Fig.~\ref{fig:BE}, it can be found that increasing the coupling parameter $\xi$ broadens the frequency range of stable solutions from the perspective of binding energy. Starting from $\omega=1$, the differences in the binding energy of boson stars under different coupling parameters are small at larger frequencies, and gradually increase as the frequency decreases. It is worth noting that the boson stars satisfying the binding energy stability mainly exist in the first branch, while the second branch does not satisfy this condition for most parameters (even when it does, the parameter space for such solutions is relatively small). For many other boson star models, various stability analyses have shown that stable solutions typically exist only in the first branch, whereas the higher branch is unstable — see e.g., Ref.~\cite{Gleiser:1988ih,Seidel:1990jh,Balakrishna:1997ej}. Therefore, in this work, unless otherwise specified, we will only use the first-branch solutions as examples to demonstrate and explain.

The presence of the field-to-torsion coupling parameter also modulates the compactness of boson stars. Since the scalar field extends to infinity, there is no unique definition of the ``radius" of boson stars. The size of a boson star is usually defined by the radius $R_{99}$, which is the radius of the sphere enclosing 99\% of the total mass, $M_{99}$, of the star. Based on this definition of size, the compactness $C$ can be defined as:
\begin{equation}
    C=\frac{M_{99}}{R_{99}}.
\end{equation}  

\begin{figure}[!htbp]
		\centering		
		\subfigure[~$n=0$]{  
			\includegraphics[height=.18\textheight,width=.22\textheight, angle =0]{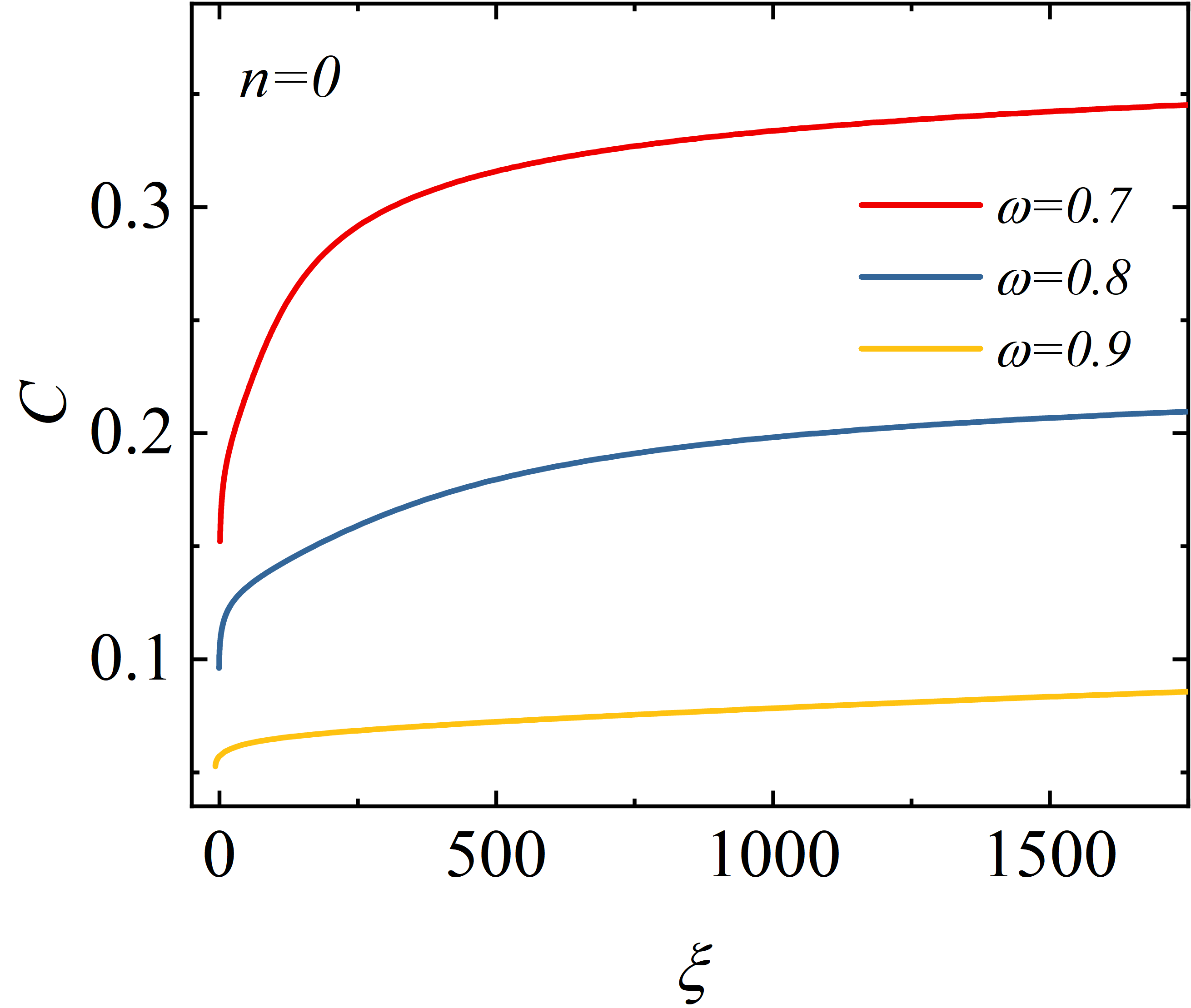}
            \label{fig:compactforkxin0}
		}
		\subfigure[~$n=1$]{
			\includegraphics[height=.18\textheight,width=.22\textheight, angle =0]{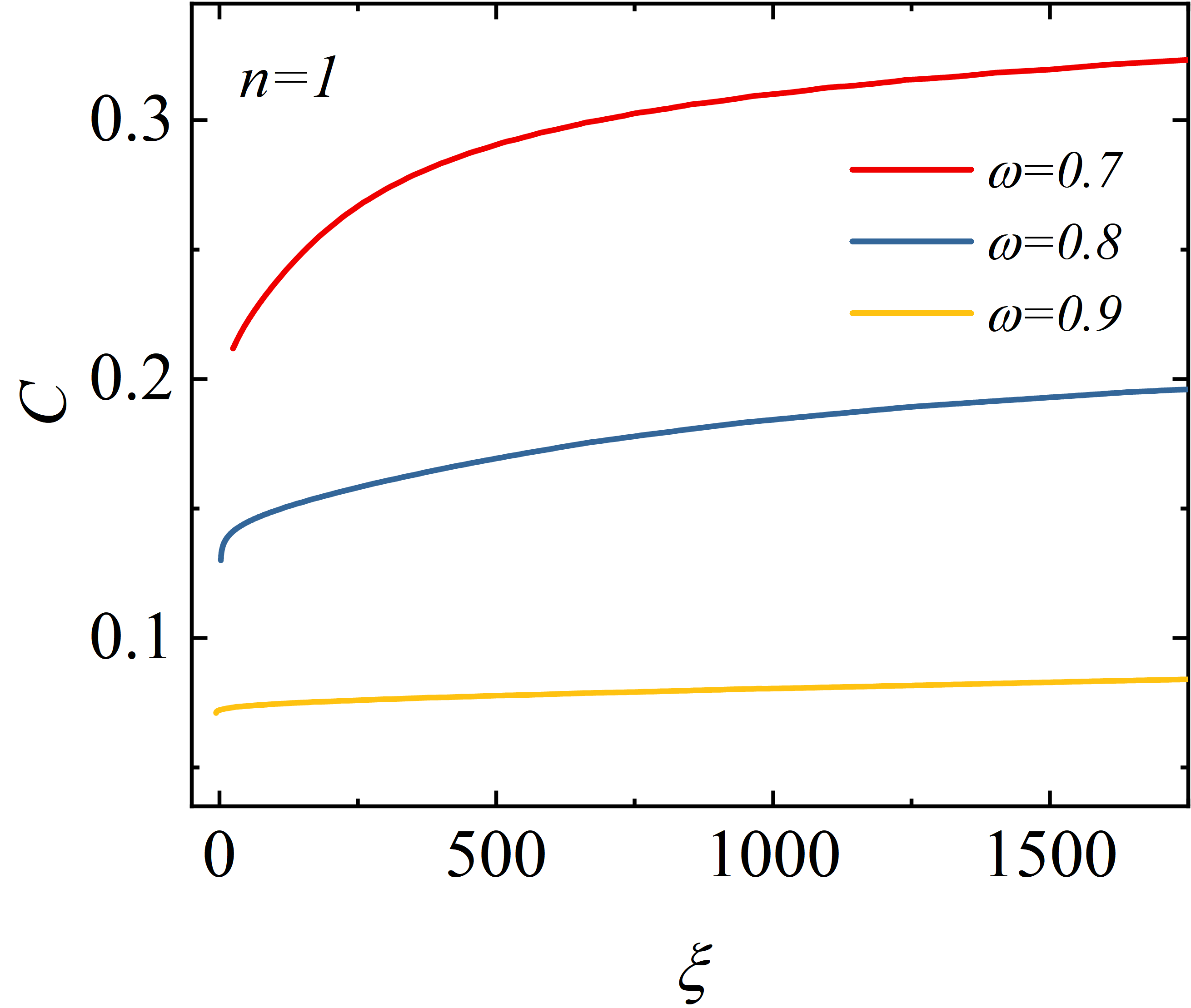}
            \label{fig:compactforkxin1}
   		} 	
     		\subfigure[~$n=2$]{
			\includegraphics[height=.18\textheight,width=.22\textheight, angle =0]{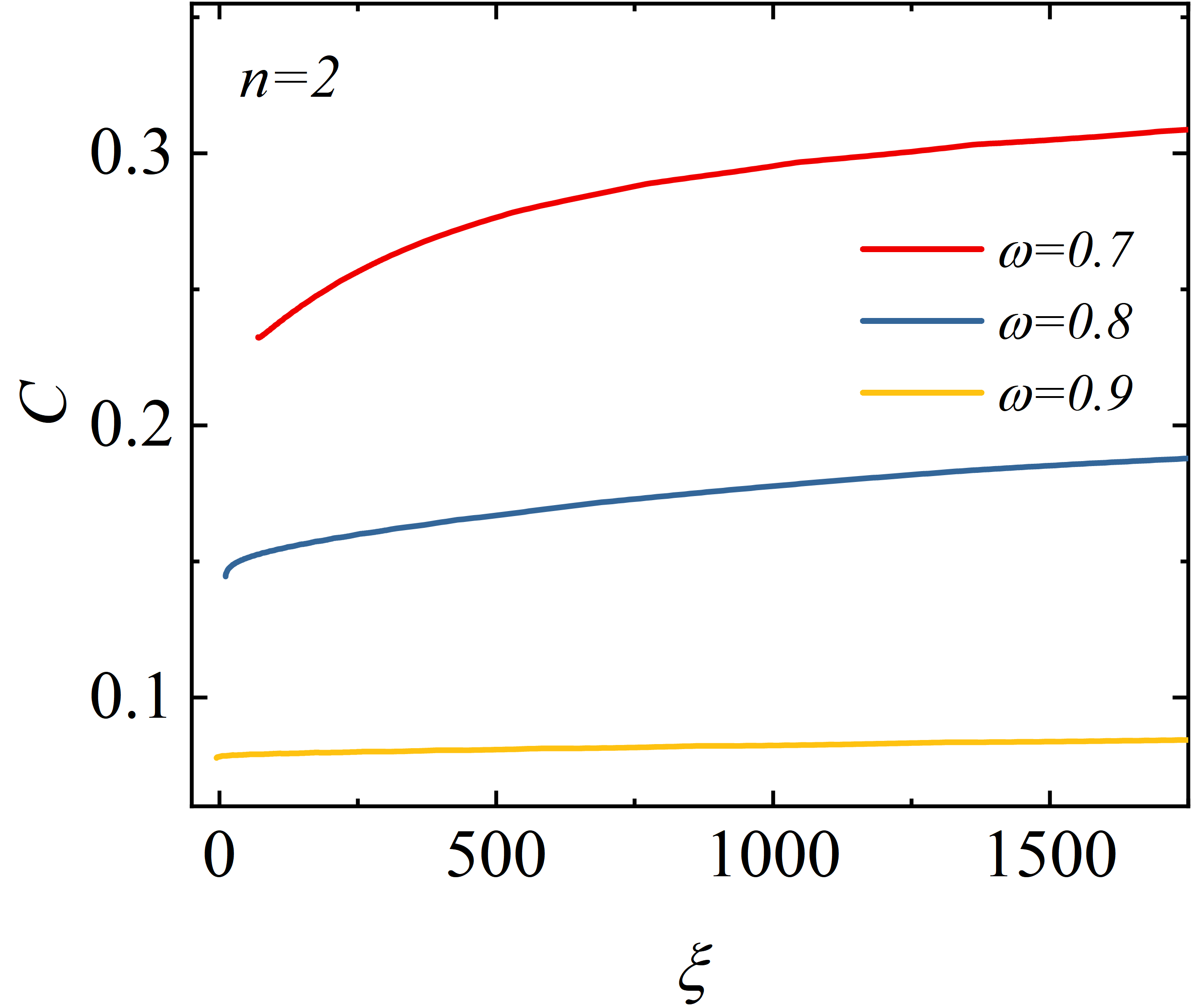}
            \label{fig:compactforkxin2}
   		} 	
		\caption{Compactness $C$ as a function of the coupling parameter $\xi$ for several frequencies, shown for the ground state (left panels) and the second excited state (right panels).}
		\label{fig:compact}		
		\end{figure}

Fig.~\ref{fig:compact} illustrates the dependence of compactness on $\xi$ for both the ground and excited states at several frequencies. It can be observed that for a fixed frequency, as $\xi$ increases, the compactness increases significantly at small $\xi$ (especially for lower frequencies or node numbers), while the variation of compactness becomes gentle for large $\xi$. In particular, when the frequency $\omega$ is relatively small (red line), the compactness of these boson stars can be higher than $0.3$. This value exceeds the compactness of neutron stars ($C \sim 0.2$), is close to that of ultracompact solitonic boson stars ($C \sim 0.3$)~\cite{Friedberg:1986tq,Boskovic:2021nfs,Cardoso:2021ehg,Collodel:2022jly}, which are among the most compact boson stars discovered to date, but remains below the Buchdahl limit ($C = 4/9 \approx 0.444$)~\cite{Buchdahl:1959zz}.

In addition, Fig.~\ref{fig:compact} shows that the curves terminate at a certain small coupling parameter. This occurs because for a given frequency, there exists a minimum value of the coupling parameter below which no solutions are found. In Tab.~\ref{tab:xi_min_full}, we list the minimum coupling parameters for both ground-state and excited-state boson stars at several different frequencies. The results show that at a given frequency, the minimum coupling parameter for the excited state is always larger than that for the ground state. Moreover, as the frequency increases, the minimum coupling parameter decreases for both ground-state and excited-state boson stars. This behavior corresponds to the increase of the minimum frequency with $\xi$ shown in Fig.~\ref{fig:massandcharged}, implying that a boson star with a lower frequency requires a larger coupling parameter $\xi$.

\begin{table}[htbp]
\centering
\begin{tabular}{|c|c|c|c|c|c|}
\hline
 & $\xi_{\min}(\omega=0.7)$ & $\xi_{\min}(\omega=0.75)$ & $\xi_{\min}(\omega=0.8)$ & $\xi_{\min}(\omega=0.85)$ & $\xi_{\min}(\omega=0.9)$ \\
\hline
$n=0$ & 1.075 & 0.258 & -0.564 & -2.202 & -7.428 \\
\hline
$n=1$ & 24.148 & 10.357 & 2.669 & -0.776 & -5.996 \\
\hline
$n=2$ & 69.727 & 29.945 & 11.108 & 0.399 & -5.24 \\
\hline
\end{tabular}
\caption{The minimum value of coupling parameter $\xi_{\min}$ for different frequencies and node numbers.}
\label{tab:xi_min_full}
\end{table}

	\begin{figure}[!htbp]
		\centering		
		\subfigure[]{  
			\includegraphics[height=.18\textheight,width=.22\textheight, angle =0]{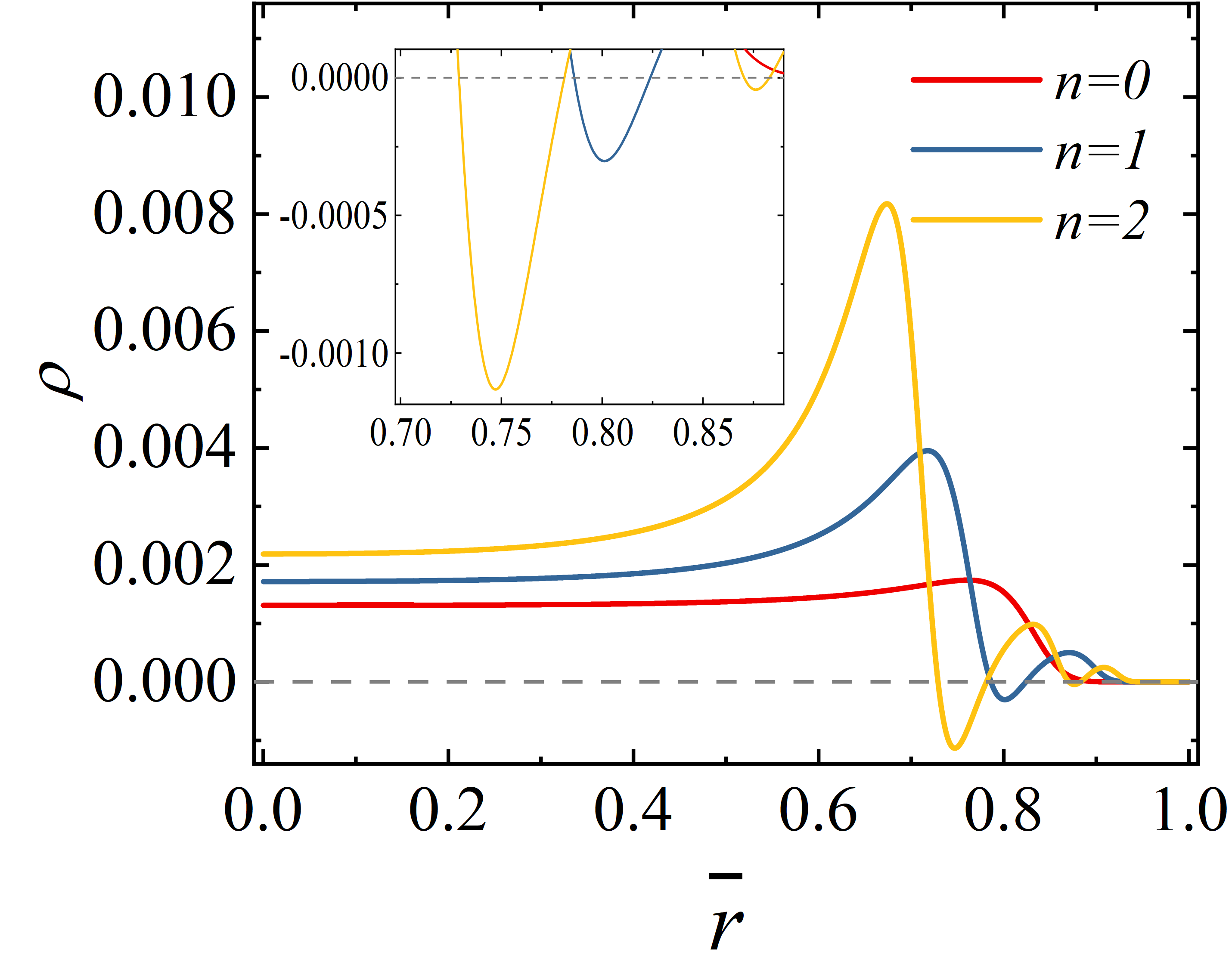}
            \label{fig:rho08xi30}
		}
		\subfigure[]{
			\includegraphics[height=.18\textheight,width=.22\textheight, angle =0]{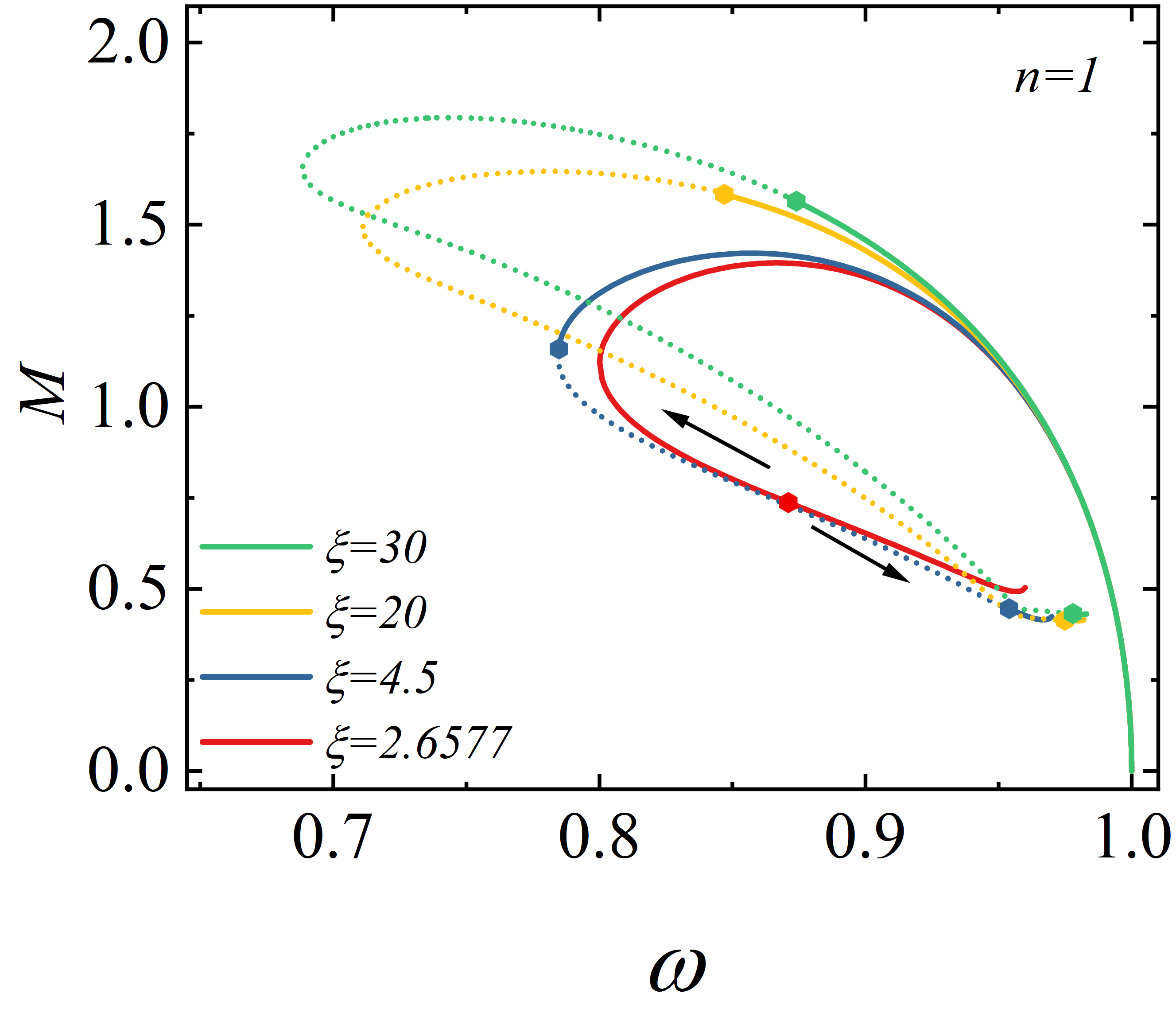}
            \label{fig:Mrhon1}
   		} 	
     		\subfigure[]{
			\includegraphics[height=.18\textheight,width=.22\textheight, angle =0]{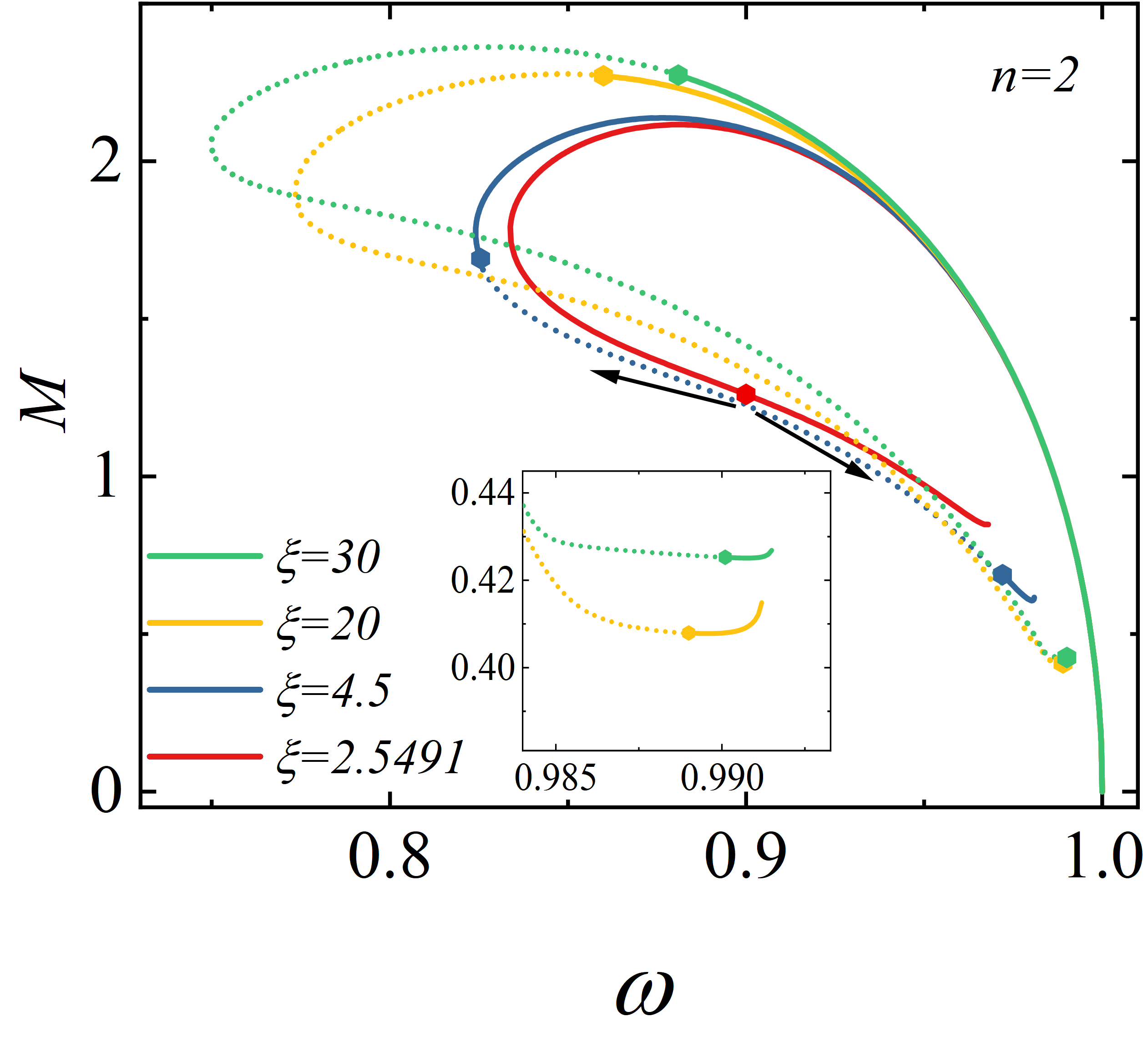}
            \label{fig:Mrhon2}
   		} 	
		\caption{Left panel: the profile of energy density $\rho$ for the boson star with different $n$, calculated for $\omega = 0.8$ and $\xi=30$. Middle and right panels: the relationship
        between $M$ and $\omega$, in which the dotted line indicates the solution where the energy density can be negative.}
		\label{fig:Mrho}		
		\end{figure}

Interestingly, as shown in Fig.~\ref{fig:Mrho}, we find that for excited states, once the coupling parameter exceeds a certain value, the minimum energy density of the boson stars can become negative at some frequencies. However, for the ground states, all solutions we have found so far exhibit a non-negative energy density. To show how these excited-state solutions with a negative minimum energy density vary with the coupling parameter, we have used dotted lines in the 
$M(\omega)$ curves of Figs.~\ref{fig:Mrhon1} and ~\ref{fig:Mrhon2} to represent such solutions for several coupling parameters. The colored hexagons in these figures mark the transition points between the two types of solution, and the red line corresponds to the smallest coupling parameter at which a negative energy density first appears. It can be observed that for the red line, only one hexagon is found, and it occurs only in the second branch. This suggests that, initially, for a small $\xi$, this type of solution occurs only once and is confined to the second branch. As the coupling parameter $\xi$ increases, these solutions with negative energy density extend along the spiral to both sides of the red hexagon, as indicated by the black arrows in this figure, causing the parameter space of such solutions to expand. Eventually, this solution can even emerge in the first branch. 

Ref.~\cite{Horvat:2014xwa} demonstrated that for all ground-state solutions they have examined, the dominant energy condition, which requires that the energy density $\rho$ is non-negative and that it is greater than or equal to the absolute values of any of the individual pressures, is consistently satisfied (i.e., $\rho\geq0$, $\rho-|p_r|\geq0$ and $\rho-|p_\bot|\geq0$). Obviously, this change in the energy density of the excited states means that the dominant energy condition is no longer satisfied.

	\begin{figure}[!htbp]
		\centering		
		\subfigure[]{  
			\includegraphics[height=.18\textheight,width=.22\textheight, angle =0]{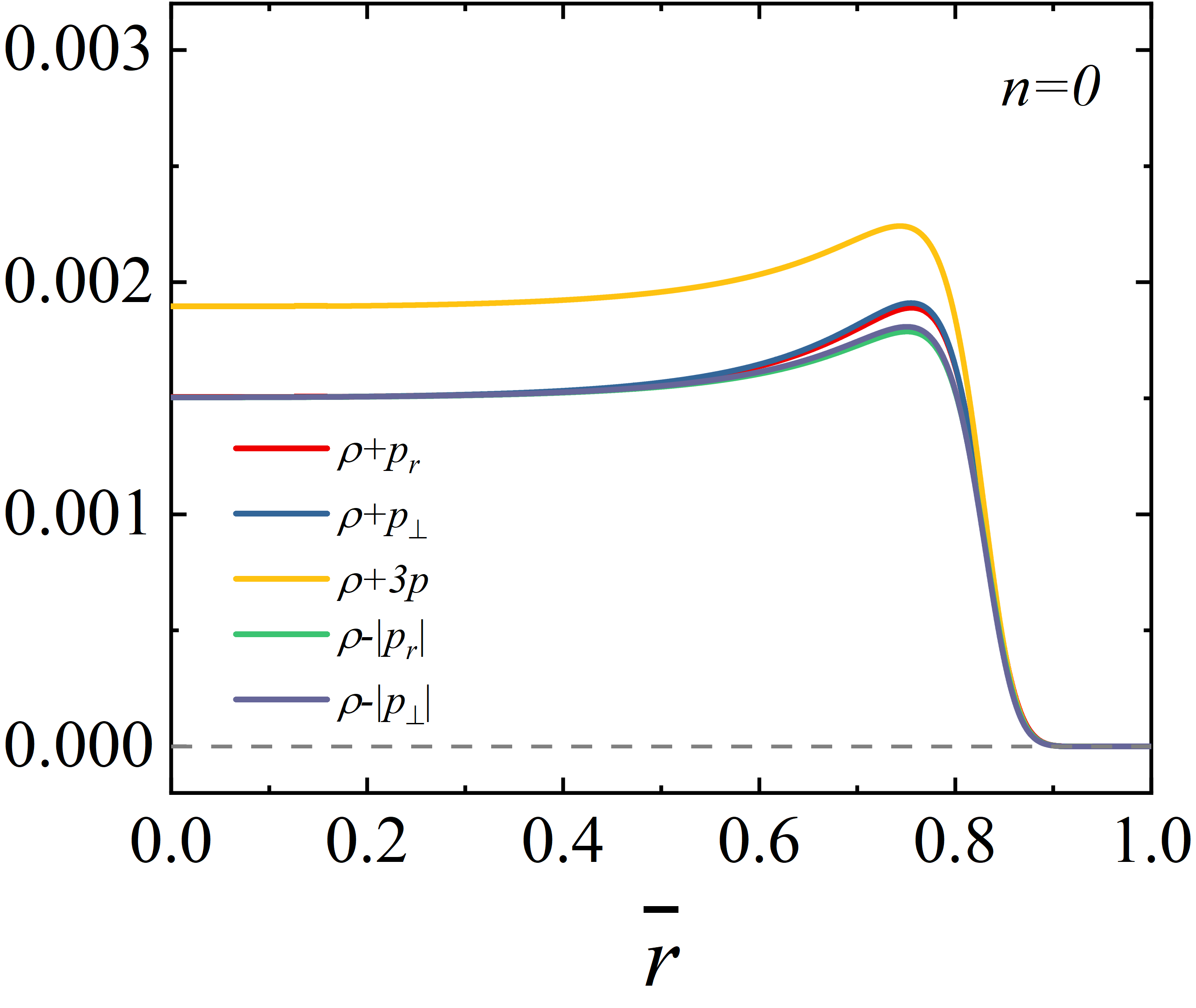}
            \label{fig:ECn0}
		}
		\subfigure[]{
			\includegraphics[height=.18\textheight,width=.22\textheight, angle =0]{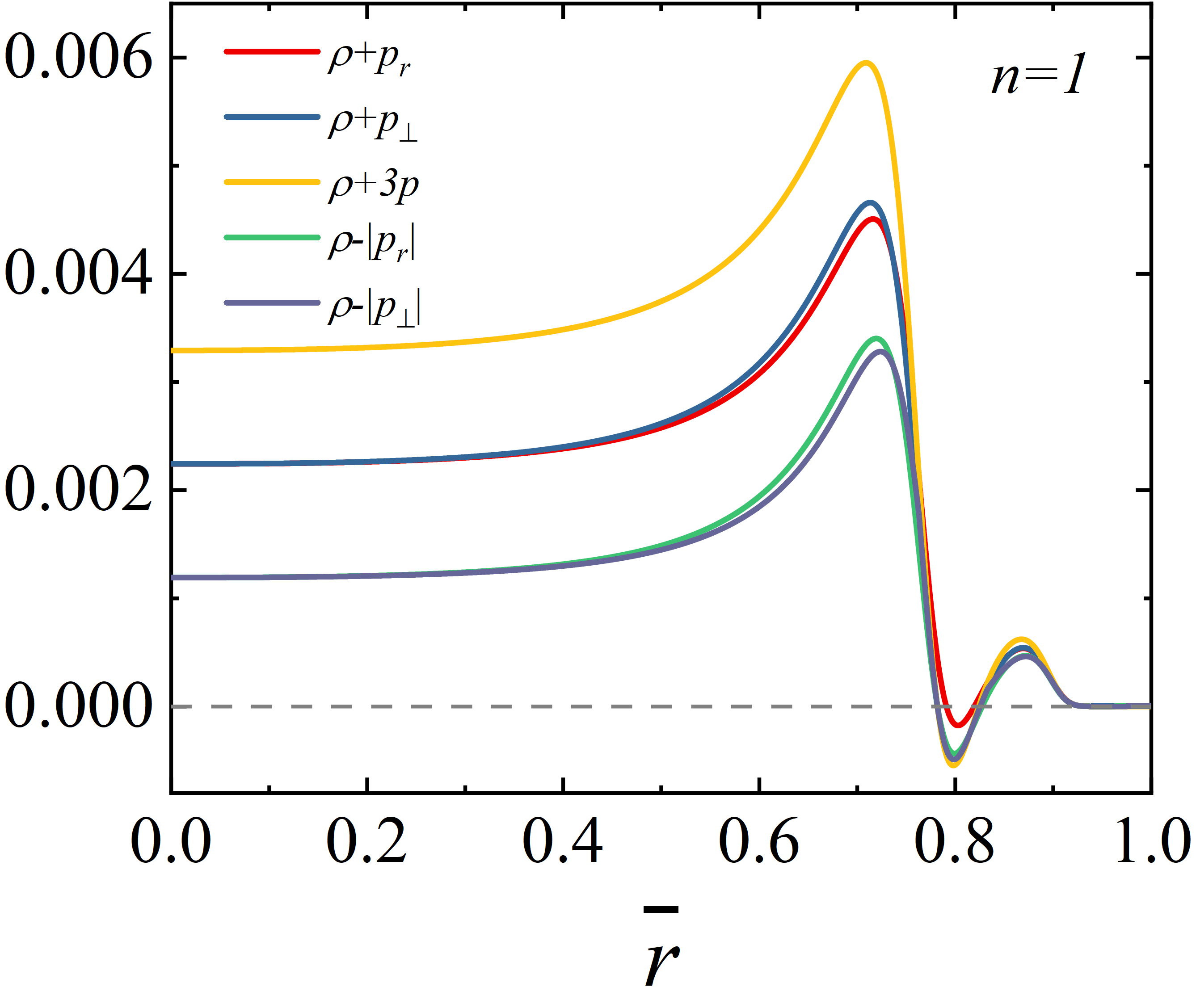}
            \label{fig:ECn1}
   		} 	
     		\subfigure[]{
			\includegraphics[height=.18\textheight,width=.22\textheight, angle =0]{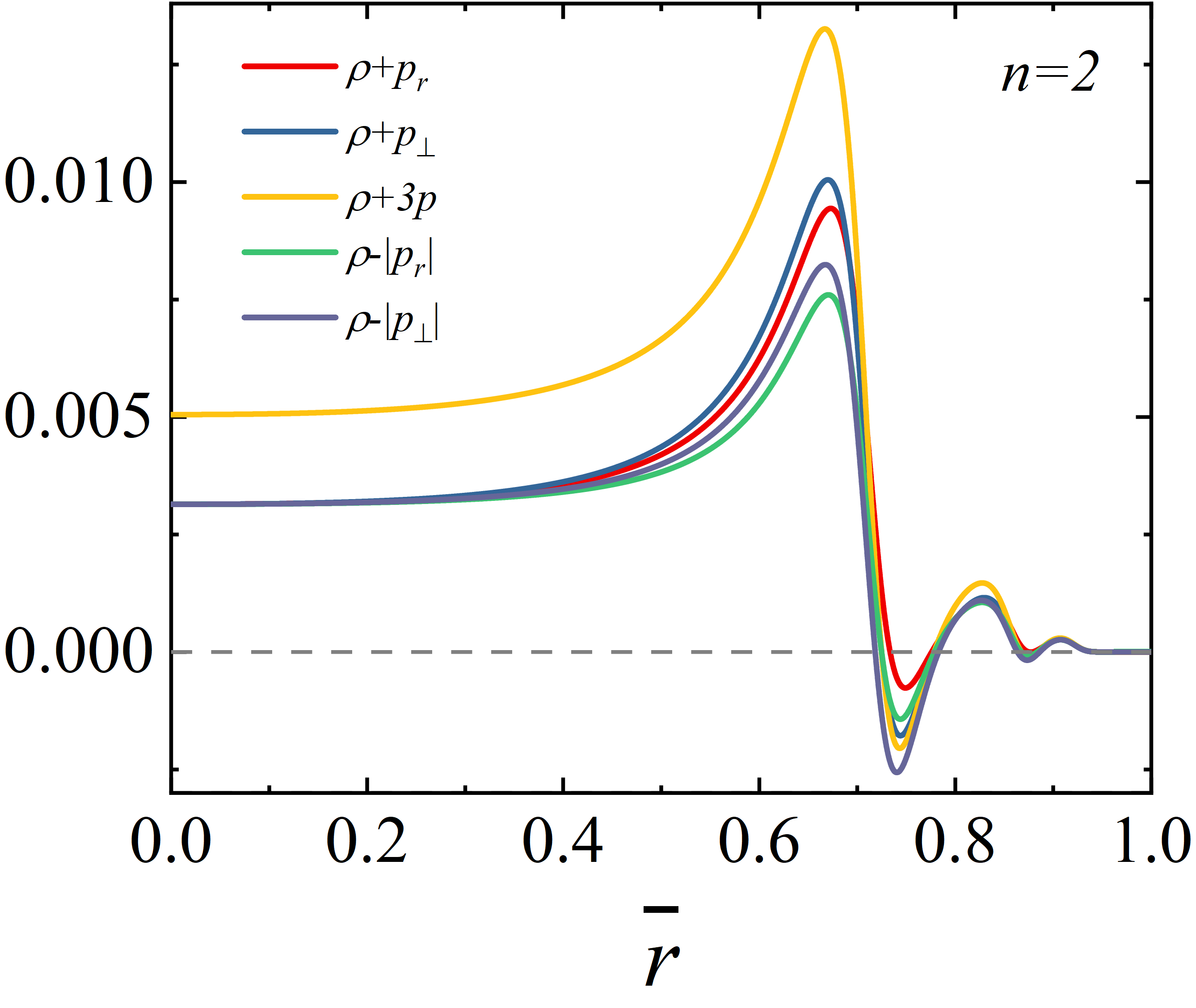}
            \label{fig:ECn2}
   		} 	
		\caption{The profile of $\rho+p_r$, $\rho+p_{\bot}$, $\rho+3p$, $\rho-|p_r|$ and $\rho-|p_{\bot}|$ for the boson star with different $n$. All solutions have $\xi=30$ and $\omega=0.8$.}
		\label{fig:EC}		
		\end{figure}

In addition to the dominant energy condition, there are three other commonly used energy conditions: (1) NEC; (2) WEC; and (3) SEC. They can be expressed in terms of energy density and pressure as follows~\cite{Zubair:2015wqx}:

\begin{itemize}
    \item NEC: the sum of the energy density and any individual pressure are non-negative, i.e., $\rho+p_r\geq0$ and $\rho+p_\bot\geq0$. It describes the energy flow along the direction of the light.

    \item WEC: both the energy density and the sum of the energy density and any individual pressure must be non-negative, i.e., $\rho\geq0$, $\rho+p_r\geq0$ and $\rho+p_\bot\geq0$. It indicates that the local energy density measured by any observer is non-negative.
    
    \item  SEC: the sum of the energy density and any individual pressure or all individual pressure (denoted as $3p$) are non-negative, i.e., $\rho+p_r\geq0$, $\rho+p_\bot\geq0$ and $\rho+3p=\rho+p_r+2p_\bot\geq0$. This actually means that gravity always attracts. It ensures that geodesics converge due to gravity, rather than spreading out.
\end{itemize}

By examining these energy conditions, we find that the excited states with negative energy density violate not only the DEC but also the NEC, WEC and SEC. In contrast, for the ground states, we have found no solutions that violate any of these four conditions. As an illustrative example, we take $\xi=30$ and $\omega=0.8$ to plot the profiles of $\rho+p_r$, $\rho+p_\bot$, $\rho+3p$, $\rho-|p_r|$ and $\rho-|p_\bot|$ in Fig.~\ref{fig:EC}. It can be observed that for the ground state, the six functions are non-negative. However, for the excited state (middle and right panels), the six functions become negative in some regions. Consequently, the four energy conditions are no longer satisfied. By contrast, this feature is absent in mini-boson stars in GR~\cite{Kaup:1968zz}, where the four energy conditions are always satisfied for both the ground and excited states. Therefore, in the model of this paper, if the energy conditions are to be satisfied, their violation by the excited states places more restrictions on the coupling parameter $\xi$.

Although the differences between the ground and excited states in terms of energy density and energy conditions are challenging to explain analytically, when considering only the energy density, their distinct behaviors for non‑negative coupling parameters $\xi$ can be intuitively understood through differences in the radial profiles of their scalar field functions (see Fig.~\ref{fig:omega09function} for an illustrative case) and Eq.~(\ref{eq:rho}). Specifically, first, as shown in Fig.~\ref{fig:omeg09grr}, because $1/g_{rr}=e^{2F_1}\leq1$, one can deduce that $(1 - e^{F_1}) \geq 0$. Therefore, according to Eq.~(\ref{eq:rho}), when $\xi \geq 0$, only the sign of the term $\phi \phi'$ can change, while all other terms are non-negative. For the ground state, since the scalar field function $\phi$ is non-negative and monotonically decreasing, it consistently follows that $-\phi \phi' \geq 0$, and thus its energy density remains strictly non-negative. In contrast, for the excited states, the presence of multiple nodes results in $\phi \phi' > 0$ in certain regions, thereby causing the energy density to potentially become negative at these locations.

\section{ORBITAL MOTION OF TEST PARTICLES AND GRAVITATIONAL WAVES FROM EMRIS}\label{sec:emri}

Our analysis shows that considering a non-minimal coupling between the scalar field and torsion causes many properties of the resulting boson star solutions to change from those of GR boson stars. These changes will leave potentially detectable imprints in astrophysical observations. 

In this respect, EMRIs, characterized by a stellar-mass compact object spiraling into a supermassive central body, offer a premier laboratory for probing gravity in the strong-field regime and serve as one of the most promising sources for low-frequency, space-based GW observatories. Unlike EMRIs around supermassive black holes, where the inspiral terminates at the horizon and the signal decays via quasinormal modes, the horizonless nature of boson stars allows the objects to continue their motion through the interior, producing unique GW signatures. In this section, by applying the adiabatic approximation together with the mass-quadrupole radiation formula, we investigate the influence of the non-minimal coupling parameter $\xi$ on GWs emitted from the EMRI system, consisting of a $10^6 M_\odot$ central boson star and a $10 M_\odot$ stellar-mass compact object modeled as a test particle. Without loss of generality, we focus on the ground-state and the first excited-state boson star. 

\subsection{The orbit of test particles} \label{sec:orbit}
The gravitational waveforms produced by this EMRI system are governed by the timelike geodesic motion of the test particle in the background of the central boson star. Therefore, we first need to analyze the orbital motion of a test particle in the boson star background, which is governed by the following Lagrangian:
\begin{equation}
    \mathcal{L}=\frac{1}{2}g_{\mu\nu}\dot{x}^\mu\dot{x}^\nu,
    \label{eq:lag}
\end{equation}
where the overdot denotes derivatives with respect to the affine parameter $\lambda$. For a massive particle, $\mathcal{L}=1/2$. In the following, since particles are less affected when farther away from the boson star and using the $\bar{r}$ coordinate will deform the shape of the orbit of the particle, we adopt the radial coordinate $r$ (dimensionless) before the conformal transformation.

	\begin{figure}[!htbp]
		\centering		
		\subfigure[~$(L=0.5,E=0.9)$]{  
			\includegraphics[width=6.5cm]{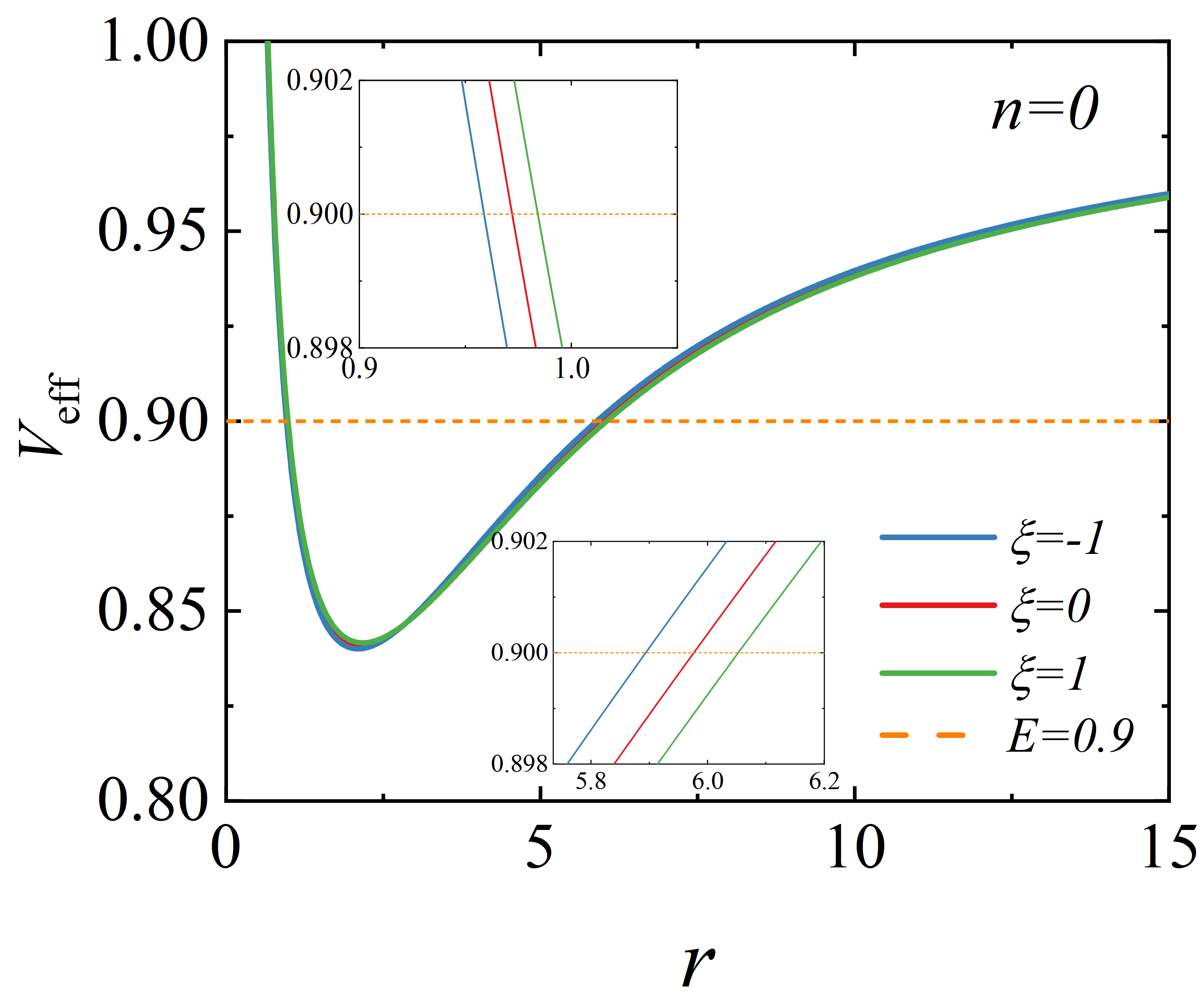}
            \label{fig:VEFFn0L05E09}
		}    
		\subfigure[~$(L=3,E=0.985)$]{
			\includegraphics[width=6.5cm]{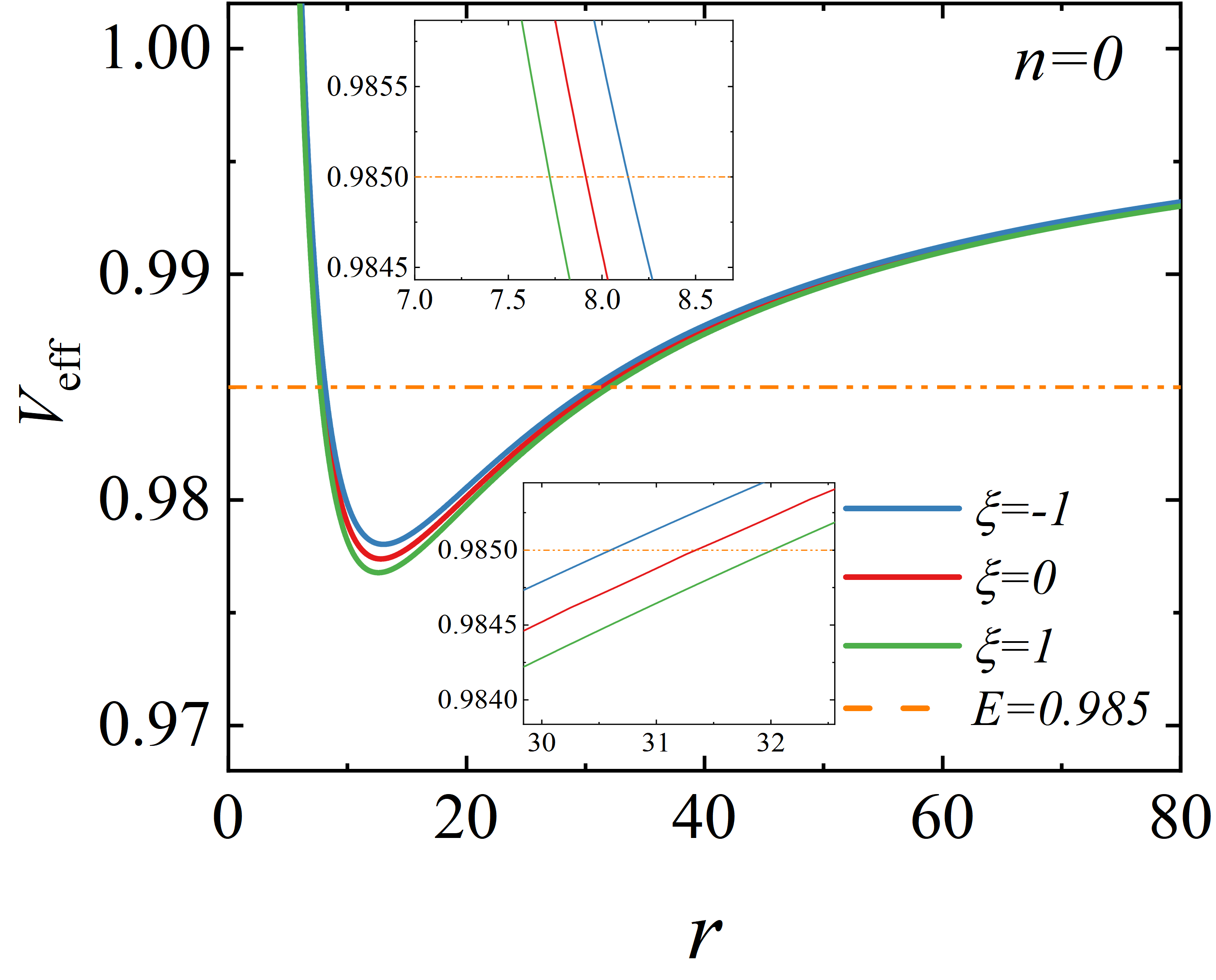}
            \label{fig:VEFFn0L3E0985}
   		} 	
     		\subfigure[~$(L=0.5,E=0.82)$]{
			\includegraphics[width=6.5cm]{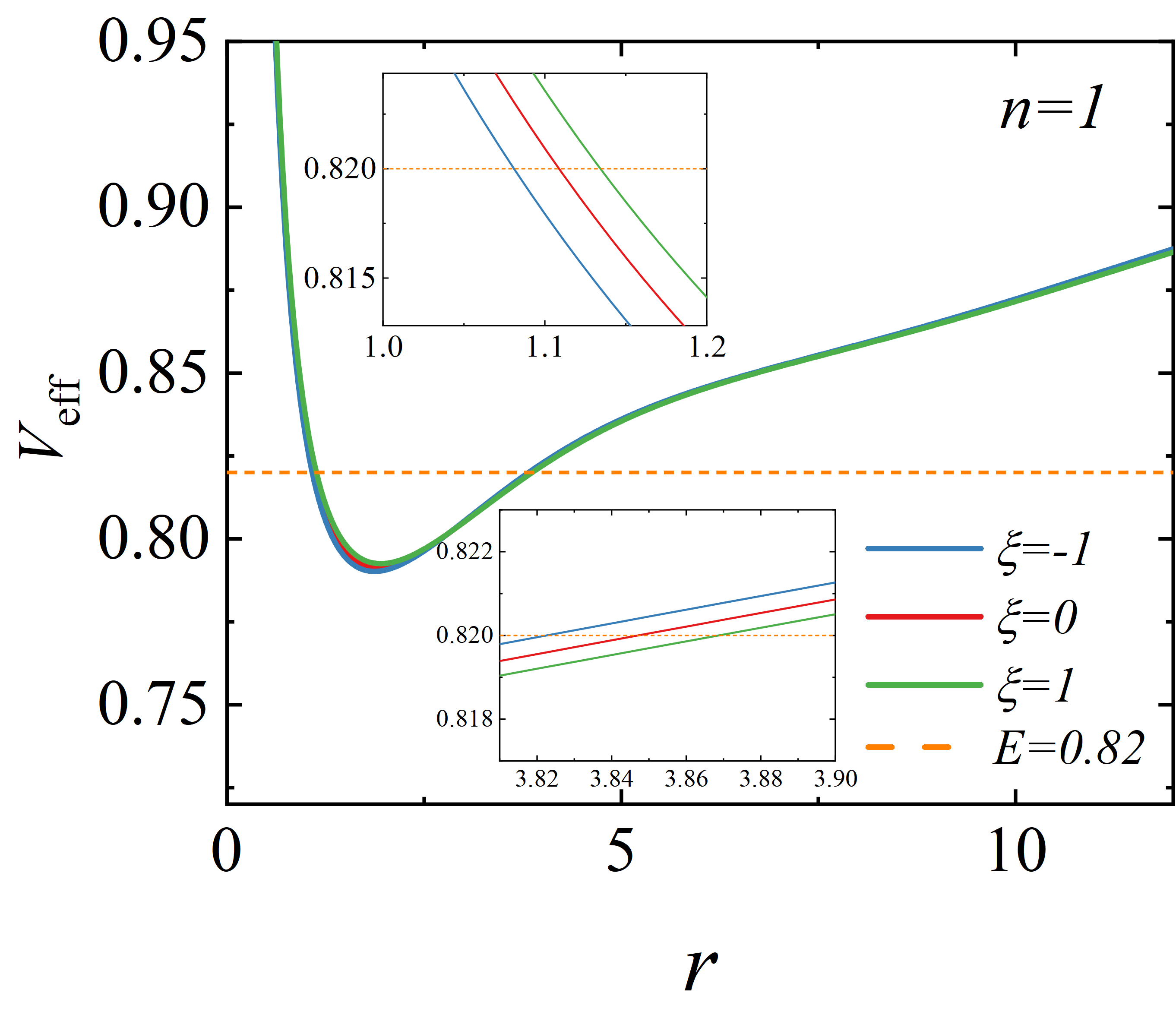}
            \label{fig:VEFFn1L082E05}
   		} 	
             \subfigure[~$(L=6,E=0.975)$]{
			\includegraphics[width=6.5cm]{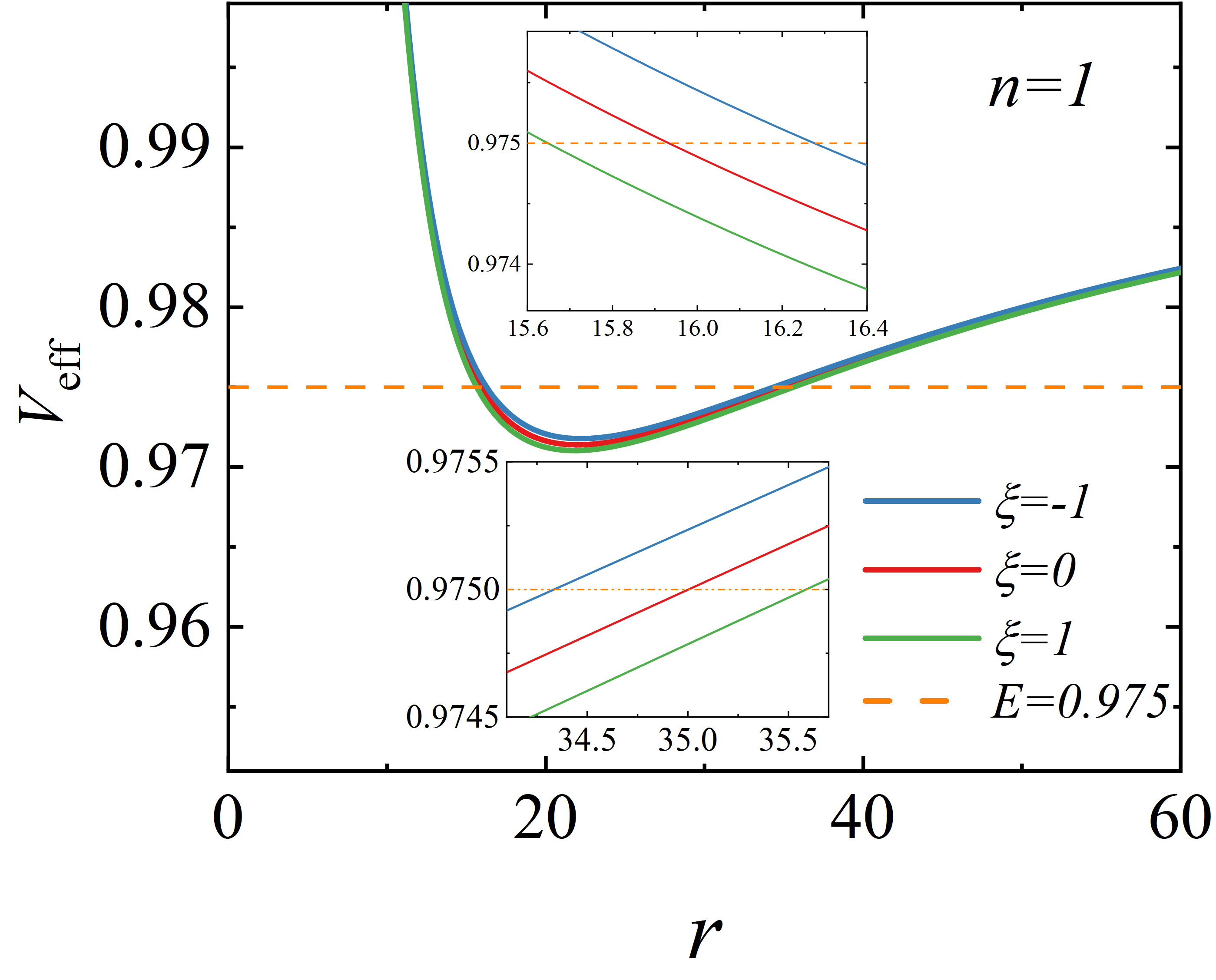}
            \label{fig:VEFFn1L6E0975}
   		} 
		\caption{The effective potential $V_{\mathrm{eff}}$ for boson stars with different coupling parameter $\xi$. The top and bottom panels correspond to the ground state and the first excited state, respectively.}
		\label{fig:EFF}		
		\end{figure}
\begin{enumerate}
    \item In addition, due to spherical symmetry, for simplicity, we only consider the orbit lying on the equatorial plane (i.e. $\theta = \pi/2$). In addition, static spherical symmetry means that the spacetime can possess a timelike Killing vector $(\partial_t)^\mu$ and a spacelike Killing vector $(\partial_\varphi)^\mu$, which are associated with two motion constants, i.e., the specific energy
\end{enumerate}
\begin{equation}
E=\frac{\partial\mathcal{L}}{\partial\dot{t}}=e^{2F_0}\dot{t},
\label{eq:mce}
\end{equation} 
and the specific angular momentum 
\begin{equation}
L=-\frac{\partial\mathcal{L}} {\partial\dot{\varphi}}=r^2\dot{\varphi}.\label{eq:mcl}
\end{equation}
Thus, based on these two motion constants and the Lagrangian (\ref{eq:lag}), we obtain the following equations  
\begin{equation}
    \dot{r}^2=\frac{E^2}{\mathrm{e}^{2F_0+2F_1}}-\frac{1}{\mathrm{e}^{2F_1}}\left(1+\frac{L^2}{r^2}\right),
    \label{eq:dotr}
\end{equation}
\begin{equation}
    \dot{\varphi}=\frac{L}{r^2}.
    \label{eq:dotr}
\end{equation}
By setting $\dot{r}=0$, the effective potential $ V_{\mathrm{eff}}$ can be defined as
\begin{equation}
    V_{\mathrm{eff}}=\mathrm{e}^{F_0}\sqrt{\left(1+\frac{L^2}{r^2}\right)}.
\end{equation}

The effective potential is very useful for understanding the bound motion of particles. As examples, Fig.~\ref{fig:EFF} shows the effective potentials of the boson stars with the same frequency $\omega=0.9$ but different parameter sets ($L, E$) and different values of the coupling parameters $\xi$. The top and bottom panels correspond to the ground and first excited states, respectively. These parameter sets correspond to two special scenarios. The smaller angular momentum ($L = 0.5$, left panels) corresponds to the test particle being able to penetrate deeper into the central region of boson stars. In contrast, the larger angular momentum ($L = 3$ for the ground state and $L=6$ for the excited state, right panels) corresponds to the test particle being farther from the center of boson stars.

\begin{figure}[htbp]
    \centering
		\subfigure[~$n=0,\xi=-1$]{  
			\includegraphics[width=0.3\textwidth, angle =0]{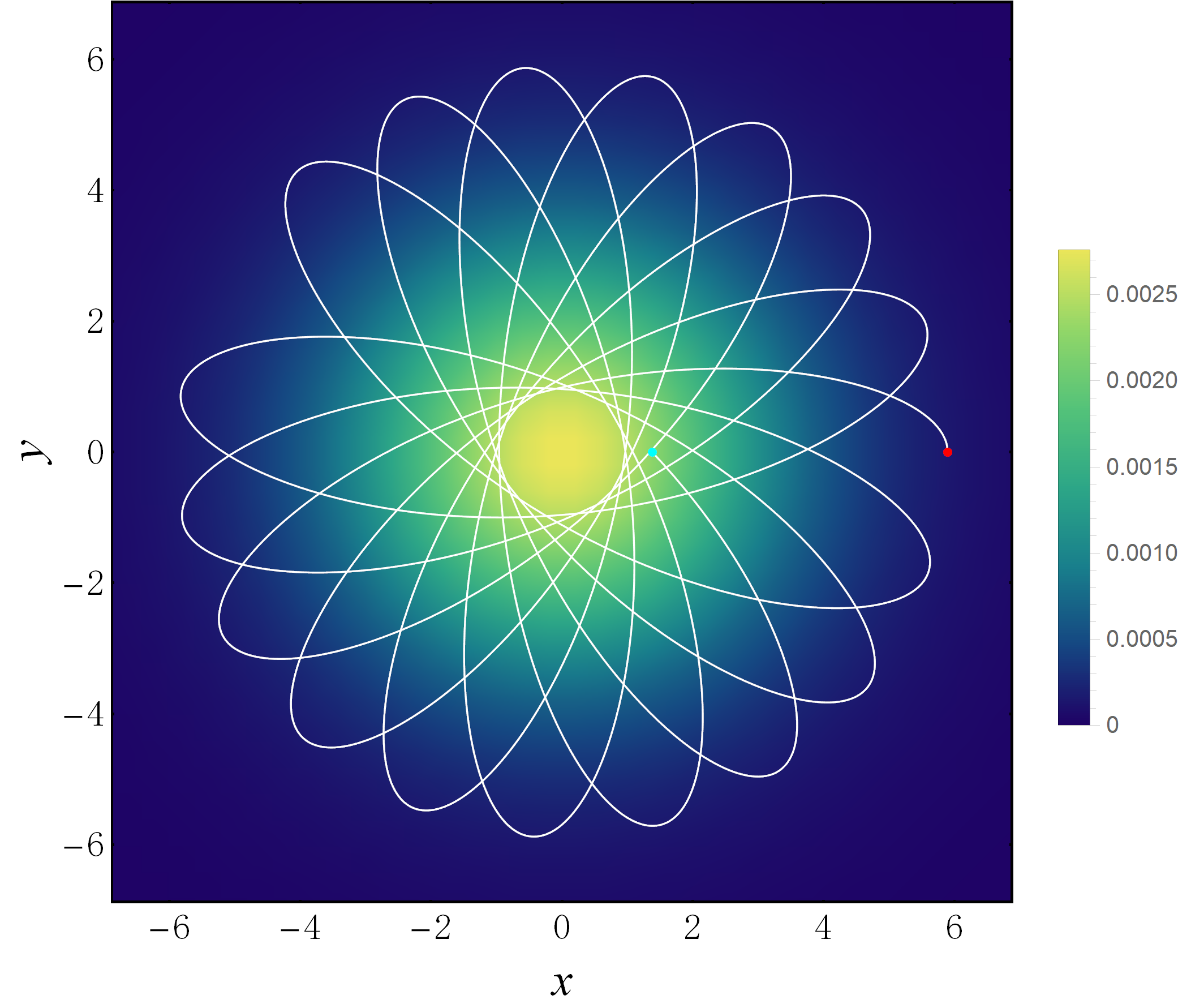}
		}
        \subfigure[~$n=0,\xi=0$]{  
			\includegraphics[width=0.3\textwidth, angle =0]{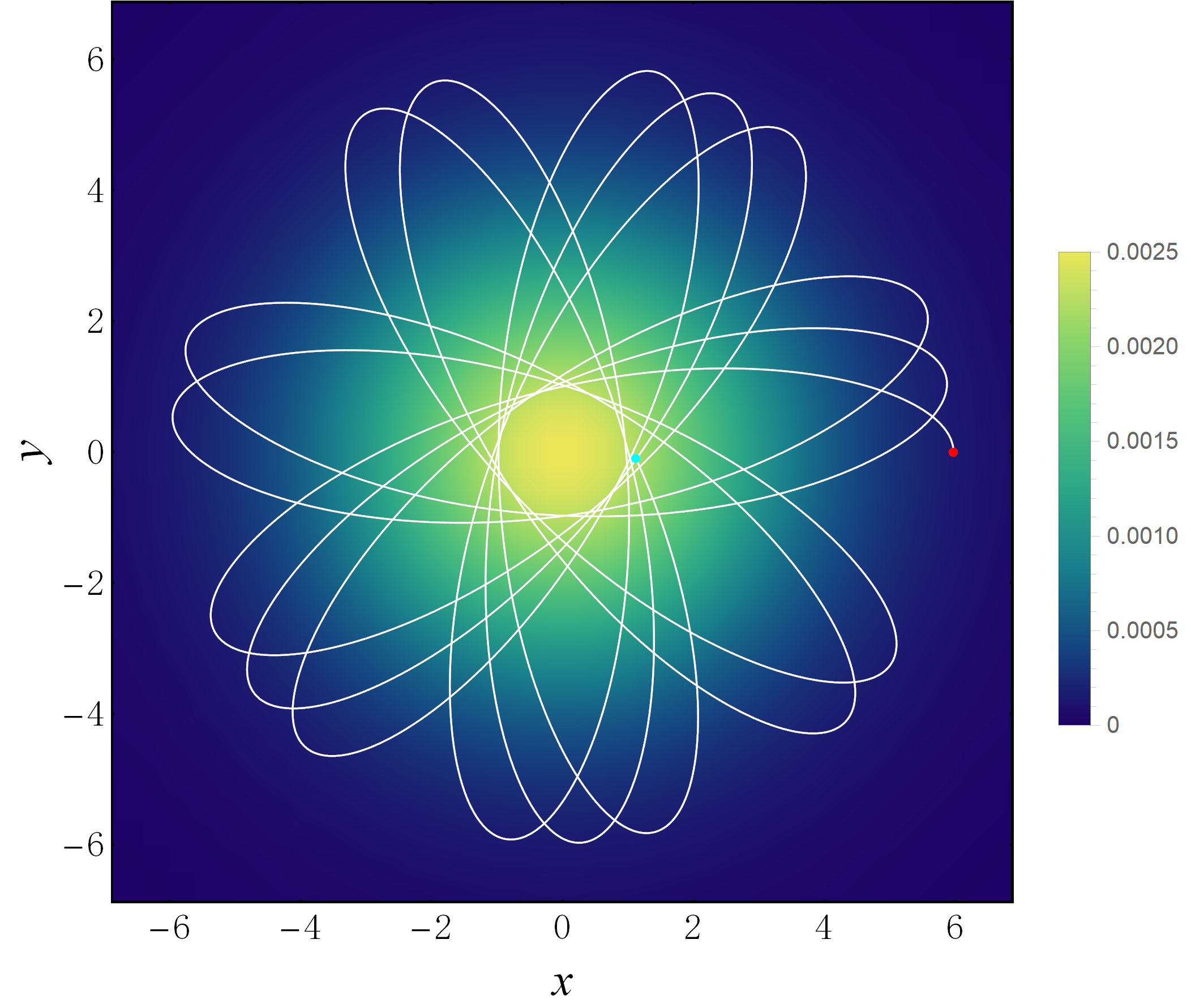}
		}
        \subfigure[~$n=0,\xi=1$]{  
			\includegraphics[width=0.3\textwidth, angle =0]{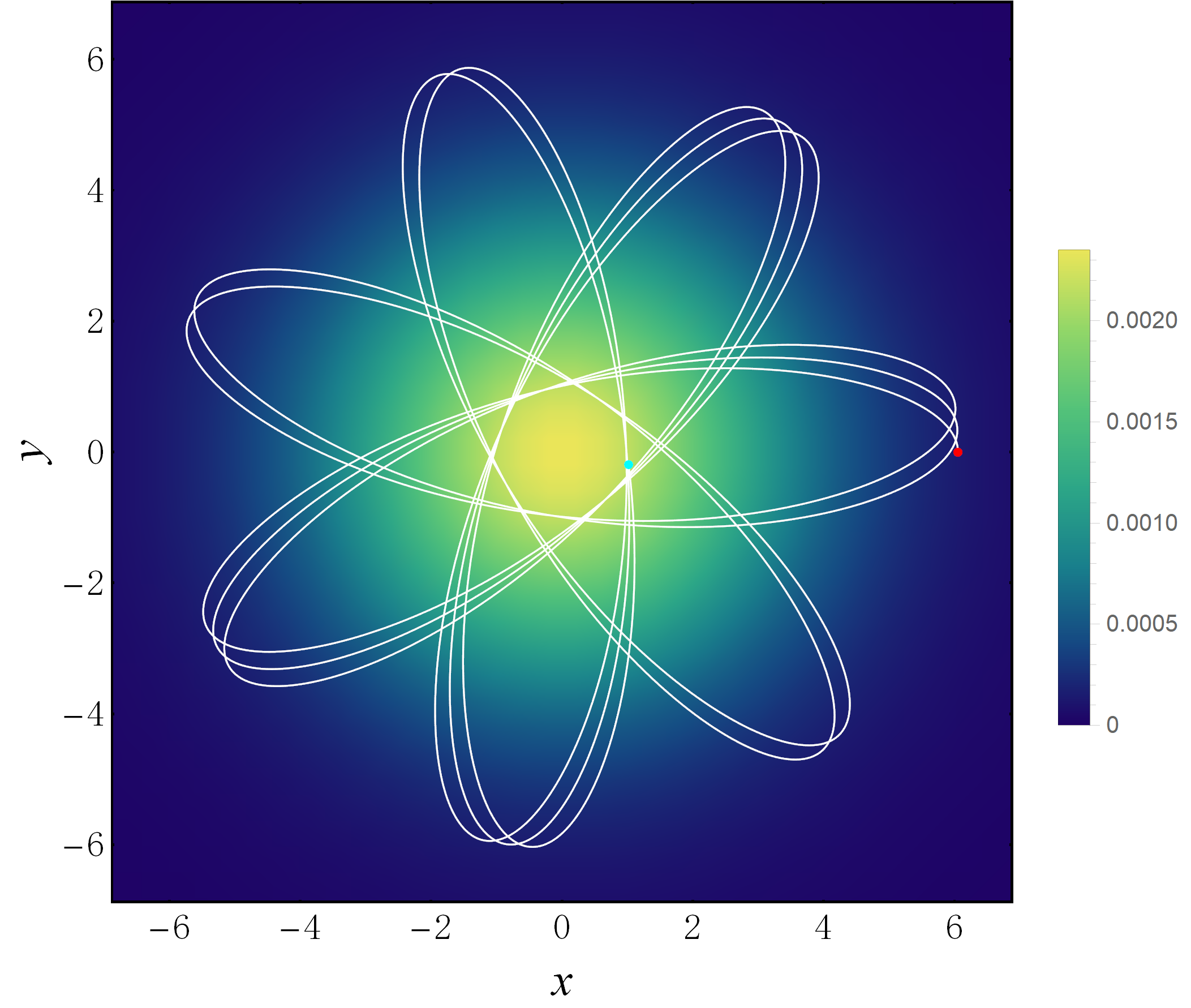}
		}
        \subfigure[~$n=0,\xi=-1$]{  
			\includegraphics[width=0.3\textwidth, angle =0]{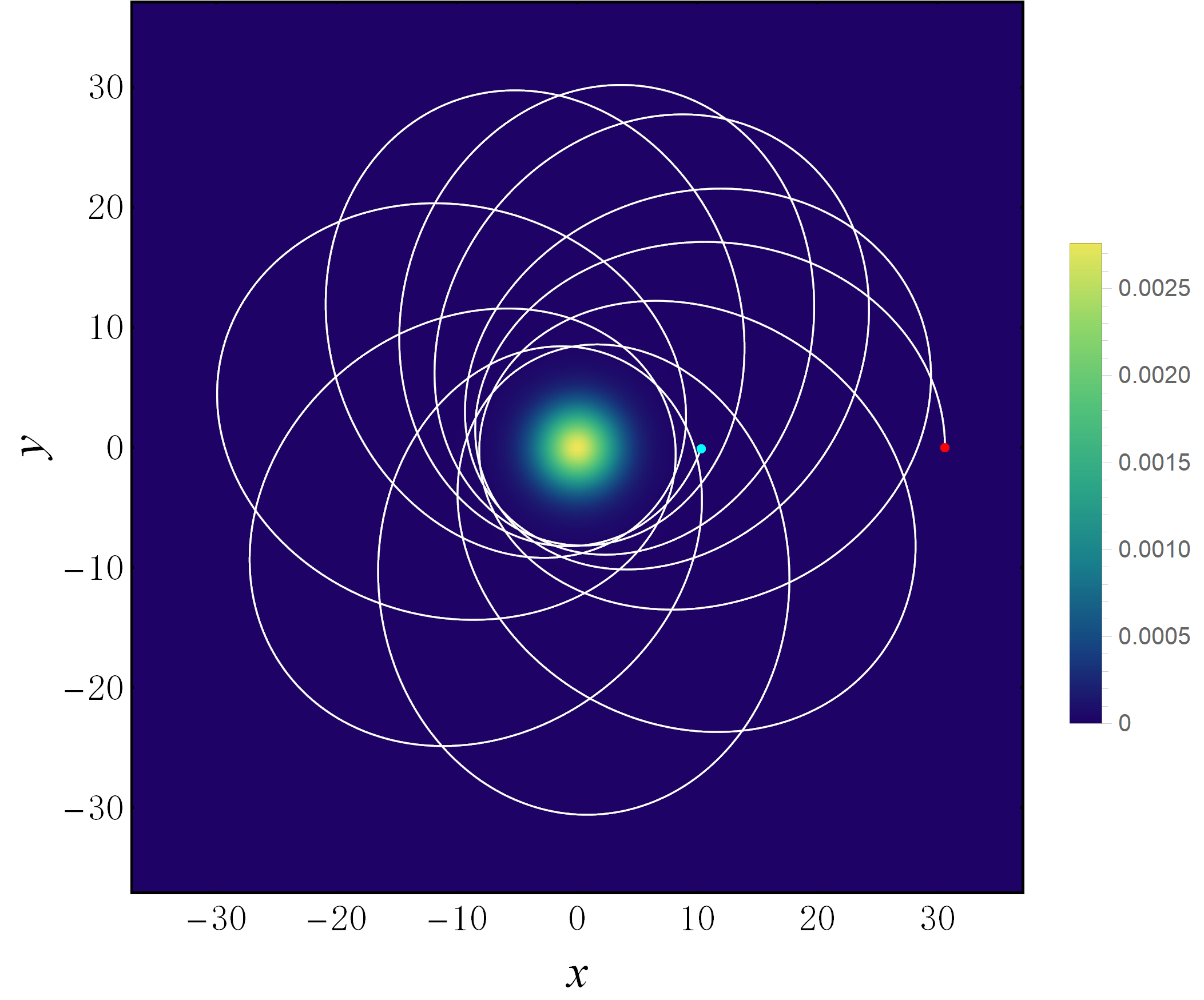}
		}
        \subfigure[~$n=0,\xi=0$]{  
			\includegraphics[width=0.3\textwidth, angle =0]{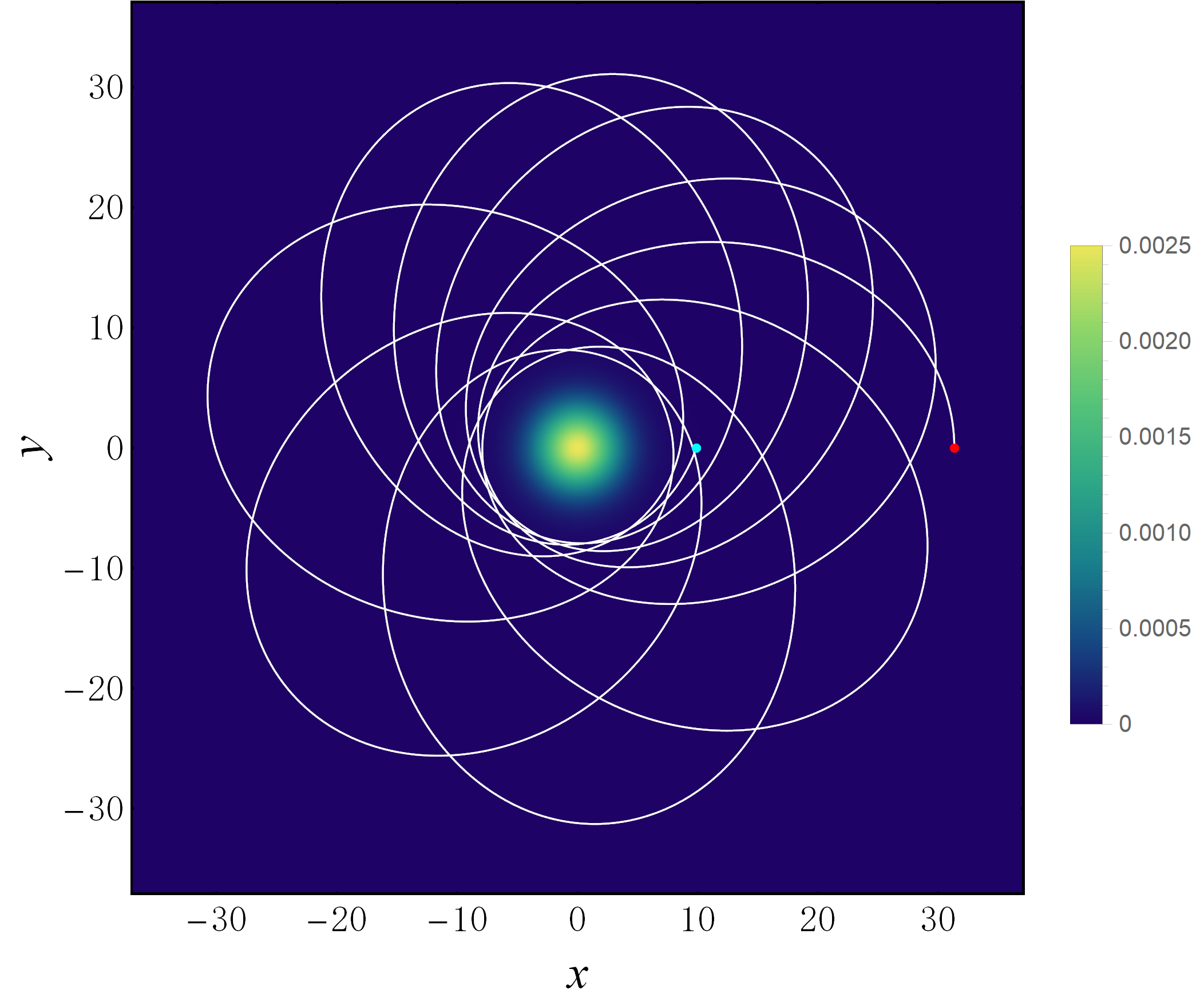}
		}
        \subfigure[~$n=0,\xi=1$]{  
			\includegraphics[width=0.3\textwidth, angle =0]{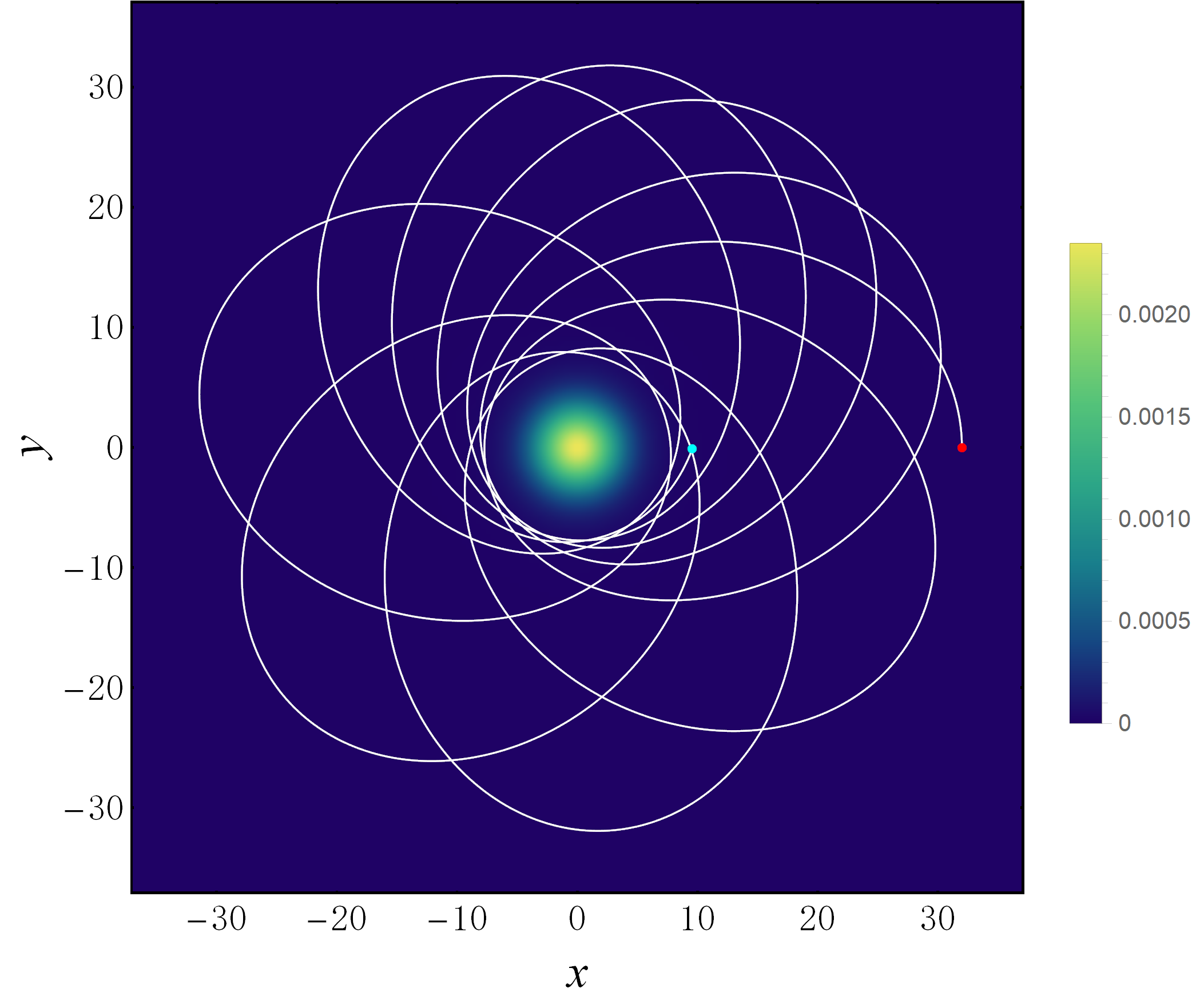}
		}
    \caption{Orbits for the ground-state boson stars with different values of the coupling parameter $\xi$, calculated for $\omega = 0.9$. The top panels ($L = 0.5$, $E = 0.9$) show penetrating orbits with smaller orbital radii, while the bottom panels ($L = 3$, $E = 0.985$) show grazing orbits with larger orbital radii. The green density map represent the distribution of the energy density $\rho$.}
    \label{fig:Orbitn0}
\end{figure}

Although Eq.~(\ref{eq:dotr}) provides an explicit expression for the square of the radial velocity, direct numerical integration of this equation requires careful handling of the sign change of $\dot{r}$ before and after the periapsis or apoapsis. To avoid this issue, using Eqs.~(\ref{eq:mce})~-~(\ref{eq:dotr}), we adopt the following second-order differential equation for the radial motion:
\begin{equation}
\frac{d^2 r}{d\tau^2} = -\Gamma^r_{\mu\nu} \frac{dx^\mu}{d\tau}\frac{dx^\nu}{d\tau}=-\Gamma^r_{tt} \left(\frac{dt}{d\tau}\right)^2-\Gamma^r_{rr} \left(\frac{dr}{d\tau}\right)^2-\Gamma^r_{\varphi\varphi} \left(\frac{d\varphi}{d\tau}\right)^2,\label{eq:dr2}
\end{equation}
where the Christoffel symbols $\Gamma^r_{\mu\nu}$ are computed from the metric coefficients $g_{tt}$ and $g_{rr}$, and their radial derivatives. Hence, the trajectories can be obtained by integrating Eqs.~(\ref{eq:mcl}) and (\ref{eq:dr2}). We employ the fourth-order Runge-Kutta method for this numerical integration, and the step size is $0.01$. The corresponding convergence test can be found in Appendix~\ref{apA}. 

\begin{figure}[htbp]
    \centering
		\subfigure[~$n=1,\xi=-1$]{  
			\includegraphics[width=0.3\textwidth, angle =0]{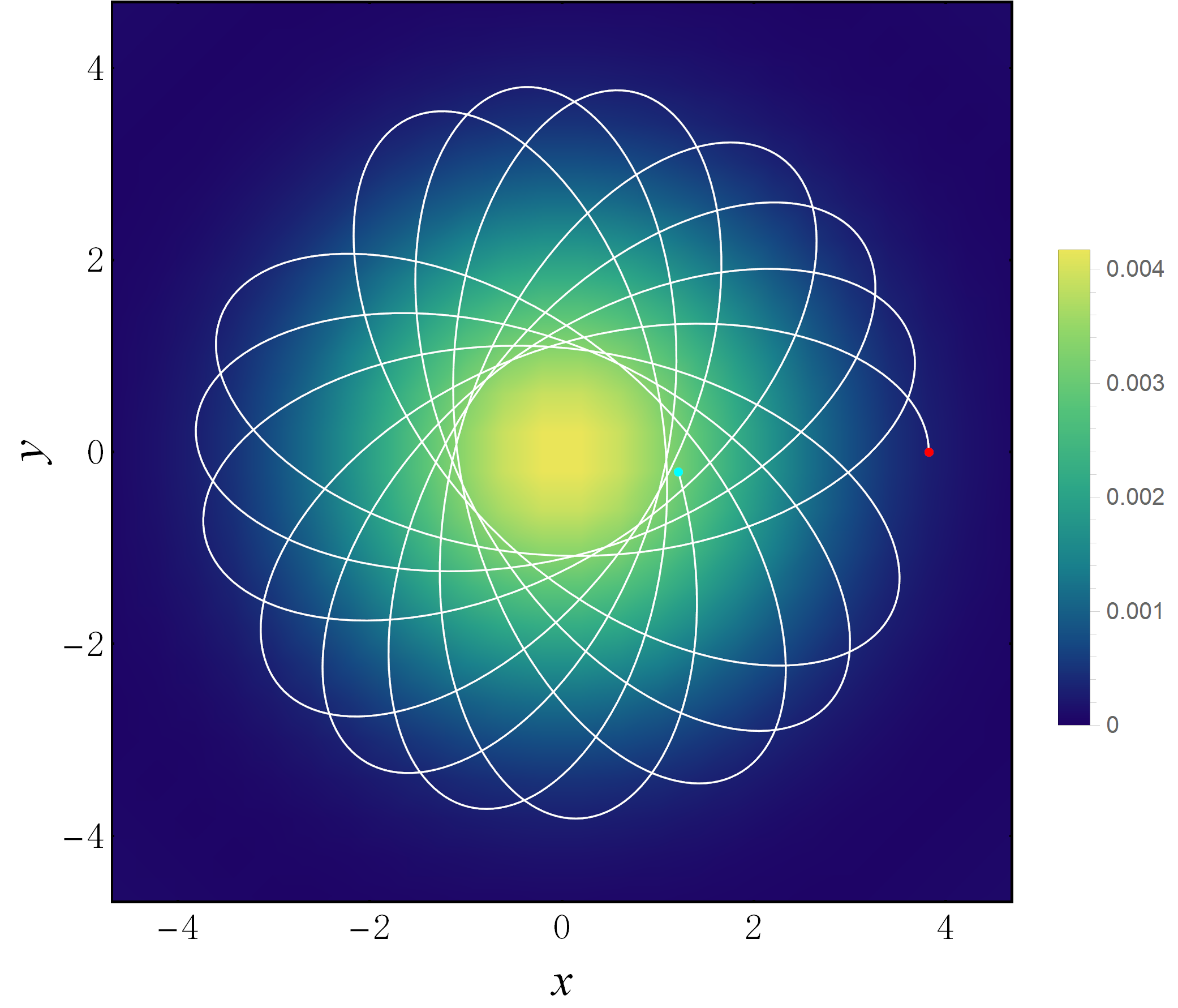}
		}
        \subfigure[~$n=1,\xi=0$]{  
			\includegraphics[width=0.3\textwidth, angle =0]{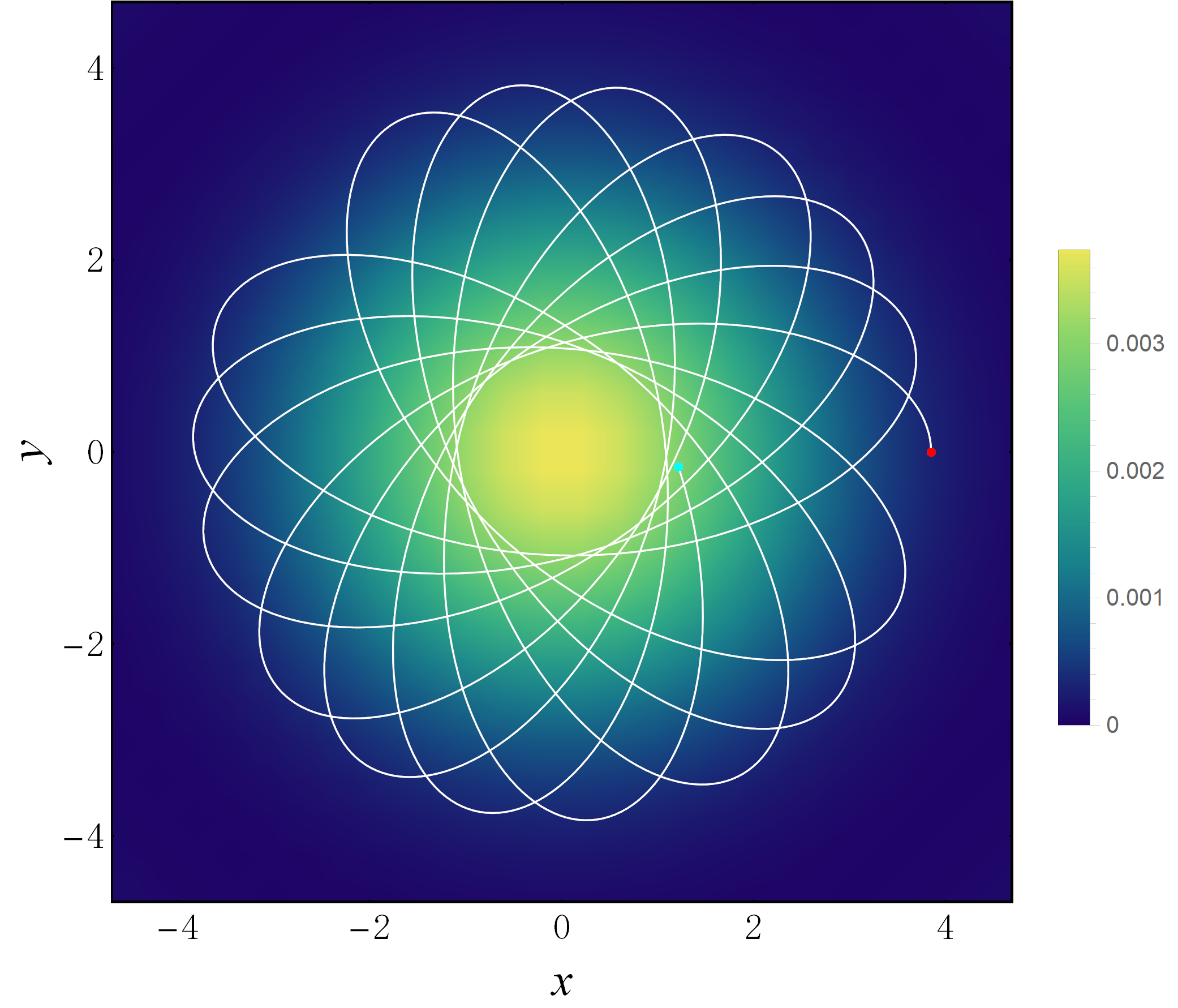}
		}
        \subfigure[~$n=1,\xi=1$]{  
			\includegraphics[width=0.3\textwidth, angle =0]{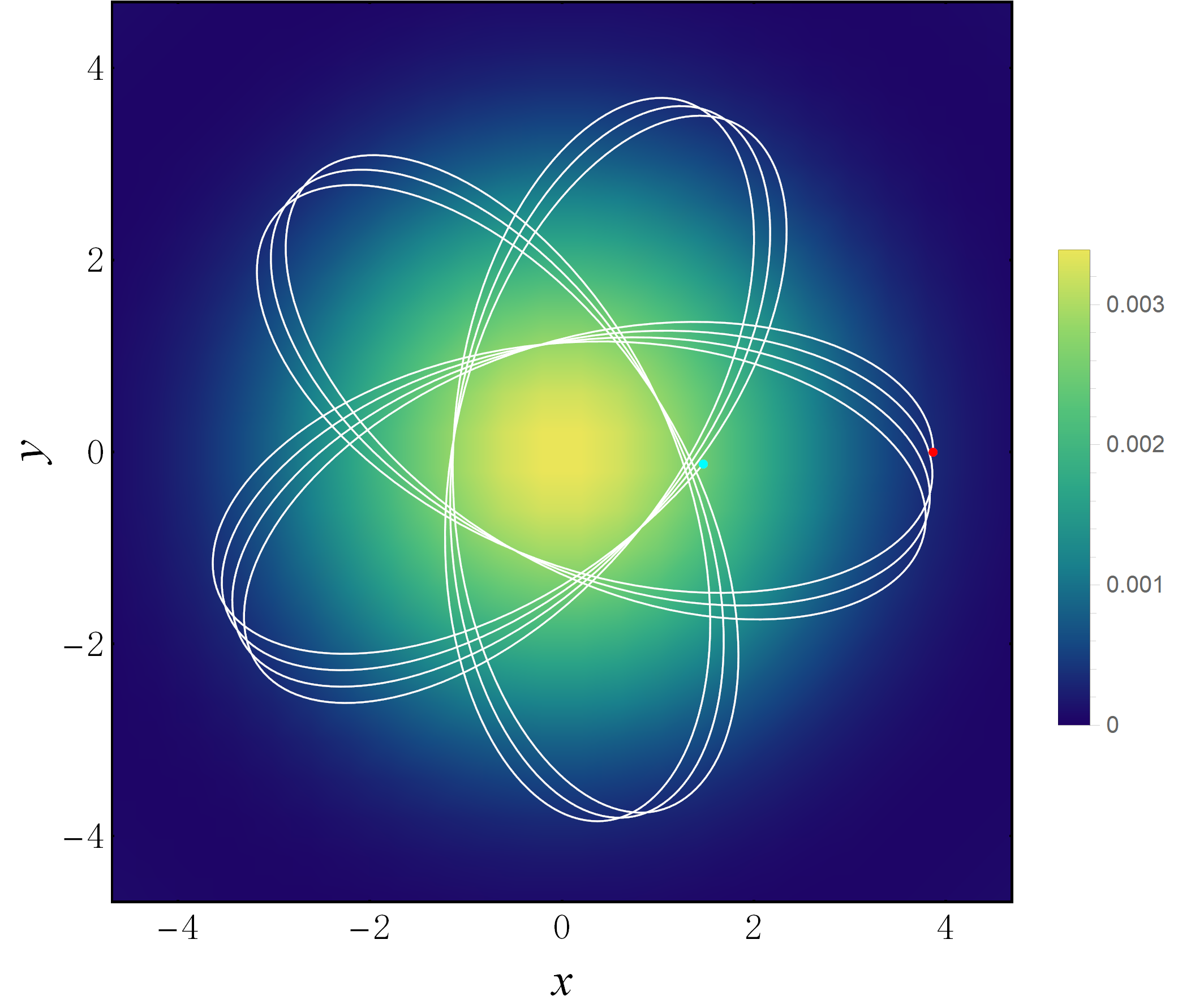}
		}
        \subfigure[~$n=1,\xi=-1$]{  
			\includegraphics[width=0.3\textwidth, angle =0]{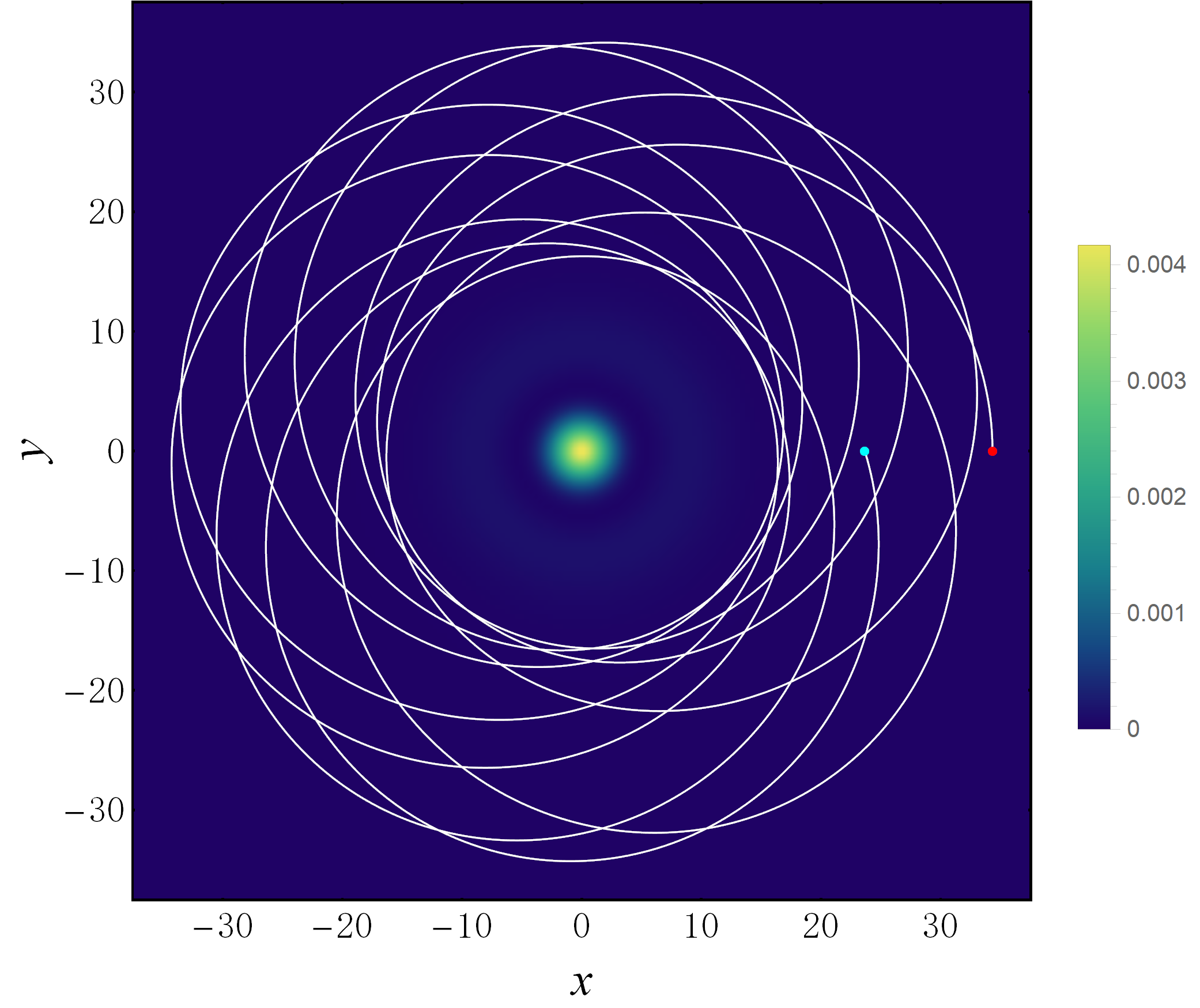}
		}
        \subfigure[~$n=1,\xi=0$]{  
			\includegraphics[width=0.3\textwidth, angle =0]{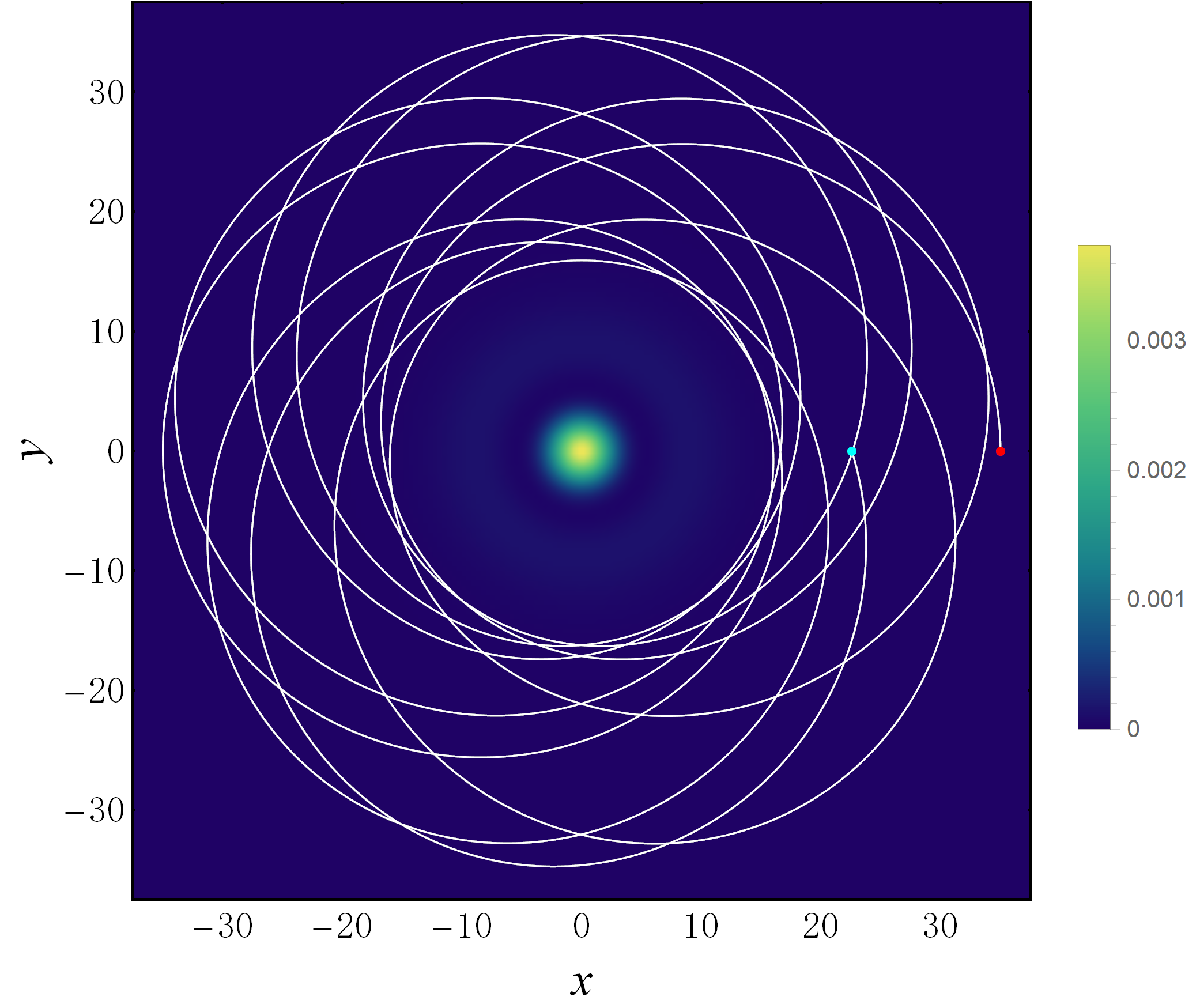}
		}
        \subfigure[~$n=1,\xi=1$]{  
			\includegraphics[width=0.3\textwidth, angle =0]{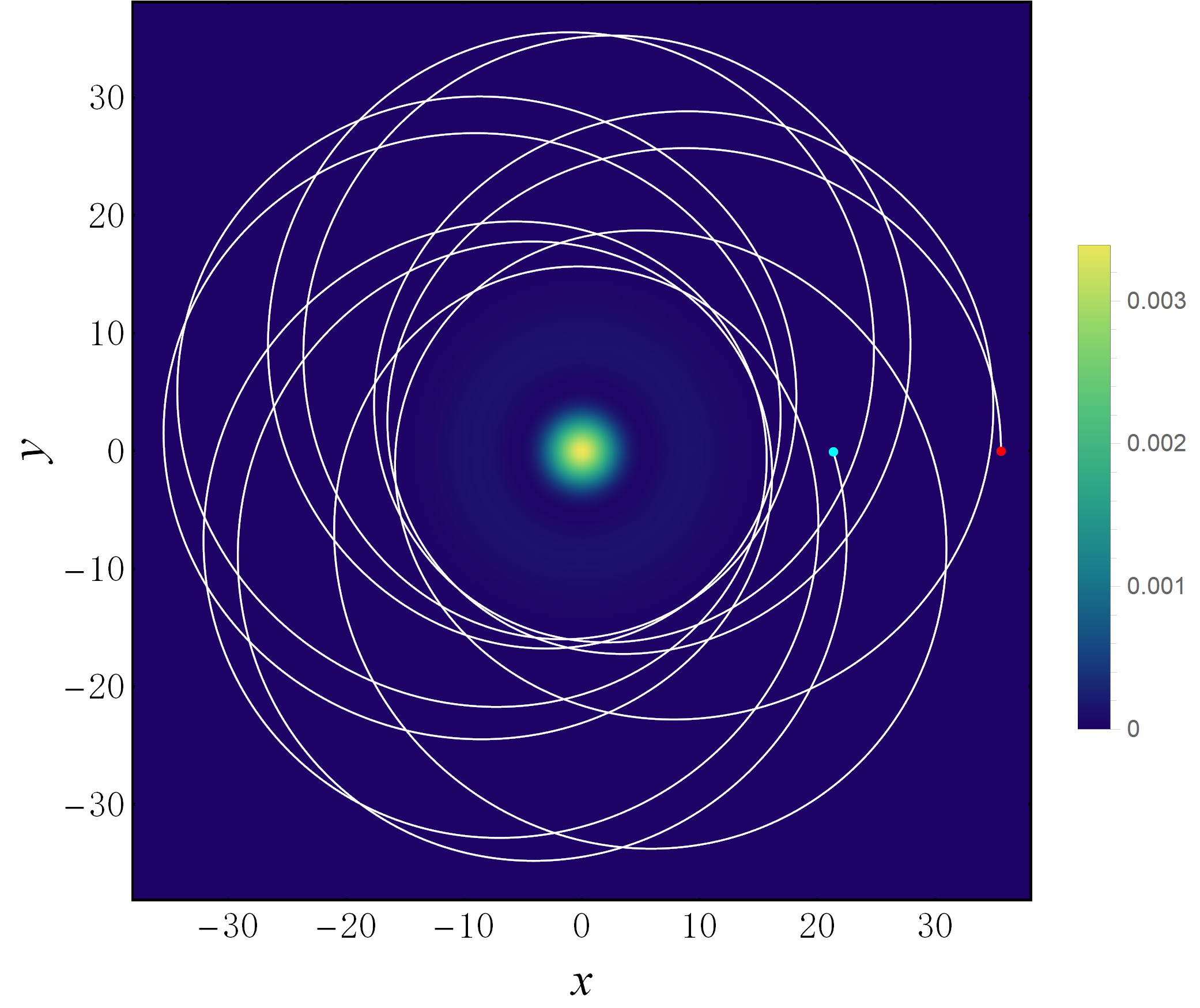}
		}
    \caption{The orbit for the first excited state boson stars with different values of the coupling parameter $\xi$, calculated for $\omega = 0.9$. The top panels ($L = 0.5$, $E = 0.82$) show penetrating orbits with smaller orbital radii, while the bottom panels ($L = 6$, $E = 0.975$) show grazing orbits with larger orbital radii. The green density map represent the distribution of the energy density $\rho$.}
    \label{fig:Orbitn1}
\end{figure}

The resulting trajectories corresponding to the effective potential shown in Fig.~\ref{fig:EFF} for the ground and excited states are presented in Figs.~\ref{fig:Orbitn0} and~\ref{fig:Orbitn1}, respectively. In each panel, the green colormap indicates the energy density $\rho$ of the boson stars, with $x=r\cos\theta\sin\varphi, y=r\sin\theta\sin\varphi$. It can be observed that these bound orbits are spiral trajectories that gradually spread out in space like a rose pattern, revealing significant periastron precession. 

Consistent with the effective potential presented in Fig.~\ref{fig:EFF}, the top panels of Figs.~\ref{fig:Orbitn0} and~\ref{fig:Orbitn1} show that for a smaller angular momentum $L$, the particle is able to penetrate deep into the core of the boson star (i.e., the region of high energy density) and form ``penetrating orbits". For the same number of nodes, the coupling parameter has a significant impact on the shape of these penetrating orbits. In contrast, as shown in the bottom panels of Figs.~\ref{fig:Orbitn0} and~\ref{fig:Orbitn1}, a larger angular momentum generates a stronger centrifugal barrier, keeping the test particle away from the core of the boson star and confining it to ``grazing orbits" with larger average radii. For a fixed $n$, the shape of the orbits is barely affected by the coupling parameter. It is worth noting that, since the penetrating orbits can traverse the interior of the star, they can more fully manifest the unique contribution of the horizonless nature of boson stars compared to grazing orbits.

The apastron $r_a$ and the periastron $r_p$ of these orbits determine the eccentricity $\epsilon=\frac{r_a-r_p}{r_a+r_p}$, which can quantify the shape of the orbit. For the ground and first excited states, we present these orbital parameters in Tabs. \ref{tab:OrbitParan0} and \ref{tab:OrbitParan1}, respectively. It can be observed that the eccentricity $\epsilon$ decreases as $\xi$ increases. This means that a larger value $\xi$ leads to more circular orbits, whereas a lower value $\xi$ results in more eccentric orbits. 

\begin{table}[htbp]
\centering
\setlength{\tabcolsep}{12.5pt} 
\begin{tabular}{|c|c|c|c|c|c|c|}
\hline
$(L,E)$ & $\xi$ & $r_p$ & $r_a$ & $\Delta r$ & $\varepsilon$ & $T_{2\pi}$ \\
\hhline{|=|=|=|=|=|=|=|} 
\multirow{3}{*}{$(0.5,\,0.9)$} & -1 & 0.9588 & 5.8939 & 4.9351 & 0.7202 & 539.284 \\
\cline{2-7}
& 0  & 0.9721 & 5.9767 & 5.0046 & 0.7202 & 548.997 \\
\cline{2-7}
& 1  & 0.9842 & 6.0527 & 5.0686 & 0.7203 & 558.042 \\
\hhline{|=|=|=|=|=|=|=|} 
\multirow{3}{*}{$(3,\,0.985)$} & -1 & 8.1393 & 30.5999 & 22.4606 & 0.5798 & 3226.26 \\
\cline{2-7}
& 0  & 7.9137 & 31.3419 & 23.4282 & 0.5968 & 3263.95 \\
\cline{2-7}
& 1  & 7.7213 & 32.0120 & 24.2907 & 0.6114 & 3299.90 \\
\hline
\end{tabular}
\caption{The orbital parameters associated with orbits for the ground state in Fig.~\ref{fig:Orbitn0}. $1^{st}$: different parameter $(L,E)$. $2^{nd}$ column: the three different models for the ground state. $3^{rd}-7^{th}$: the periastron $r_p$, apastron $r_a$, and radial range $\Delta r=r_a-r_p$, eccentricity $E$, and amzimuthal orbital period $T_{2\pi} $.}
\label{tab:OrbitParan0}
\end{table}

\begin{table}[htbp]
\centering
\setlength{\tabcolsep}{12pt} 
\begin{tabular}{|c|c|c|c|c|c|c|}
\hline
$(L,E)$ & $\xi$ & $r_p$ & $r_a$ & $\Delta r$ & $\varepsilon$ & $T_{2\pi}$ \\
\hhline{|=|=|=|=|=|=|=|} 
\multirow{3}{*}{$(0.5,\,0.82)$} & -1 & 1.0809 & 3.8226 & 2.7417 & 0.5591 & 391.47 \\
\cline{2-7}
& 0  & 1.1087 & 3.8472 & 2.7384 & 0.5526 & 393.45 \\
\cline{2-7}
& 1  & 1.1343 & 3.8687 & 2.7345 & 0.5466 & 401.65 \\
\hhline{|=|=|=|=|=|=|=|} 
\multirow{3}{*}{$(6,\,0.975)$} & -1 & 16.2738 & 34.3660 & 18.0921 & 0.3573 & 3397.14 \\
\cline{2-7}
& 0  & 15.9320 & 35.0319 & 19.0999 & 0.3748 & 3419.37 \\
\cline{2-7}
& 1  & 15.6472 & 35.6210 & 19.9737 & 0.3896 & 3440.06 \\
\hline
\end{tabular}
\caption{The orbital parameters associated with orbits for the first excited state in Fig.~\ref{fig:Orbitn0}. $1^{st}$: different parameter $(L,E)$. $2^{nd}$ column: the three different models for the first excited states. $3^{rd}-7^{th}$: the periastron $r_p$, apastron $r_a$, and radial range $\Delta r=r_a-r_p$, eccentricity $E$, and amzimuthal orbital period $T_{2\pi} $.}
\label{tab:OrbitParan1}
\end{table}

 \subsection{Gravitational waves from EMRIs}
In this subsection, we will analyze the gravitational waveforms generated by the two types of orbits discussed above. Our goal is to elucidate how the non-minimal torsion coupling parameter $\xi$ modifies standard boson stars in GR ($\xi=0$). 

To obtain the gravitational waveforms emitted by periodic orbits in boson stars, we adopt the Kludge method~\cite{Babak:2006uv}, which has already been applied to boson stars in modified gravity~\cite{Liu:2025swi}.  In this method, once the trajectory of the small body is obtained, an effective trajectory can be constructed by projecting the spherical polar coordinates onto a pseudo-Euclidean space, as follows:
\begin{equation}
    x=r\sin\theta\cos\varphi,\quad y=r\sin\theta\sin\varphi,\quad z=r\cos\theta.
    \label{eq:Ecoordinate}
\end{equation}
The gravitational waveform is obtained up to quadratic order by using the quadrupole relation
\begin{equation}
    h_{ij}=\frac{2}{D_L}\frac{d^2I_{ij}(t^\prime)}{dt^{\prime 2}}\Big|_{t^\prime=t-D_L}, \label{eq:hij}
\end{equation}
here, $D_L$ is the luminosity distance from the EMRI system to the detector, and the symmetric trace-free mass quadrupole moment $I_{ij}(t^\prime)$ is defined as
\begin{equation}
    I_{ij}(t^\prime)=\int\rho(t^\prime, \bm{x}^\prime)\left(x^{\prime i}x^{\prime j}-\frac{1}{3}\delta^{ij}r^{\prime2}\right)d^3x^\prime,
    \label{eq:Iij}
\end{equation}
where $x^ {\prime i}$ is the position of the test particle (small star) along the corresponding geodesic trajectory described by Eq.~\ref{eq:Ecoordinate}, and $\rho(t^\prime,\bm{x}^\prime)$ is the mass density.

 Under the approximation of the ``point-mass", the mass density of the orbiting body with mass $m_b$ on a trajectory $\bm{Z}(t)$ is
 \begin{equation}
     \rho(t,\bm{x})=m_b\delta^3(\bm{x}-\bm{Z}(t)). \label{eq:rhoz}
 \end{equation}

Following the method in~\cite{Babak:2006uv}, we use Cartesian coordinates $(x, y, z)$ as the basis for generating gravitational waveforms. Thus, combining Eqs.~(\ref{eq:hij})~-~(\ref{eq:rhoz}) and evaluating the integrals and derivatives, one can obtain the explicit expression of the metric perturbations describing the GWs as~\cite{Yang:2024lmj}
\begin{equation}
h_{i j}=\frac{2 m_{b}}{D_{\mathrm{L}}}\left[a_i x_j+a_j x_i+2 v_i v_j-\frac{2}{3} \delta_{i j}\left(\boldsymbol{a} \cdot \boldsymbol{x}+v^2\right)\right],\label{eq:hijfinal}
\end{equation}
where $a_i$ and $v_i$ denote the Cartesian components of the acceleration and velocity of the orbiting body, respectively.

To construct the observable polarizations, we construct a detector-adapted coordinate system $(X, Y, Z)$ with the origin coinciding with that of the original $(x, y, z)$ coordinate system, both centered on the supermassive object. The basis vectors of this
frame are expressed in the original coordinates $(x,y,z)$ as~\cite{Poisson:2014} 
\begin{equation}
\begin{aligned}
& \boldsymbol{e}_X=[\cos \zeta,-\sin \zeta, 0], \\
& \boldsymbol{e}_Y=[\cos \iota \sin \zeta, \cos \iota \cos \zeta,-\sin \iota], \\
& \boldsymbol{e}_Z=[\sin \iota \sin \zeta, \sin \iota \cos \zeta, \cos \iota],
\end{aligned}
\end{equation}
where $\iota$ is the inclination angle of the orbital plane with respect to the $X-Y$ plane, and $\zeta$ is the longitude of the pericenter measured in the orbital plane.
 
Subsequently, the corresponding physical polarizations $h_{+}$ and $h_{\times}$ from Eq.~(\ref{eq:hijfinal}) are obtained as~\cite{Babak:2006uv}:
\begin{equation}
\begin{aligned}
h_{+} & =\frac{1}{2}\left(\boldsymbol{e}_X^i \boldsymbol{e}_X^j-\boldsymbol{e}_Y^i \boldsymbol{e}_Y^j\right) h_{i j}=\frac{1}{2}\left(h_{\zeta \zeta}-h_{\iota \iota}\right), \\
h_{\times} & =\frac{1}{2}\left(\boldsymbol{e}_X^i \boldsymbol{e}_Y^j+\boldsymbol{e}_Y^i \boldsymbol{e}_X^j\right) h_{i j}=h_{\iota \zeta},
\end{aligned}
\end{equation}
in which the components $h_{\zeta \zeta}$, $h_{\iota \iota}$, $h_{\iota \zeta}$ are given by
\begin{equation}
\begin{aligned}
h_{\zeta \zeta} & =h_{x x} \cos ^2 \zeta-h_{x y} \sin 2 \zeta+h_{y y} \sin ^2 \zeta, \\
h_{\iota \iota} & =\cos ^2 \iota\left[h_{x x} \sin ^2 \zeta+h_{x y} \sin 2 \zeta+h_{y y} \cos ^2 \zeta\right]+h_{z z} \sin ^2 \iota-\sin 2 \iota\left[h_{x z} \sin \zeta+h_{y z} \cos \zeta\right], \\
h_{\iota \zeta} & =\cos \iota\left[\frac{1}{2} h_{x x} \sin 2 \zeta+h_{x y} \cos 2 \zeta-\frac{1}{2} h_{y y} \sin 2 \zeta\right]+\sin \iota\left[h_{y z} \sin \zeta-h_{x z} \cos \zeta\right] .
\end{aligned}
\end{equation}

Now, let us consider a fictitious EMRI system composed of the boson stars with mass $M = 10^6M_\odot$ and a compact stellar-mass object orbiting it with mass $M_b = 10M_\odot$ at a distance $D_L = 0.1 \mathrm{Gpc}$, inclination angle $\iota = \pi/3$ and the longitude of the pericenter $\zeta = \pi/3$.

\begin{figure}[htbp]
\subfigure[~$h_+$]  {\label{Fig_GW_n0_0_plus_L3E0985}
\includegraphics[width=15cm]{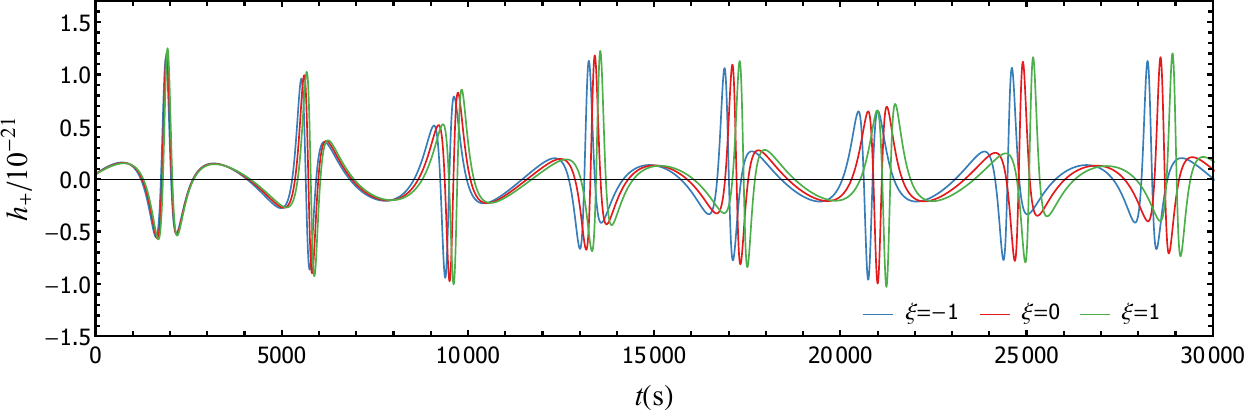}}
\subfigure[~$h_\times$]  {\label{Fig_GW_n0_0_cross_L3E0985}
\includegraphics[width=15cm]{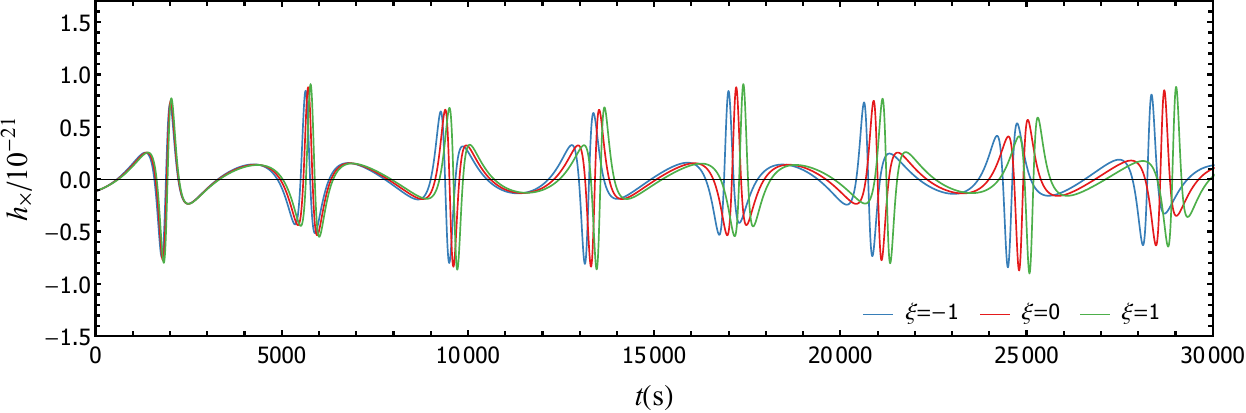}}
\caption{GW polarizations $h_+$ and $h_\times$ with different values of the coupling parameter $\xi$ corresponding to trajectories for $(L=3, E=0.985)$, where $n=0$ and $\omega=0.9$.}
\label{fig:n0Gravitational_Waveforms_L3E0985}
\end{figure}

\begin{figure}[t]
\subfigure[~$h_+$]  {\label{Fig_GW_n1_0_plus_L6E0975}
\includegraphics[width=15cm]{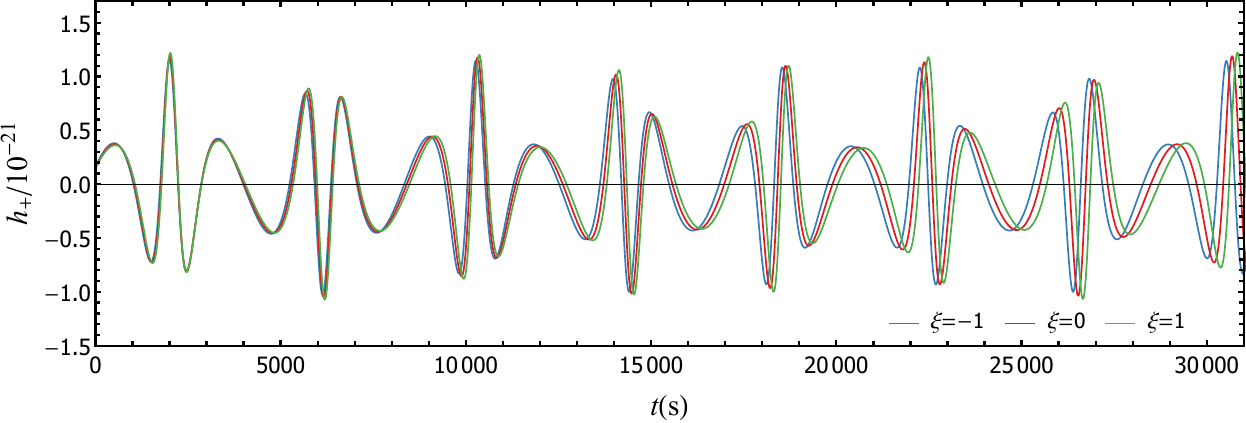}}
\subfigure[~$h_\times$]  {\label{Fig_GW_n1_0_cross_L6E0975}
\includegraphics[width=15cm]{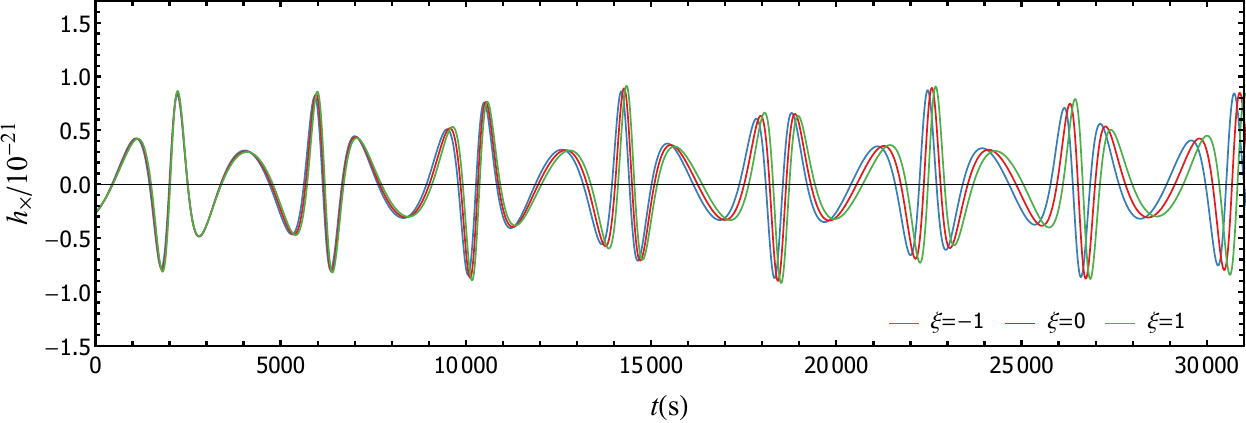}}
\caption{GW polarizations $h_+$ and $h_\times$ with different values of the coupling parameter $\xi$ corresponding to trajectories for the parameter set $(L=6, E=0.975)$, where $n=1$ and $\omega=0.9$}
\label{fig:n1Gravitational_Waveforms_L6E0975}
\end{figure}

Based on different nodes $n$ and parameter sets $(E,L)$ (corresponding to the orbits in subsection~\ref{sec:orbit}), Figs.~\ref{fig:n0Gravitational_Waveforms_L3E0985} to~\ref{fig:n1Gravitational_Waveforms_L6E0975} present the polarization components $h_{+}$ and $h_{\times}$ as functions of the coordinate time, showing how $\xi$ influences the GW. In all of these figures, the red waveform corresponds to the boson star case in GR, whereas the blue and green waveforms correspond to the cases with parameters $\xi = -1$ and $\xi = 1$, respectively.

\begin{figure}[htbp]
\subfigure[~$h_+$]  {\label{Fig_GW_n0_plus_L05E09}
\includegraphics[width=15cm]{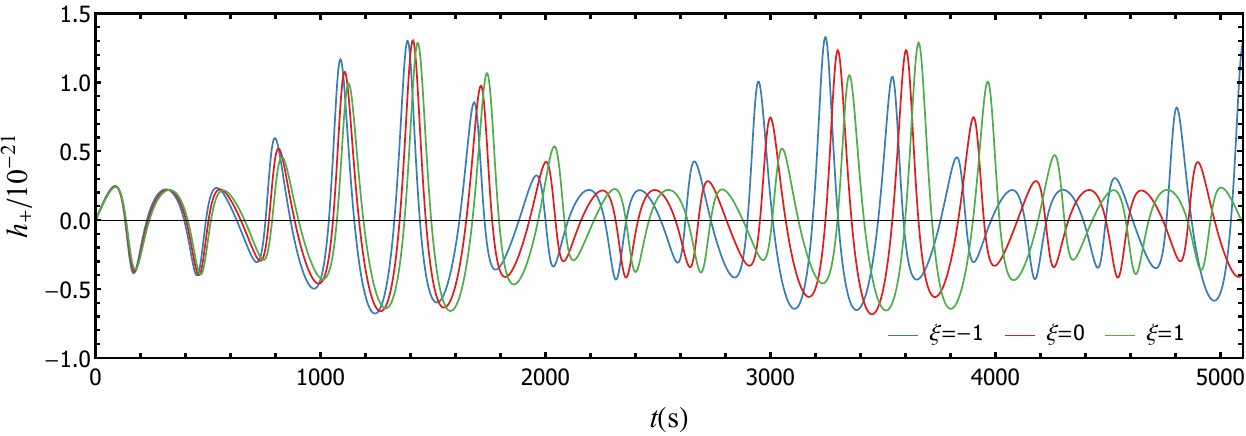}}
\subfigure[~$h_\times$]  {\label{Fig_GW_n0_cross_L05E09}
\includegraphics[width=15cm]{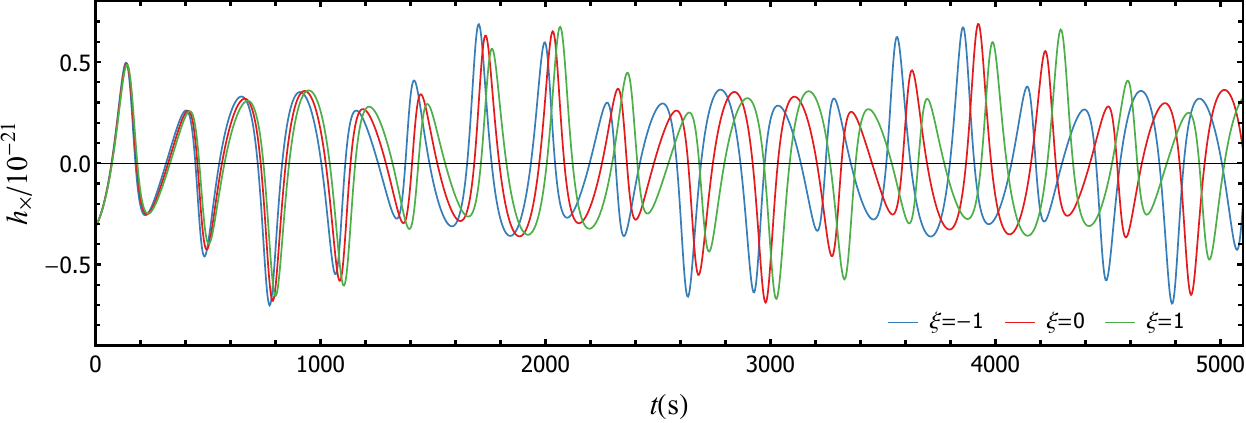}}
\caption{GW polarizations $h_+$ and $h_\times$ with different values of the coupling parameter $\xi$ corresponding to trajectories for the parameter set $(L=0.5, E=0.9)$, where $n=0$ and $\omega=0.9$.}
\label{fig:n0Gravitational_Waveforms_L05E09}
\end{figure}

The gravitational waves corresponding to the grazing orbits are shown in Figs.~\ref{fig:n0Gravitational_Waveforms_L3E0985} and \ref{fig:n1Gravitational_Waveforms_L6E0975}. It can be observed that  the waveform structure exhibits intermittent burst characteristics. This waveform structure resembles the signal morphology of the black hole EMRIs: The peaks in the gravitational waveform correspond to the ``whirl" period of the black hole GW signal, while the relatively flat periods correspond to the ``zoom" period. The comparison of different coupling parameters reveals that waveforms with non-zero coupling parameters $\xi$ are very similar to the GR case (i.e., $\xi=0$) in the early stage. As the system evolves further, noticeable deviations gradually appear in both the amplitude and frequency of the waveform. To quantitatively characterize this divergence, we employ the normalized strain deviation, defined as~\cite{Liu:2025swi}
\begin{equation}
\Delta h\left(t\right)=\frac{\left|h_\xi\left(t\right)-h_{\mathrm{GR}}\left(t\right)\right|}{h_{\mathrm{peak}}},
\end{equation}
where $h_\xi$ represents the strain amplitude of either the plus
or cross polarization for boson stars with $\xi\neq0$, and $h_{\mathrm{peak}}$ is the peak strain amplitude of the corresponding GR. We adopt a threshold of $\Delta h = 10\%$ to define the onset of discernible de‑phasing. In the case of the grazing orbits, the threshold is approximately reached at $t = 1705$\,s for $n=0$ and $t = 5772$\,s for $n=1$.

\begin{figure}[t]
\subfigure[~$h_+$]  {\label{Fig_GW_n1_plus_L05E082.pdf}
\includegraphics[width=15cm]{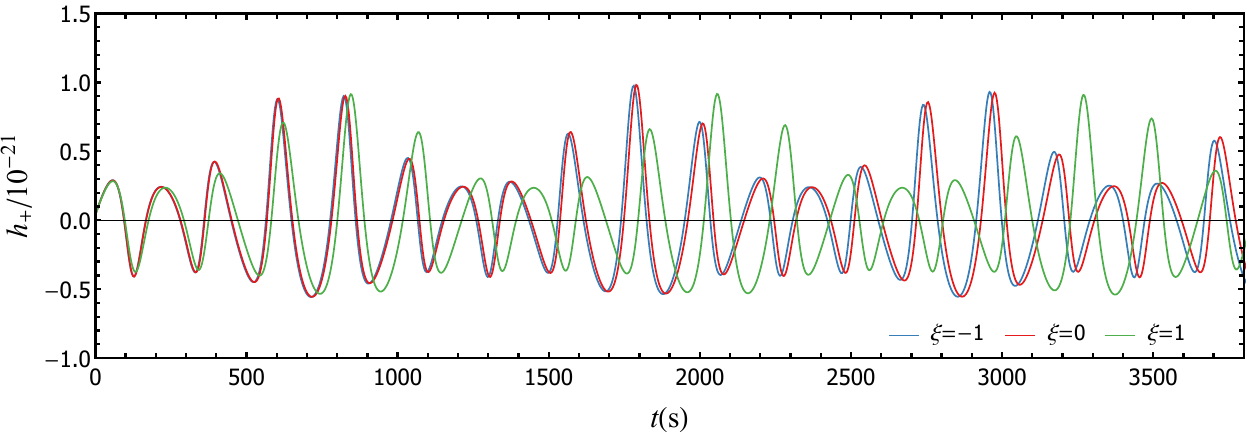}}
\subfigure[~$h_\times$]  {\label{Fig_GW_n1_cross_L05E082}
\includegraphics[width=15cm]{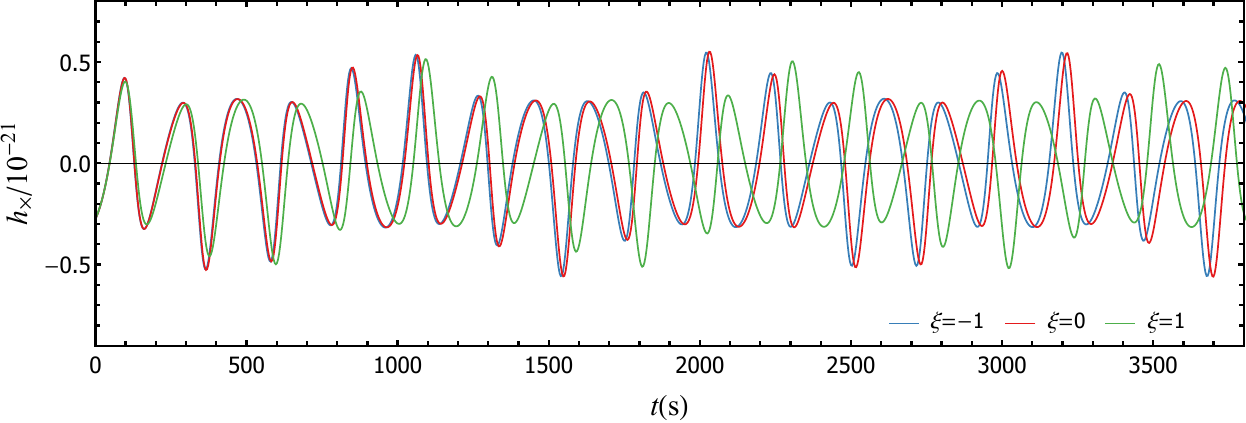}}
\caption{GW polarizations $h_+$ and $h_\times$ with different values of the coupling parameter $\xi$ corresponding to trajectories for parameter set $(L=0.5, E=0.82)$, where $n=1$ and $\omega=0.9$.}
\label{fig:n1Gravitational_Waveforms_L05E082}
\end{figure}

Unlike EMRI systems associated with supermassive central black holes, where the inspiral terminates at the event horizon and the signal decays in the form of quasinormal modes, for the boson stars, the penetrating orbits can enter the stellar core, leading many distinctive features to arise compared to black holes. Due to the effective potential being deeper and narrower compared to grazing orbits, particles are tightly bound near a certain radius, resulting in relatively uniform motion. Therefore, as shown in Figs.~\ref{fig:n0Gravitational_Waveforms_L05E09} and~\ref{fig:n1Gravitational_Waveforms_L05E082}, the GW radiation appears as continuous oscillations that are amplitude-modulated, rather than intermittent bursts separated by relatively quiet intervals. Compared to grazing orbits, the effects of non-minimal coupling on the amplitude and frequency of such orbital waveforms manifest at later times. For the ground state, the threshold of discernible de‑phasing occurs approximately at $434$\,s, while for the excited state, approximately at $121$\,s. It is worth noting that, as with gravitational waves from grazing orbits, a lower coupling parameter causes the frequency of gravitational waves generated by penetrating orbits to increase, whereas a larger one causes the frequency to decrease. In particular, for penetrating orbits, it can also be seen that the difference between the gravitational waveforms of the excited state at $\xi=0$ and those at $\xi\neq0$ is larger than the difference between the gravitational waveforms of the ground state at $\xi=0$ and those at $\xi\neq0$.

Next, we discuss the detectability of the GW signals we obtained. For this purpose, we calculate the frequency-domain characteristic strain $h_c(f)$ from the discrete Fourier transforms $\tilde{h}_{+}$ and $\tilde{h}_{\times}$ of the time-domain signals $h_{+}$ and $h_{\times}$~\cite{Robson:2018ifk}:

\begin{equation}
h_c(f) = 2 f \sqrt{|\tilde{h}_{+}(f)|^2 + |\tilde{h}_{\times}(f)|^2}.
\end{equation}

\begin{figure}[t]
\centering
\subfigure[~$(n=0,L=3, E=0.985)$]  {\label{Fig_hc_L3E0985}
\includegraphics[width=7cm,height=5cm]{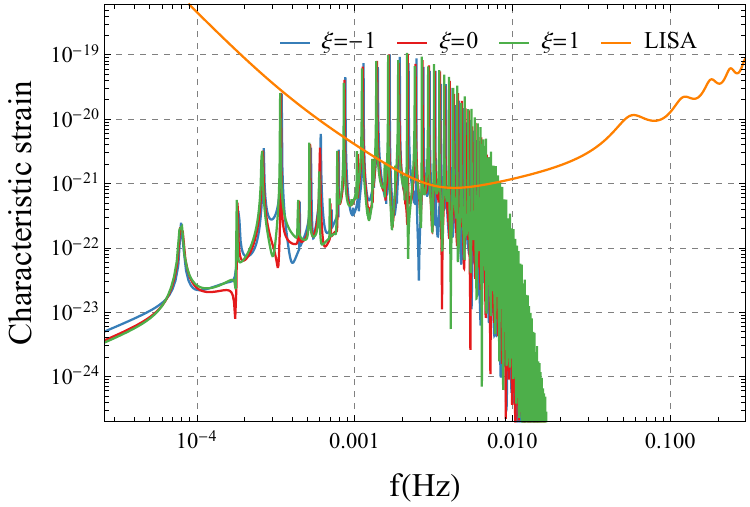}}
\subfigure[~$(n=0,L=0.5, E=0.9)$]  {\label{Fig_hc_L05E09}
\includegraphics[width=7cm,height=5cm]{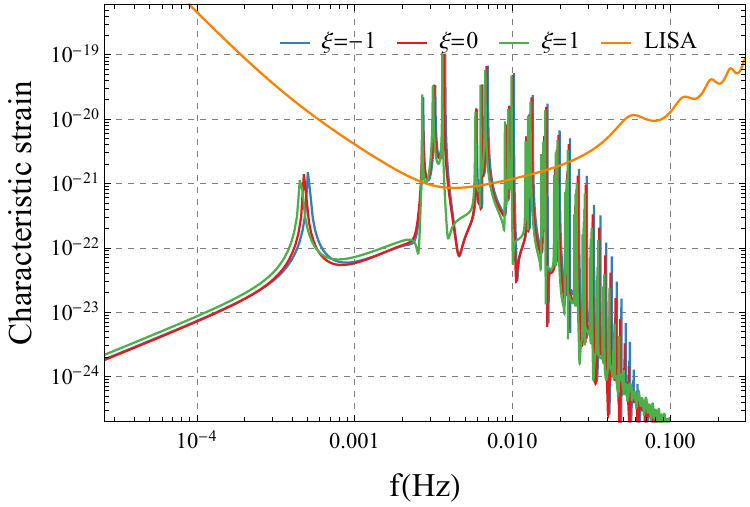}}
\quad
\subfigure[~$(n=1,L=6, E=0.975)$]  {\label{Fig_hc_L6E0975}
\includegraphics[width=7cm,height=5cm]{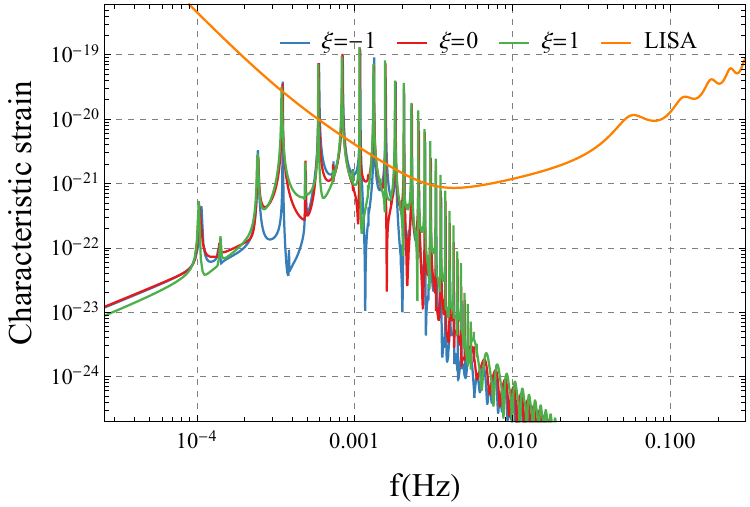}}
\subfigure[~$(n=1,L=0.5, E=0.82)$]  {\label{Fig_hc_L05E082}
\includegraphics[width=7cm,height=5cm]{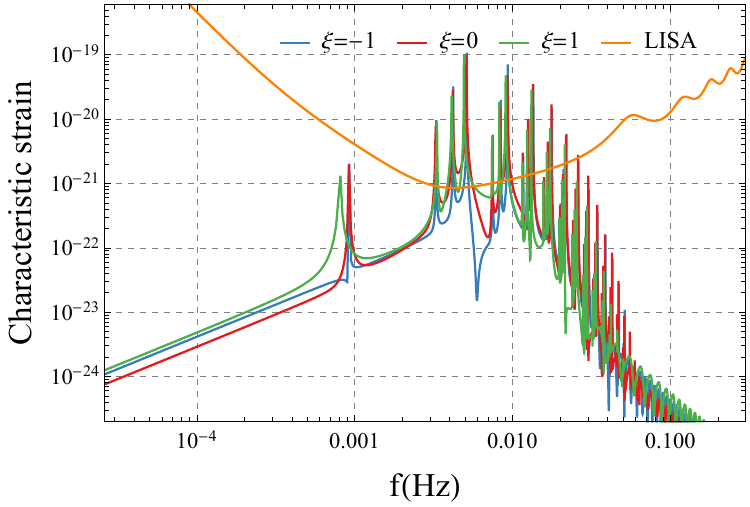}}
\caption{The characteristic strain curve for different values of the  parameter $\xi$. Top and bottom panels correspond to the ground state and the excited state, respectively. Left and right panels correspond to grazing and penetrating orbits, respectively. The LISA sensitivity curve is shown in orange for comparison.}
\label{fig:Characteristic_strain}
\end{figure}

Fig.~\ref{fig:Characteristic_strain} shows the characteristic strain of the EMRI systems for ground-state and excited-state boson stars, respectively. The organe curves represent the LISA sensitivity threshold, while the left and right panels correspond to grazing and penetrating orbits, respectively. In terms of spectral distribution for both the ground and excited states, grazing orbits yield a dominant GW frequency of approximately $0.01$\,Hz, whereas for penetrating orbits it is around $0.1$\,Hz. This difference corresponds exactly to the orbital periods listed in Tabs.~\ref{tab:OrbitParan0} and \ref{tab:OrbitParan1}: the period of grazing orbits ($T_{2\pi}\in (300,600)$ s) is longer than that of penetrating orbits ($T_{2\pi}\in (3200,3500)$ s), leading to a lower dominant frequency. By comparing the colored curves representing different non-minimal coupling parameters $\xi$, it is evident that the corrective effects of the non-minimal coupling on the signals are more pronounced in the low-frequency range.

Although the peak amplitudes in all cases lie above the LISA sensitivity threshold, suggesting potential detectability, the sensitivity to $\xi$ varies significantly between the two types of orbit. For grazing orbits, although the harmonic peaks for different $\xi$ values all exceed the sensitivity threshold, they are overlapping highly in the spectrum, making them difficult to distinguish effectively. Conversely, penetrating orbits display discernible frequency shifts and intensity modulations near $f \sim 0.003$ Hz, a feature consistently observed in both ground-state and excited-state scenarios. Consequently, compared to grazing orbits, the unique modulation features of penetrating orbits are more likely to offer opportunities for future space-based detectors to identify non-minimal coupling effects, and hold the potential to provide a prospective observational channel for constraining the non-minimal coupling parameter $\xi$ between curvature and torsion.

\section{CONCLUSION}\label{sec:conclusion}

In this work, we investigate a system featuring a complex scalar field non-minimally coupled to torsion within the framework of teleparallel gravity, obtaining ground-state and excited-state boson stars. Our results show that the compactness $C$ of boson stars increases with $\xi$. In other words, compared to boson stars in GR (corresponding to $\xi=0$), the introduction of a non‑minimal coupling with a positive coupling parameter makes boson stars more compact. Interestingly, for the excited states, we found that once the coupling parameter exceeds a certain threshold, the minimum value of the energy density can become negative, and the four commonly used energy conditions are no longer satisfied. In contrast, for the ground states, the energy density remains positive for all solutions we have considered, and all four energy conditions are consistently satisfied. In other words, the generation of nodes in the scalar field can lead to the violation of the energy conditions, and the satisfaction of the energy conditions imposes certain constraints on the coupling parameter for excited states. This phenomenon has not been observed in standard GR boson star models and can be regarded as a unique effect induced by torsion.


Given the importance of astrophysical observations, we also explored the gravitational waves from EMRIs composed of boson stars. The results show that for both ground-state and excited-state boson stars, the gravitational waveforms generated by grazing orbits confined to larger radii exhibit features similar to those of black holes: intermittent GW bursts separated by long quiescent intervals. In contrast, penetrating orbits that go through the core of boson stars fully illustrate the unique property of the absence of event horizons in boson stars. They produce sustained, amplitude-modulated oscillations with almost no quiescent periods. Moreover, for both types of orbit, a lower coupling parameter $\xi$ causes the frequency of gravitational waves to increase, whereas a larger coupling parameter causes it to decrease.

Based on the frequency-domain characteristic strain of the waveforms for both types of orbit, we found that the peak signals in all cases exceed the LISA sensitivity threshold, which means that these GW signals may be detectable by future space-based GW detectors. However, there is a distinct difference in the sensitivity to $\xi$ between these orbits. For grazing orbits, the portions of the GW strain that exceed the LISA sensitivity are highly overlapping and difficult to distinguish. In comparison, penetrating orbits exhibit significant frequency shifts and intensity modulations in this region.

It is worth noting that the analysis of binding energy shows that a larger coupling parameter broadens the frequency range of stable solutions and reduces the value of the binding energy. However, although binding energy is a useful clue, it does not guarantee that these solutions remain stable under small perturbations which are inevitable in any realistic astrophysical environment. For these boson stars to be physically viable, they must be able to withstand small perturbations and maintain stability. In this respect, some ground-state mini boson star solutions are stable, while all currently discovered excited-state mini boson star solutions are unstable under generic perturbations~\cite{Balakrishna:1997ej}. However, extended boson star models may change the stability of mini boson stars. For example, ground-state mini boson stars, though stable in four dimensions, become unstable in higher dimensions~\cite{Blazquez-Salcedo:2019qrz,Franzin:2024jhg}, and sufficiently large self-interactions can make excited-state mini boson stars stable~\cite{Sanchis-Gual:2021phr,Brito:2023fwr}. Considering that changing the parameter $\xi$ alters various physical properties of mini boson stars, we speculate that the perturbative stability of boson stars with a torsion-coupled field is also likely to change accordingly. A more in-depth investigation of this kind will be left for future work.


    \section*{ACKNOWLEDGEMENTS}
	This work is supported by the National Natural Science Foundation of China (Grant Nos.~12275110, 12247101 and 12475062) and the National Key Research and Development Program of China (Grant No.~2022YFC2204101 and 2020YFC2201503). K. Yang also acknowledges the support of the Natural Science Foundation of Chongqing (Grant No.~CSTB2024NSCQ-MSX0358).
    
\appendix
\section{}\label{apA}

To ensure the accuracy and robustness of the boson star solutions and the orbital integrals, we perform a convergence test of our code for each of them separately in this appendix. 

In principle, the numerical results of a system should obey the following convergence rule:
\begin{equation}
\left|A_h-A\right|\propto h^p,
\label{eq:converse}
\end{equation}
where $A_h$ is the numerical result and $A$ is the exact solution of the system, $h$ corresponds to numerical resolution, and $p$ is the convergence order. When taking three different numerical accuracies from coarse to fine: $h_l$, $h_i$, $h_h$, according to Eq.~\ref{eq:converse}, we have:
\begin{equation}
\left|\frac{A_{h_l}-A_{h_i}}{A_{h_i}-A_{h_h}}\right|=\left|\frac{h^p_l-h^p_i}{h^p_i-h^p_h}\right|=Q,
\label{eq:order}
\end{equation}
here, $Q$ denotes the convergence factor.

	\begin{figure}[!htbp]
		\begin{center}
		\subfigure{ 
			\includegraphics[width=6.5cm, angle =0]{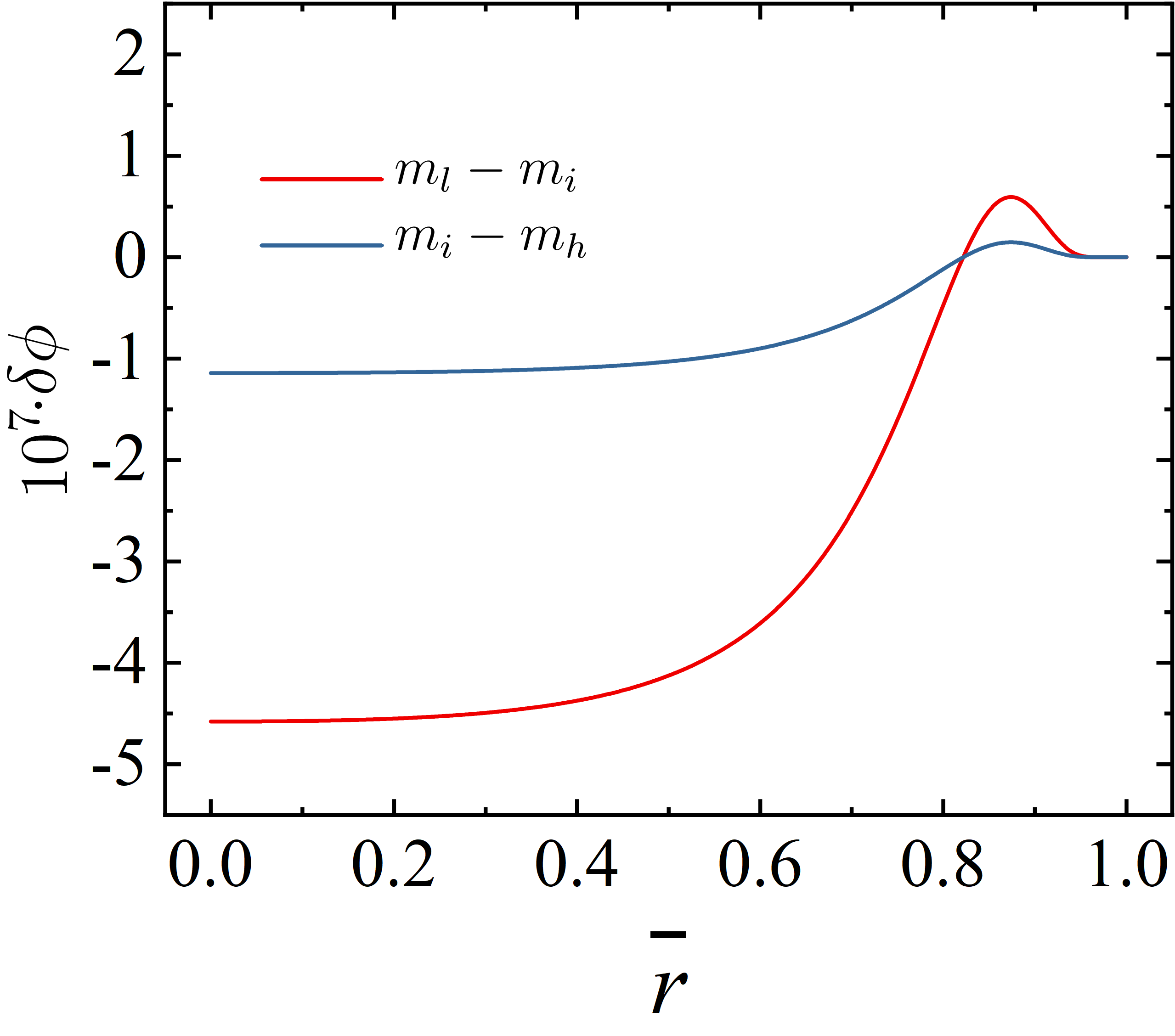}
			\label{fig:conphi}
		}	 
  		\subfigure{  
			\includegraphics[width=6.5cm, angle =0]{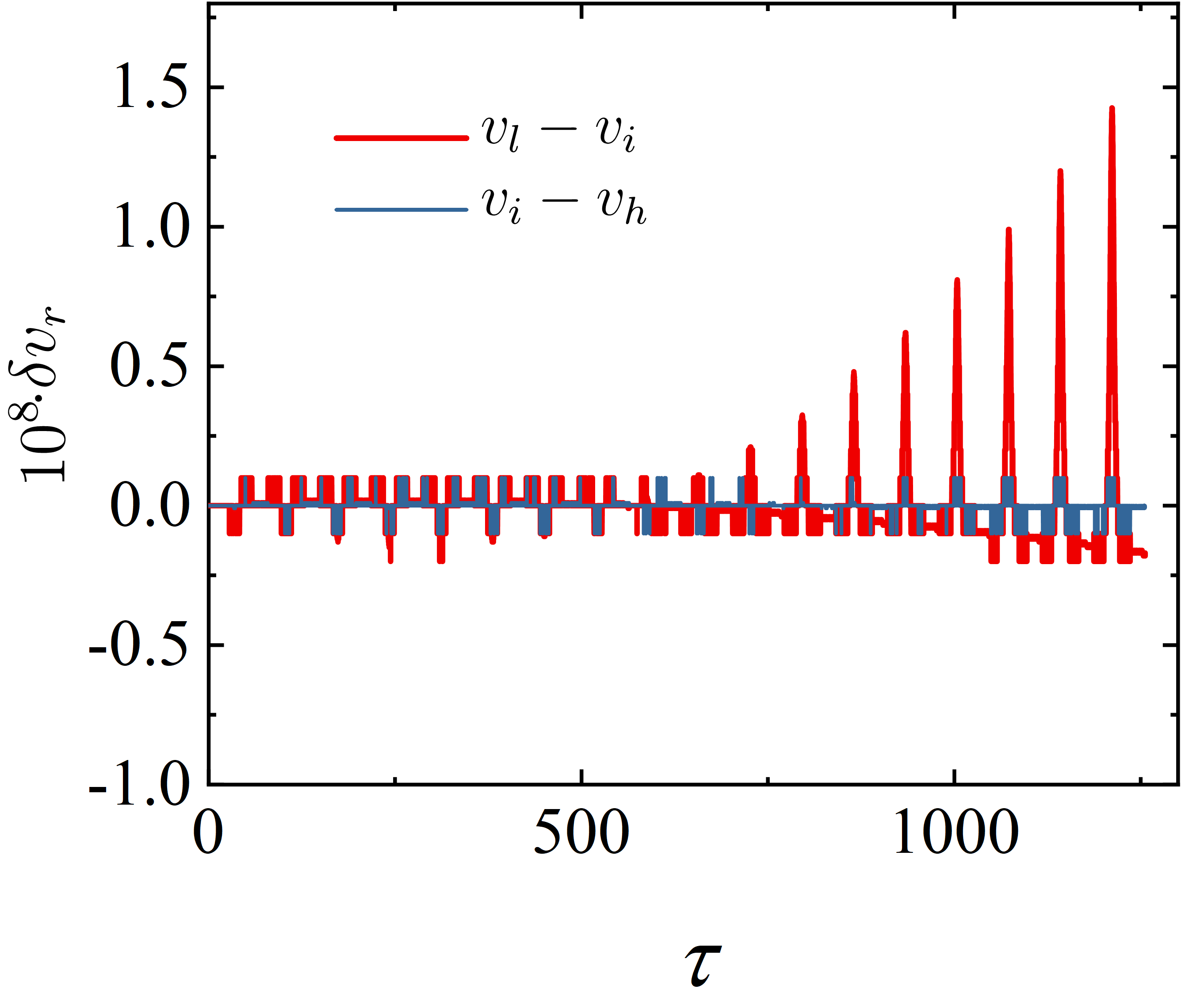}
			\label{fig:conorbit}
		}	 	
  		\end{center}	
		\caption{Left: Convergence test for the scalar function $\phi(r)$; Right: for the radial velocity $v_r$ with ($L=0.5, E=0.9$). In each panel, $n=0$, $\omega=0.9$, $\xi=1$ and the red and blue curves show the residuals between successive resolutions: coarse–intermediate ($\phi_l - \phi_i$ or $v_{r,l} - v_{r,i}$) and intermediate–high ($\phi_i - \phi_h$ or $v_{r,i} - v_{r,h}$), respectively.}	
		\end{figure}

For numerical solutions of boson stars, we carry out a convergence test through the scalar function $\phi(x)$ with three different grid resolutions, characterized by the number of grid points $N$: coarse ($N_l=1000$), medium ($N_m=2000$), and fine ($N_h=4000$). The numerical resolution $h$ can be defined as $h=1/N$. As illustrated in Fig.~\ref{fig:conphi}, the calculated convergence factor $Q\approx4$, which indicates that the numerical error is well under control and the results obtained with 1000 grid points are reliable. According to Eq.~\ref{eq:order}, the convergence order is approximately two.

For the orbital integration part, although the fourth-order Runge–Kutta method employed in our code has a theoretical convergence order of four, the interpolation schemes used for orbit calculation affect the convergence behavior. We perform a convergence test for the radial velocity $v_r$. The three proper time steps we adopt are: coarse ($h_l$: $\delta\tau = 0.04$), intermediate ($h_i$: $\delta\tau = 0.02$), and high ($h_h$: $\delta\tau = 0.01$). Fig.~\ref{fig:conorbit} shows the evolution of the velocity $v_r$ between successive resolutions during the simulation duration. It can be seen that as the size of the steps is gradually refined, the residuals decrease systematically. The calculated maximum convergence factor $Q =\left|v_{r,l}-v_{r,i}\right|/\left|v_{r,i}-v_{r,h}\right|\approx 14$. This indicates that the convergence order of our code is $p \approx 3.8$.

\end{document}